\documentclass[twocolumn,twocolappendix]{aastex701}
\usepackage{amsmath}
\usepackage{lipsum}
\usepackage{footmisc}

\newdimen\digitwidth    
\setbox1=\hbox{0}       
\digitwidth=\wd1        
\catcode`"=\active      
\def"{\kern\digitwidth}

\newcommand{\ha}{\mbox{H$\alpha$}}
\newcommand{\hb}{\mbox{H$\beta$}}
\newcommand{\hg}{\mbox{H$\gamma$}}

\newcommand{\heiia}{He{\sevenrm\,II}\,$\lambda$4686}

\newcommand{\hei}{He{\sevenrm\,I}}
\newcommand{\heia}{He{\sevenrm\,I}\,$\lambda$5876}
\newcommand{\feii}{Fe{\sevenrm\,II}}

\newcommand{\OIIIb}{[O{\sevenrm\,III}]\,$\lambda$5007}
\newcommand{\OIIIc}{[O{\sevenrm\,III}]\,$\lambda\lambda$4959,\,5007}
\newcommand{\OIIId}{[O{\sevenrm\,III}]\,$\lambda$4363}
\newcommand{\OIII}{[O{\sevenrm\,III}]}

\newcommand{\NII}{[N{\sevenrm\,II}]}
\newcommand{\NIIa}{[N{\sevenrm\,II}]\,$\lambda$6548}
\newcommand{\NIIb}{[N{\sevenrm\,II}]\,$\lambda$6584}
\newcommand{\NIIc}{[N{\sevenrm\,II}]\,$\lambda\lambda$6548,\,6584}
\newcommand{\SII}{[S{\sevenrm\,II}]}

\newcommand{\SIIc}{[S{\sevenrm\,II}]\,$\lambda\lambda$6716,\,6731}
\newcommand{\OI}{[O{\sevenrm\,I}]}
\newcommand{\OIa}{[O{\sevenrm\,I}]\,$\lambda$6300}
\newcommand{\OIb}{[O{\sevenrm\,I}]\,$\lambda$6364}
\newcommand{\OIc}{[O{\sevenrm\,I}]\,$\lambda\lambda$6300,\,6364}
\newcommand{\HI}{H{\sevenrm\,I}}

\newcommand{\mbh}{\mbox{$M_\mathrm{BH}$}}

\newcommand{\rfeii}{\mbox{${R}_\mathrm{Fe\,II}$}}
\newcommand{\kms}{{\mbox{$\mathrm{km\,s^{-1}}$}}}
\newcommand{\Msun}{\mbox{$M_\odot$}}

\newcommand{\galspec}{\mbox{\texttt{GalSpec}}}
\newcommand{\cf}{\mbox{$C_f$}}

\newcommand{\taua}{\mbox{$\tau_{0,\mathrm{H}\alpha}$}}
\newcommand{\taub}{\mbox{$\tau_{0,\mathrm{H}\beta}$}}
\newcommand{\taug}{\mbox{$\tau_{0,\mathrm{H}\gamma}$}}
\newcommand{\ewa}{\mbox{$\mathrm{EW}_{\mathrm{H}\alpha}$}}
\newcommand{\ewb}{\mbox{$\mathrm{EW}_{\mathrm{H}\beta}$}}

\newcommand{\fwb}{\mbox{$\mathrm{FWHM}_{\mathrm{H}\beta}$}}

\newcommand{\ph}[1]{\phantom{#1}}

\newcommand{\lEdd}{$\lambda_{\rm Edd}$}

\font\sevenrm=cmr7 scaled 1000

\begin{document}

\title{The Extreme Rarity and Physical Properties of Low-redshift AGNs with Balmer Absorption}

\author[orcid=0000-0002-4569-9009,sname='Shangguan']{Jinyi Shangguan}
\affiliation{The Kavli Institute for Astronomy and Astrophysics, Peking University, Beijing 100871, China}
\email[show]{shangguan@pku.edu.cn}  
\author[orcid=0009-0003-4721-177X,sname='Chen']{Chang-Hao Chen}
\affiliation{Department of Astronomy, School of Physics, Peking University, Beijing 100871, People's Republic of China}
\email{}  
\author[orcid=0000-0001-6947-5846,sname='Ho']{Luis C. Ho}
\affiliation{The Kavli Institute for Astronomy and Astrophysics, Peking University, Beijing 100871, China}
\affiliation{Department of Astronomy, School of Physics, Peking University, Beijing 100871, People's Republic of China}
\email{}  
\author[orcid=0000-0001-6762-5599]{Jiwei Liao}
\affiliation{Department of Astronomy, School of Physics, Peking University, Beijing 100871, People's Republic of China}
\email{}  
\author[orcid=0009-0007-9075-5333]{Yanqing Liu}
\affiliation{Department of Astronomy, School of Physics, Peking University, Beijing 100871, People's Republic of China}
\email{}  
\author{Chengzhou Wu}
\affiliation{Department of Astronomy, School of Physics, Peking University, Beijing 100871, People's Republic of China}
\email{}  
\author[orcid=0000-0001-8496-4162]{Ruancun Li}
\affiliation{Max Planck Institute for extraterrestrial Physics, Giessenbachstra{\ss}e~1, 85748 Garching, Germany}
\email{}  
\author[orcid=0000-0001-9840-4959,sname='Inayoshi']{Kohei Inayoshi}
\affiliation{The Kavli Institute for Astronomy and Astrophysics, Peking University, Beijing 100871, China}
\email{}  
\author[orcid=0000-0003-4176-6486,sname='Jiang']{Linhua Jiang}
\affiliation{The Kavli Institute for Astronomy and Astrophysics, Peking University, Beijing 100871, China}
\email{}  

\begin{abstract}
Balmer absorption lines are increasingly observed in the little red dots (LRDs) discovered by the James Webb Space Telescope, potentially tracing dense circumnuclear gas around rapidly accreting black holes. Motivated by this connection, we search for Balmer absorption using homogeneously analyzed spectra of a representative parent sample of 14{,}584 low-redshift ($z<0.35$) type~1 active galactic nuclei selected from the Sloan Digital Sky Survey. We identify seven sources with robust Balmer absorption (occurrence $\sim 0.05\%$) and model them with a partially covering absorber model, accounting for the spectral resolution. By fitting \ha, \hb, and \hg\ simultaneously and tying their optical-depth ratios to theoretical values, we constrain optical depth at the line center ($\tau_0$) and the covering factor (\cf). All sources with robust modeling require optically thick \ha\ absorption and typically moderate covering factors ($C_f\approx 0.2-0.6$), while the LRD analog J1025 shows $C_f \gtrsim 0.8$ consistent with recent measurements of high-redshift LRDs. The absorbers have modest velocity offsets ($\sim 150-850$~km~s$^{-1}$) and narrow intrinsic widths ($\sim 20-200$~km~s$^{-1}$). Multi-epoch spectroscopy of three sources reveals Balmer-absorption variability on both year and month timescales. Three objects exhibit exceptionally weak \feii\ emission, high Eddington ratio, and low gas-phase metallicity, an atypically rare combination of properties that might elevate the incidence of Balmer-absorption to $\sim$10\%. We argue that low-metallicity conditions may suppress disk winds and help retain dense neutral gas along the line-of-sight in systems of high accretion rate.
\end{abstract}


\section{Introduction} 
\label{sec:intro}

Deep James Webb Space Telescope (JWST) spectroscopy has revealed a rapidly growing population of compact, optically red sources at $z\gtrsim 4$, often referred to as ``little red dots'' (LRDs). These objects show unusually red rest-frame UV/optical continua, frequently accompanied by a pronounced turnover near the Balmer limit in at least a subset of the population, as well as broad emission lines with full width at half maximum (FWHM) $\gtrsim 1000\ {\rm km\ s^{-1}}$ that are often treated as signatures of accretion-powered activity (e.g., \citealt{Kocevski2023,Labbe2023,Greene2024,Kokorev2024,Matthee2024}). Intriguingly, Balmer absorption troughs that appear on top of the broad emission lines have now been reported in 10\%--20\% \citep{Matthee2024,Lin2024,Kocevski2025} or more \citep{Zhuang2026,Chen2026} of the JWST-selected broad-line LRD/AGN samples. The superposition of absorption troughs over emission line profiles differs significantly from the stellar absorption features observed in evolved stellar populations, providing direct evidence for dense, predominantly neutral gas along the line-of-sight to the continuum and broad-line emission regions. A key theoretical development motivated by these discoveries is the idea that LRDs may host rapidly accreting black holes (BHs) embedded in a dense, optically thick gas envelope \citep{Inayoshi2025a,Kido2025,Rusakov2026,Inayoshi2025c}. In such scenarios, the envelope can simultaneously redden the emergent continuum, imprint Balmer break-like features and Balmer absorption, and—if coupled to powerful outflows—help explain the weak variability and X-ray faintness characteristic of LRDs \citep[e.g.,][]{Inayoshi2025b}.

Nonstellar Balmer absorption is exceptionally rare in broad-line AGNs and has largely been identified through serendipitous discoveries instead of systematic searches \citep{Hutchings2002,Aoki2006,Aoki2010,Hall2007,Wang2008, Ji2012,Ji2013,Wang2015,Zhang2015,Zhang2018,Schulze2018,Burke2021}. Physically, producing Balmer absorption requires a substantial population of neutral hydrogen in the $n=2$ level. Two mechanisms are commonly invoked. Pure collisional excitation demands extremely high densities, approaching a critical density of $n_{\rm crit}\approx10^{11}\,\mathrm{cm^{-3}}$ \citep[e.g.,][]{Hall2007}, which would imply very short recombination times (e.g., $\sim 40$~s for $T=10^4$~K) and would seem to make long-lived, readily detectable absorption unlikely. Alternatively, Ly$\alpha$ trapping can maintain an enhanced $n=2$ population at more moderate densities, provided that the column density is high; typical estimates require $N_{\rm H}(n=2)\gtrsim10^{17}\,\mathrm{cm^{-2}}$ \citep[e.g.,][]{Hall2007,Wang2015,Burke2021}. Consistent with the need for large column density, Balmer absorption is reported more frequently in iron low-ionization broad absorption-line quasars (FeLoBALQs; e.g., \citealt{Aoki2006, Hall2007,Aoki2010,Ji2012,Zhang2015}). Building on this connection, \cite{Leighly2025} compiled a sample of 14 FeLoBALQs with Balmer absorption and highlighted their potential similarity to LRDs, including a sub-population with weak \feii\ that more closely resembles the LRD phenomenology \citep{Trefoloni2025}.

At the same time, the emerging LRD context suggests that Balmer absorption may not be confined to classical FeLoBALQs and may instead trace a broader set of physical conditions linked to rapid accretion and circumnuclear gas structure. A particularly compelling example is J1025+1402: originally reported as a low-$z$ Balmer-absorption AGN \citep{Burke2021}, it has since been recognized as a local analog of LRDs \citep{Lin2026a,Ji2025b}. While its optical spectrum shows little or no detectable \feii\ pseudo-continuum, higher-quality LBT/MODS and Magellan/FIRE spectroscopy reveals abundant narrow \feii\ emission and absorption features \citep{Lin2026a}. Moreover, several Balmer-absorption AGNs are not classified as FeLoBALQs \citep[e.g.,][]{Hutchings2002,Wang2015,Schulze2018}, motivating a more unified view of the phenomenon and its diversity. This raises three closely related questions. How rare are Balmer-absorption AGNs within the overall low-$z$ type~1 population? How do their absorber properties relate to global AGN properties, particularly accretion rate, as suggested by recent LRD models \citep{Liu2025,Rusakov2026}? And what parameter space do they preferentially occupy? We address these questions by identifying a small sample of low-$z$ type~1 AGNs with robust Balmer absorption and performing uniform spectral decomposition and absorption-line modeling that enforces the atomic-physics constraints among \ha, \hb, and \hg, then placing the resulting measurements in the context of the parent AGN population to quantify both the incidence and the parameter-space preference of Balmer absorption.

The paper is organized as follows. Section~\ref{sec:sample} describes the parent sample, target selection, and spectroscopic and photometric data. Section~\ref{sec:fit} presents our continuum and emission-line decomposition and the absorber model, including the treatment of instrumental resolution. Section~\ref{sec:res} summarizes the main measurements, including absorber optical depths, covering factors, kinematics, and broad-band spectral energy distribution (SED) properties. In Section~\ref{sec:disc}, we discuss the number statistics and selection effects, the Balmer decrements of the broad and narrow components, the observed multi-epoch variability, and the physical implications for circumnuclear gas in high-accretion, potentially low-metallicity systems. We summarize our main conclusions in Section~\ref{sec:sum}. This work adopts the following parameters for a $\Lambda$CDM cosmology: $\Omega_m = 0.308$, $\Omega_\Lambda = 0.692$, and $H_{0}=67.8$ km s$^{-1}$ Mpc$^{-1}$ \citep{Planck2016}.

\begin{figure*}
\centering
\includegraphics[width=0.95\textwidth]{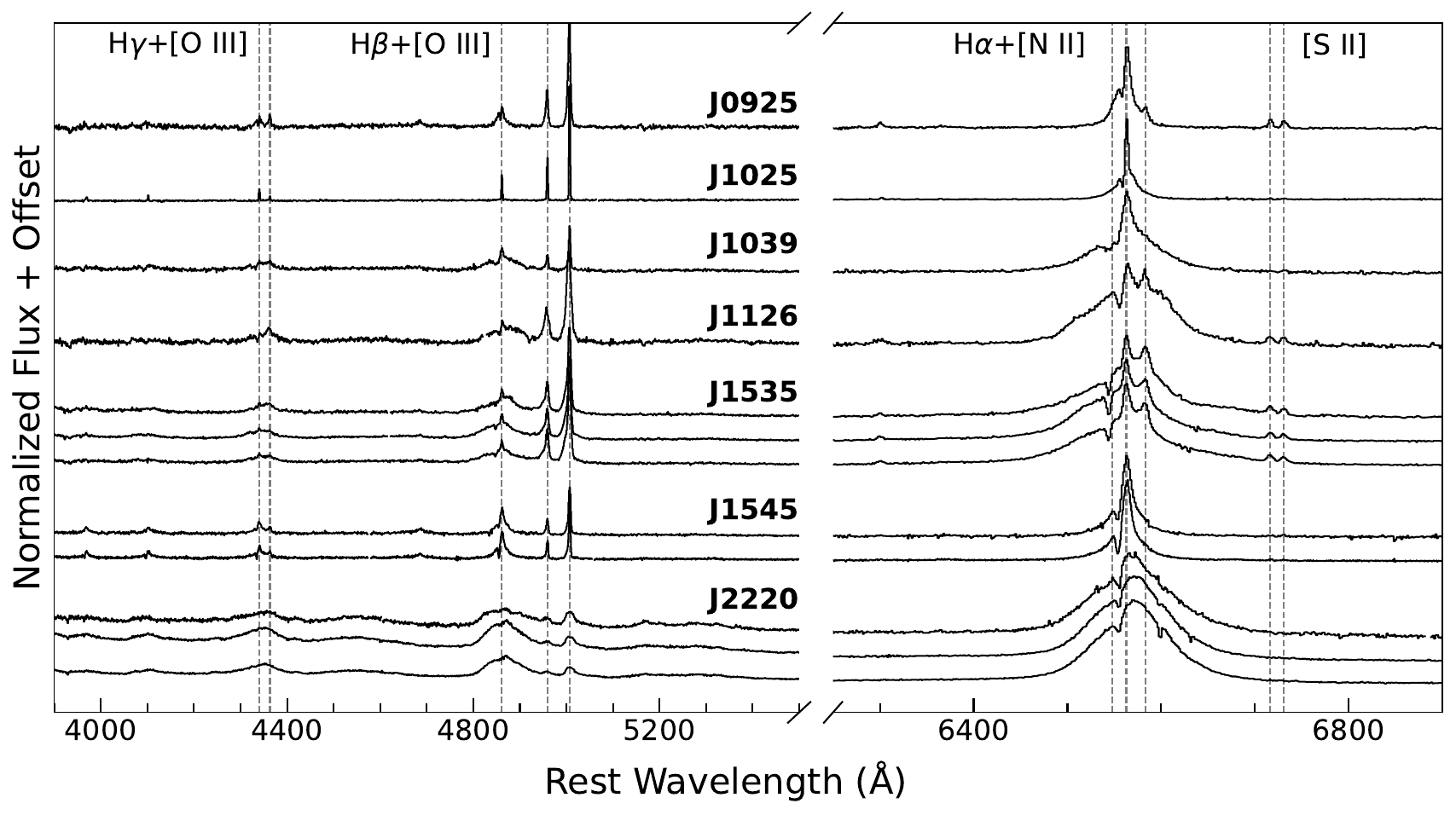}
\caption{Spectra of the selected AGNs (Table~\ref{tab:sample}), with the 
target names labeled at the center of each panel. For J1535 and J2220, three 
SDSS spectra are available, and DESI provides a second-epoch spectrum for J1545. 
The strongest Balmer and forbidden lines are labeled and indicated by vertical 
dashed lines. From short to long wavelengths, we mark \hg\ and \OIIId, \hb\ and 
\OIIIc, \ha\ and \NIIc, and \SIIc.}
\label{fig:spec}
\end{figure*}


%
%

\begin{deluxetable*}{cccccccccc}
\tabletypesize{\footnotesize}
\tablecaption{Properties of the Sample \label{tab:sample}}
\tablecolumns{10}
\tablewidth{0pt}
\tablehead{
\colhead{Name} &
\colhead{SDSS Name} &
\colhead{Date} &
\colhead{$z$} &
\colhead{$\log\,L_{5100}$} &
\colhead{$\log\,L_{\mathrm{H\alpha}}$} &
\colhead{$\mathrm{FWHM_{H\alpha}}$} &
\colhead{$\log\,M_{\mathrm{BH}}$} &
\colhead{\rfeii} &
\colhead{$\log \lambda_{\mathrm{Edd}}$} \\
\colhead{} &
\colhead{} &
\colhead{} &
\colhead{} &
\colhead{($\mathrm{erg\,s^{-1}}$)} &
\colhead{($\mathrm{erg\,s^{-1}}$)} &
\colhead{($\mathrm{km\,s^{-1}}$)} &
\colhead{($M_\odot$)} &
\colhead{} &
\colhead{}
}
\colnumbers
\startdata
J0925   & J092537.83+640921.7 & 2008-01-29 & 0.052956 & $42.66$ & $41.07$ & $1101 \pm 48$" & $5.88$ & $0.22 \pm 0.05$ &      $-0.31$ \\
J1025   & J102530.29+140207.4 & 2004-02-21 & 0.100957 & $42.87$ & $41.87$ & "$936 \pm 14$" & $6.17$ & $<0.08$         &      $-0.40$ \\
J1039   & J103939.31+100253.1 & 2003-04-05 & 0.161617 & $43.87$ & $42.63$ & $2047 \pm 187$ & $7.29$ & $0.26 \pm 0.02$ &      $-0.52$ \\
J1126   & J112611.63+425246.4 & 2004-02-27 & 0.156368 & $43.90$ & $42.73$ & $4501 \pm 51$" & $8.05$ & $0.53 \pm 0.03$ &      $-1.25$ \\
J1535-0 & J153539.25+564406.4 & 2001-06-12 & 0.207771 & $44.44$ & $43.24$ & $3627 \pm 50$" & $8.13$ & $0.21 \pm 0.01$ &      $-0.79$ \\
J1535-1 & J153539.25+564406.4 & 2013-05-17 & 0.207771 & $44.35$ & $43.20$ & $4270 \pm 37$" & $8.26$ & $0.24 \pm 0.01$ &      $-1.01$ \\
J1535-2 & J153539.25+564406.4 & 2018-04-17 & 0.207771 & $44.32$ & $43.19$ & $4005 \pm 45$" & $8.20$ & $0.27 \pm 0.02$ &      $-0.98$ \\
J1545-0 & J154511.30+223856.1 & 2005-07-05 & 0.218694 & $44.13$ & $42.77$ & "$712 \pm 55$" & $6.42$ & $0.21 \pm 0.01$ & $\ph{-}0.61$ \\
J1545-1 & J154511.30+223856.1 & 2021-06-07 & 0.218694 & $44.03$ & $42.73$ & "$660 \pm 23$" & $6.33$ & $0.23 \pm 0.01$ & $\ph{-}0.60$ \\
J2220-0 & J222024.59+010931.3 & 2001-08-19 & 0.212220 & $44.75$ & $43.39$ & $4941 \pm 48$" & $8.49$ & $0.79 \pm 0.03$ &      $-0.84$ \\
J2220-1 & J222024.59+010931.3 & 2017-09-18 & 0.212220 & $44.75$ & $43.40$ & $4324 \pm 31$" & $8.38$ & $0.59 \pm 0.01$ &      $-0.73$ \\
J2220-2 & J222024.59+010931.3 & 2017-11-10 & 0.212220 & $44.82$ & $43.63$ & $4155 \pm 45$" & $8.47$ & $0.54 \pm 0.03$ &      $-0.75$ \\
\enddata
\tablecomments{
Col. (1): Target name.
Col. (2): Full SDSS name.
Col. (3): Observation date of the spectrum.
Col. (4): Redshift derived from spectral fitting of the narrow emission lines.
Col. (5): Monochromatic continuum luminosity at 5100~\AA.
Col. (6): Broad H$\alpha$ luminosity; typical uncertainty of $L_{5100}$ and $L_{\mathrm{H\alpha}}$ from the spectral decomposition is $< 5\%$.
Col. (7): Broad H$\alpha$ FWHM.
Col. (8): Black hole mass based on $L_{\mathrm{H\alpha}}$ and $\mathrm{FWHM_{H\alpha}}$ (Table~3).
Col. (9): Equivalent width ratio of Fe\,{\sc ii} to H$\beta$.
Col. (10): Eddington ratio calculated from Cols. (5) and (8). 
}
\end{deluxetable*}

\section{Sample and Data}
\label{sec:sample}

The primary motivation of this work is to identify low-redshift, type~1 AGNs exhibiting Balmer absorption lines and to characterize their properties with respect to the general AGN population. We therefore begin with a representative low-$z$ broad-line AGN sample from \cite{Liu2019}, who provide a homogeneous parent catalog for subsequent inspection and comparisons.

The \cite{Liu2019} sample is particularly well suited for this purpose because it is based on careful spectral decomposition in the \ha\ and \hb\ regions, which are critical for measuring broad-line properties and for inferring AGN physical parameters. Unlike other Sloan Digital Sky Survey (SDSS) quasar catalogs \citep[e.g.,][]{Wu2022}, which are largely dominated by luminous, quasar-targeted objects, \cite{Liu2019} mined the full SDSS spectroscopic database of both galaxies and quasars at $z<0.35$ and selected AGNs by the presence of broad Balmer emission (especially \ha). This strategy recovered a substantial population of lower luminosity type~1 AGNs that can be missed or diluted in quasar-style selections. As a result, the catalog spans parameter ranges down to very low broad-line luminosities and Eddington ratios, and thus constitutes a more complete census of broad-line AGN activity in the low-$z$ Universe than a quasar-only subset. This point is essential for our search, because Balmer absorption-line AGNs may preferentially reside outside the luminous quasar regime, in analogy to the low-redshift counterparts of LRDs. To characterize the broad-band SEDs of the selected targets, we use the integrated GALEX \citep{Martin2005}, SDSS, 2MASS \citep{Skrutskie2006}, and WISE \citep{Wright2010} photometry compiled by \cite{Liu2019} (Section~\ref{ssec:sed}).

Balmer absorption features in these systems are typically very narrow and are likely unresolved in SDSS spectra, which have a velocity resolution of $\sigma < 70$~\kms. Consequently, the absorption signature can be confined to only a few spectral pixels. With moderate signal-to-noise ratios, this makes it challenging to identify such features using fully automated approaches, for example via metrics such as the ``Balnicity index'' \citep{Weymann1991} or the ``absorption index'' \citep{Hall2002}. We therefore perform, instead, a visual inspection of the SDSS spectra in the \cite{Liu2019} parent sample. In practice, we primarily focus on absorption signatures on the blue wing of \ha, which is the strongest Balmer line and thus provides the most sensitive diagnostic. The key requirement is to distinguish genuine absorption from the local trough produced by the nearby, blended \NIIa\ emission line. This is generally straightforward, because the \NIIb\ line is a factor of 3 stronger than \NIIa,  and the doublet wavelength separation is precisely known. Among the 14{,}584 type~1 AGNs, we identified seven sources that exhibit robust Balmer absorption in \ha\ and in other broad Balmer lines. Their basic information is listed in Table~\ref{tab:sample}.

We searched for multi-epoch spectroscopy of the seven targets using the SPectra Analysis and Retrievable Catalog Lab \citep[SPARCL;][]{Juneau2025}. Only single-epoch spectra are available for J0925, J1025, J1039, and J1126, whereas J1535 and J2220 were observed 3 times by SDSS in 2001, 2013, and 2018, and J1545 was observed by SDSS in 2005 and then in 2021 by the Dark Energy Spectroscopic Instrument \citep[DESI;][]{DESI2026}. The available spectra are shown in Figure~\ref{fig:spec}. In addition to Balmer absorption on the blue wings of the broad Balmer emission lines, we also detect prominent, blueshifted \hei\ $\lambda3889$ absorption in J1039, J1126, J1535, J1545, and J2220. Neutral helium absorption arising from the metastable level, most notably the \hei\ multiplets at $\lambda\lambda3189,\,3889,\,10830$, provides a powerful diagnostic of the absorber’s geometry and physical conditions \citep[e.g.,][]{Liu2015}. A detailed analysis, however, is beyond the scope of this work.

\begin{figure*}
\centering
\includegraphics[width=0.95\textwidth]{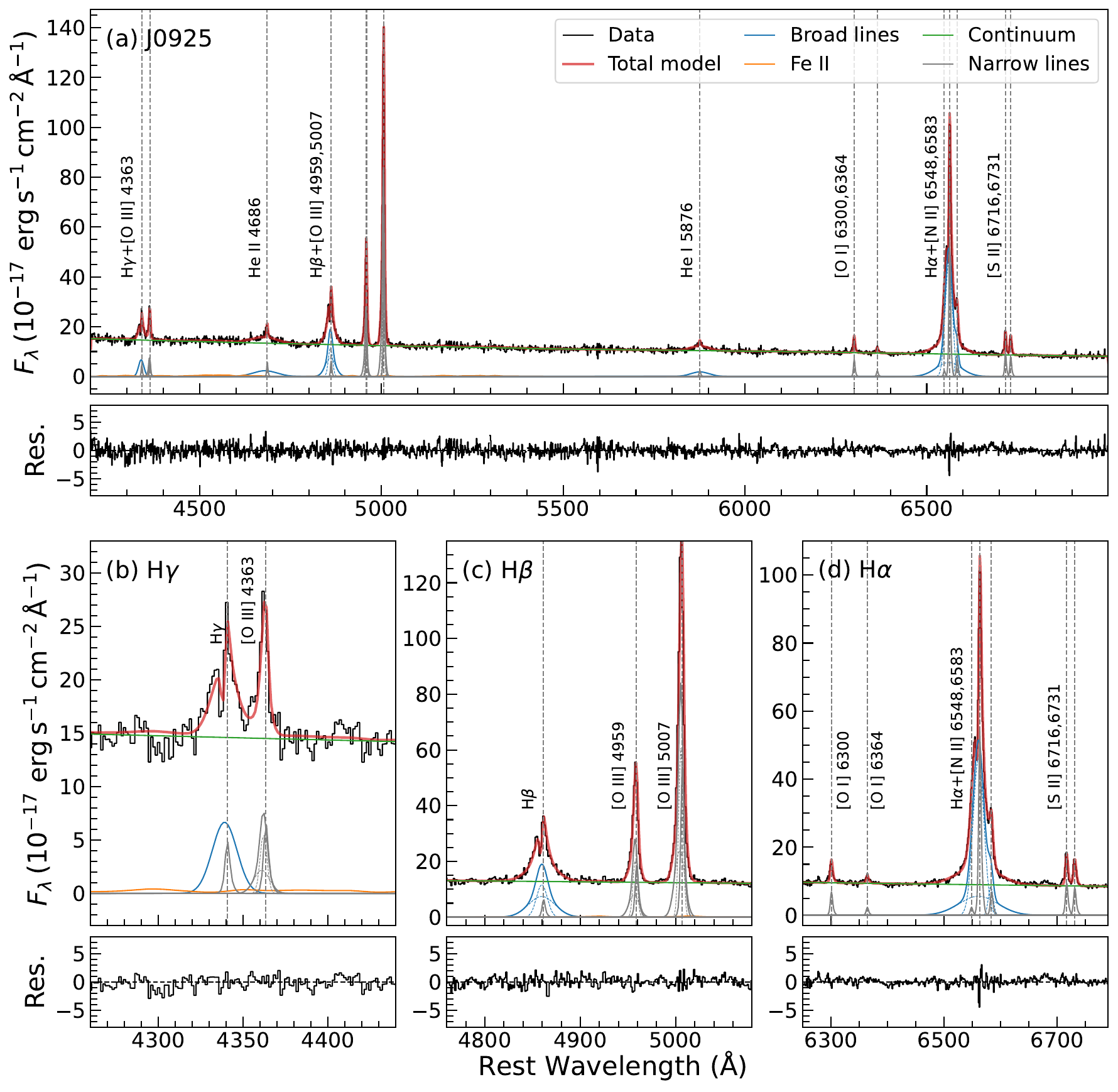}
\caption{Spectral decomposition of J0925 for (a) the full spectrum and the regions around (b) \hg, (c) \hb, and (d) \ha. The data and the total model is in black and red, respectively. The broad emission lines are shown by colored curves. The residuals between the data and model are shown in the lower panels. Only the fitted regions are displayed in panel (a), while the lower three panels zoom into the three Balmer line regions.  The absorption lines are only applied to the total model. The fitting results of the other spectra are provided in the online figure set.}
\label{fig:fit0925}
\end{figure*}

\figsetstart
\figsetnum{2}
\figsettitle{Spectral decomposition of the Balmer Absorption AGNs}

\figsetgrpstart
\figsetgrpnum{2.1}
\figsetgrptitle{Spectral decomposition of J0925}
\figsetplot{fig2_0.pdf}
\figsetgrpnote{(a) the full spectrum and the regions around (b) Hg, (c) Hb, and (d) Ha. The data and the total model is in black and red, respectively. The broad emission lines are shown by colored curves. The residuals between the data and model are shown in the lower panels. Only the fitted regions are displayed in panel (a), while the lower three panels zoom into the three Balmer line regions.  The absorption lines are only applied to the total model. The fitting results of the other spectra are provided in the online figure set.}
\figsetgrpend

\figsetgrpstart
\figsetgrpnum{2.2}
\figsetgrptitle{Spectral decomposition of J1025}
\figsetplot{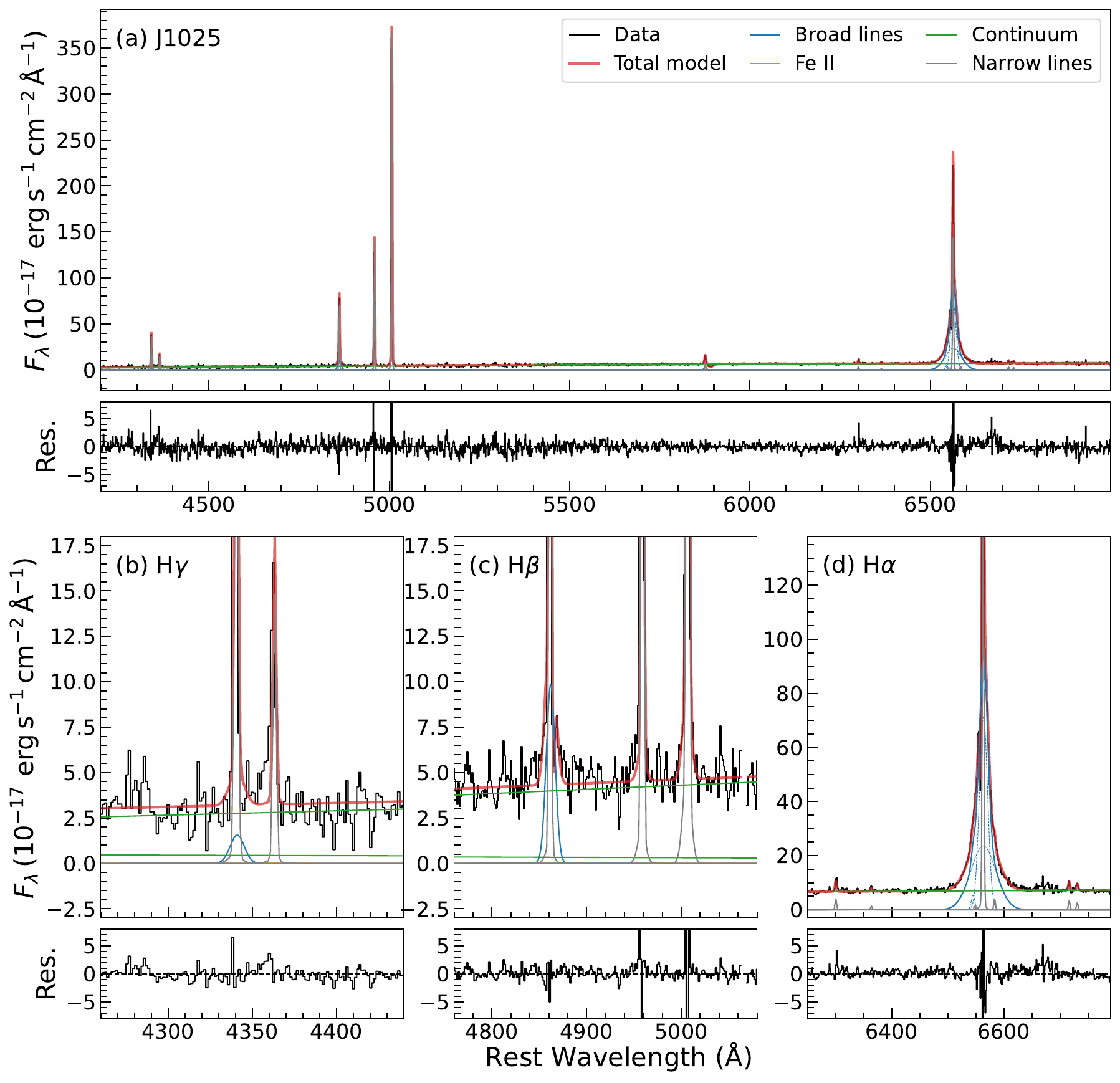}
\figsetgrpnote{(a) the full spectrum and the regions around (b) Hg, (c) Hb, and (d) Ha. The data and the total model is in black and red, respectively. The broad emission lines are shown by colored curves. The residuals between the data and model are shown in the lower panels. Only the fitted regions are displayed in panel (a), while the lower three panels zoom into the three Balmer line regions.  The absorption lines are only applied to the total model. The fitting results of the other spectra are provided in the online figure set.}
\figsetgrpend

\figsetgrpstart
\figsetgrpnum{2.3}
\figsetgrptitle{Spectral decomposition of J1039}
\figsetplot{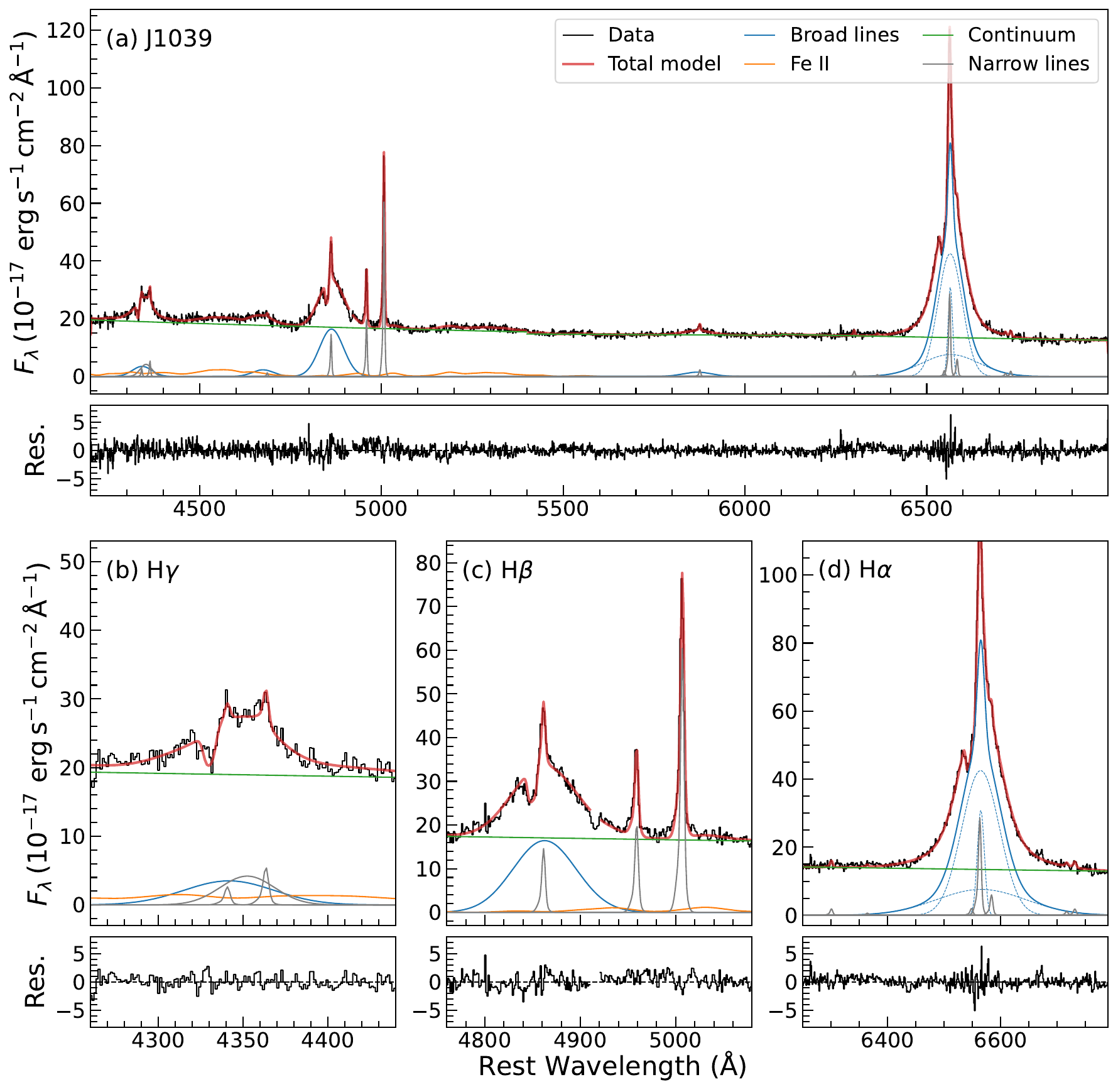}
\figsetgrpnote{(a) the full spectrum and the regions around (b) Hg, (c) Hb, and (d) Ha. The data and the total model is in black and red, respectively. The broad emission lines are shown by colored curves. The residuals between the data and model are shown in the lower panels. Only the fitted regions are displayed in panel (a), while the lower three panels zoom into the three Balmer line regions.  The absorption lines are only applied to the total model. The fitting results of the other spectra are provided in the online figure set.}
\figsetgrpend

\figsetgrpstart
\figsetgrpnum{2.4}
\figsetgrptitle{Spectral decomposition of J1126}
\figsetplot{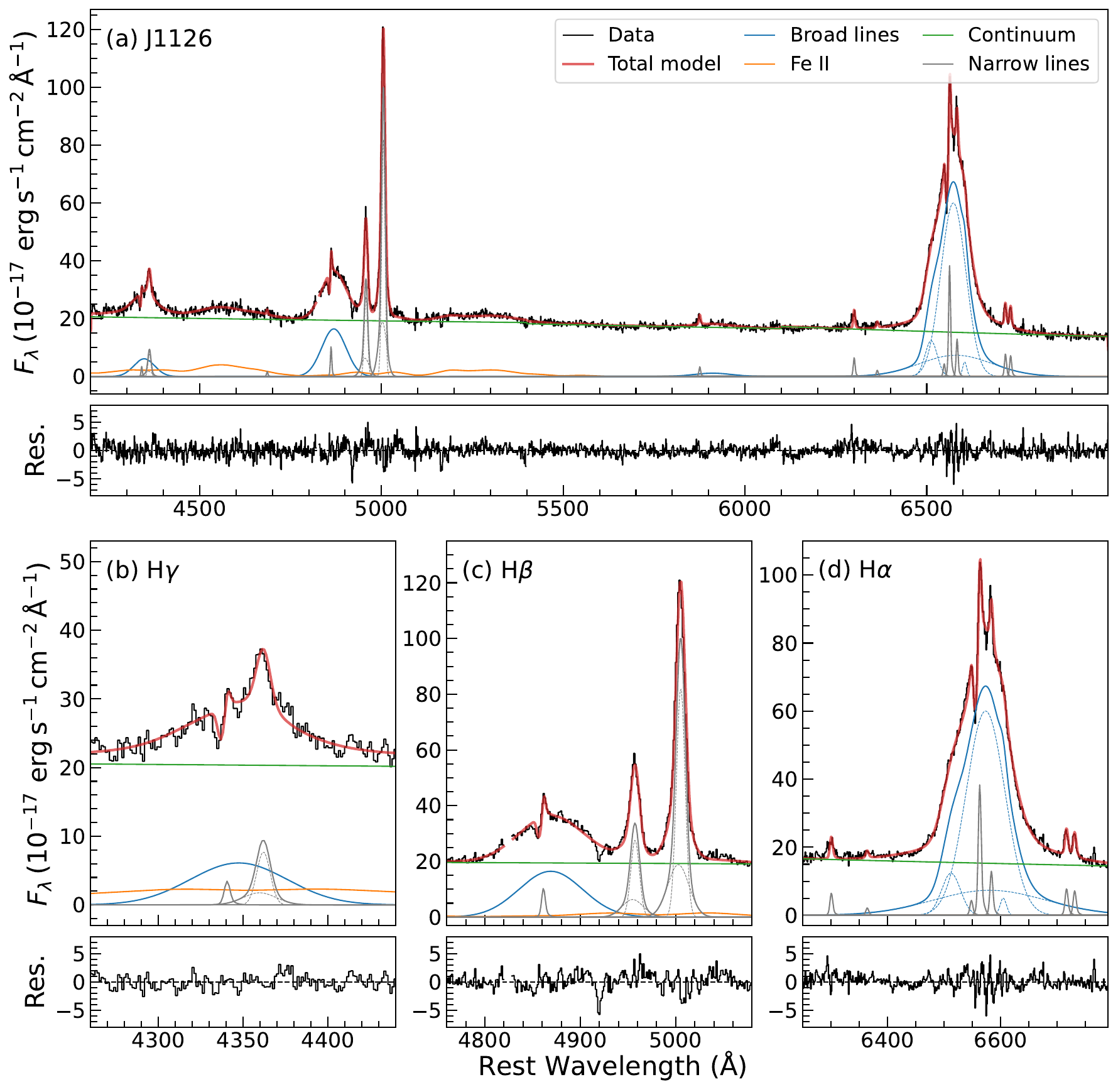}
\figsetgrpnote{(a) the full spectrum and the regions around (b) Hg, (c) Hb, and (d) Ha. The data and the total model is in black and red, respectively. The broad emission lines are shown by colored curves. The residuals between the data and model are shown in the lower panels. Only the fitted regions are displayed in panel (a), while the lower three panels zoom into the three Balmer line regions.  The absorption lines are only applied to the total model. The fitting results of the other spectra are provided in the online figure set.}
\figsetgrpend

\figsetgrpstart
\figsetgrpnum{2.5}
\figsetgrptitle{Spectral decomposition of J1535-0}
\figsetplot{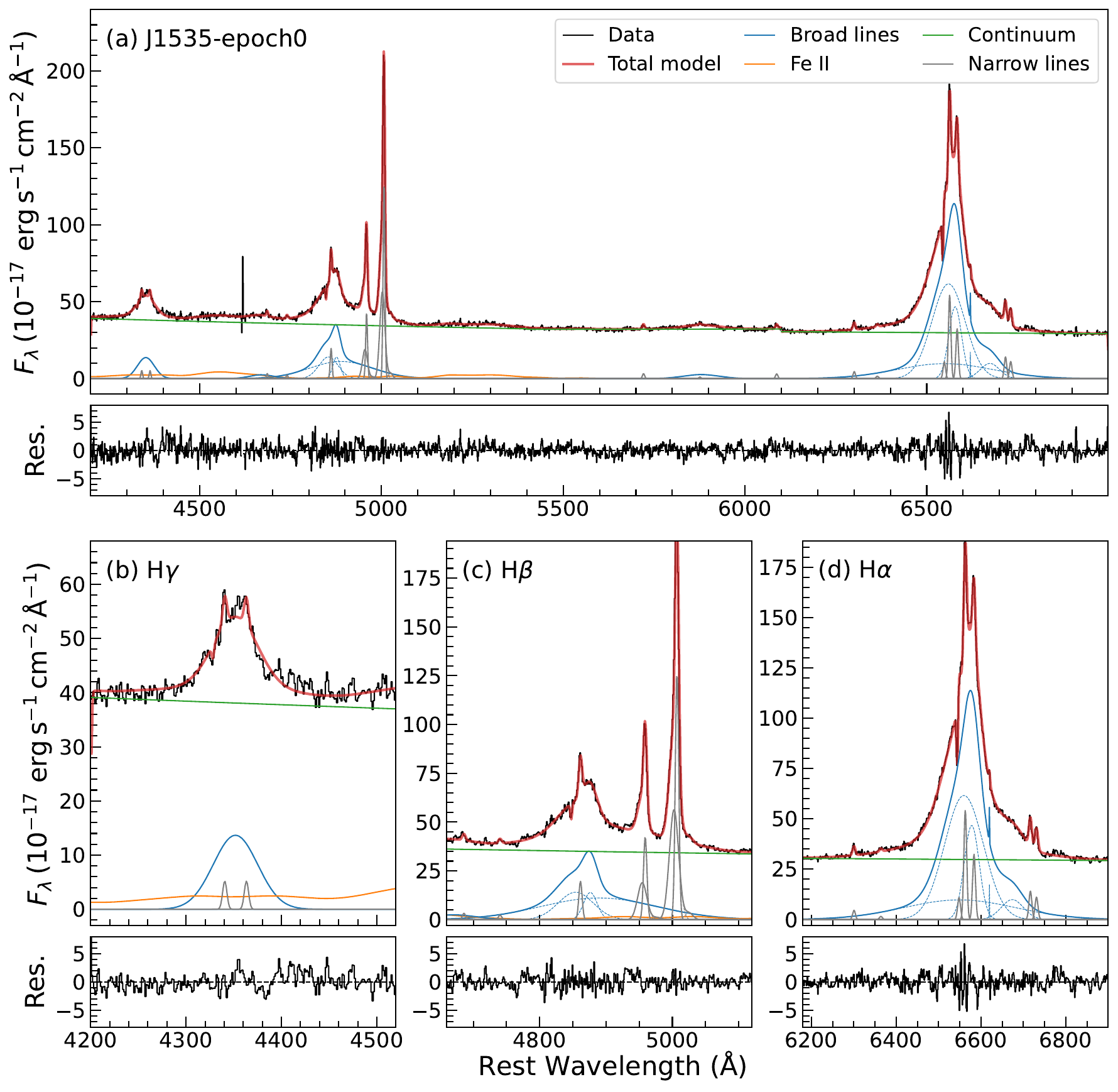}
\figsetgrpnote{(a) the full spectrum and the regions around (b) Hg, (c) Hb, and (d) Ha. The data and the total model is in black and red, respectively. The broad emission lines are shown by colored curves. The residuals between the data and model are shown in the lower panels. Only the fitted regions are displayed in panel (a), while the lower three panels zoom into the three Balmer line regions.  The absorption lines are only applied to the total model. The fitting results of the other spectra are provided in the online figure set.}
\figsetgrpend

\figsetgrpstart
\figsetgrpnum{2.6}
\figsetgrptitle{Spectral decomposition of J1535-1  }
\figsetplot{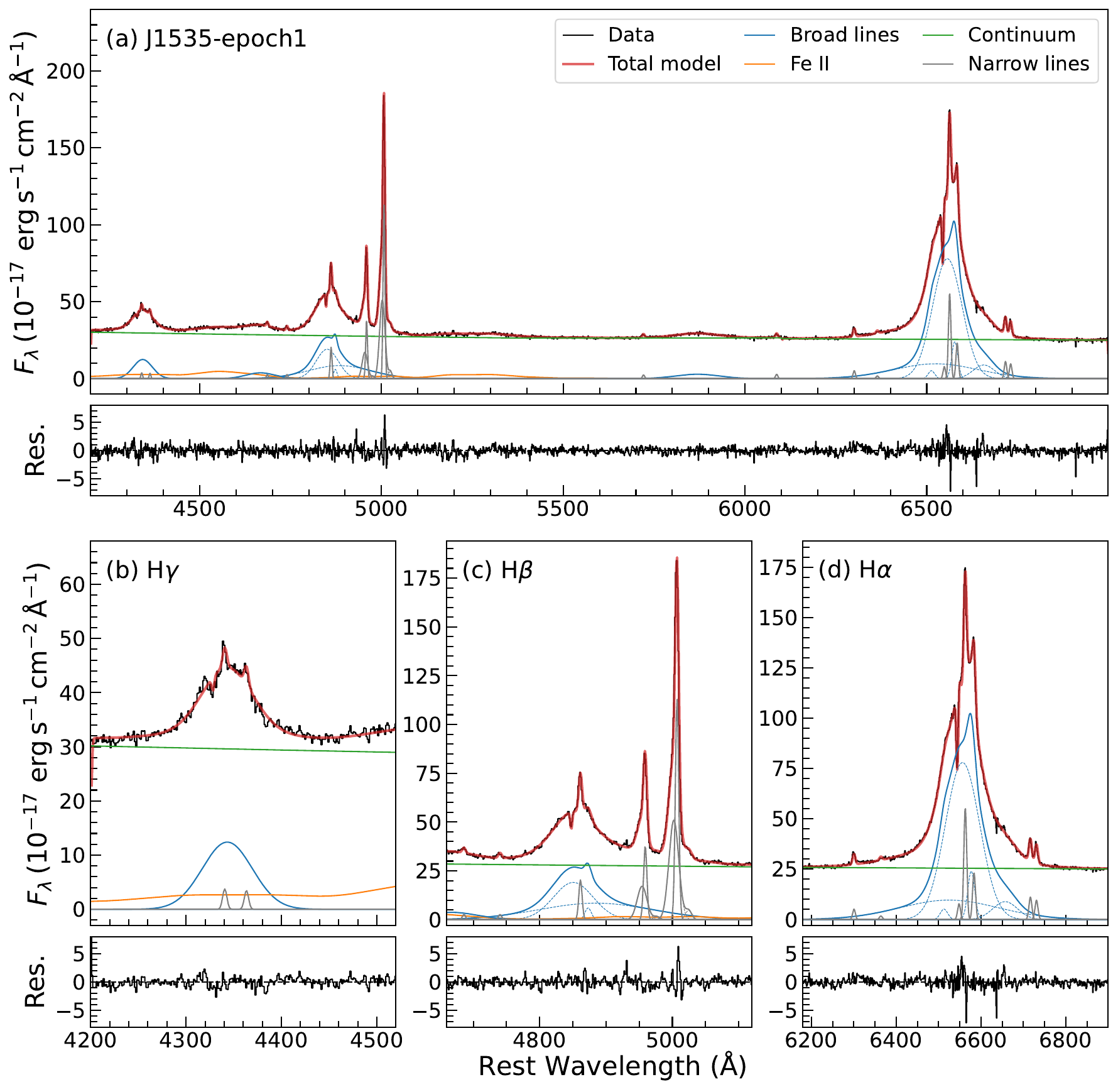}
\figsetgrpnote{(a) the full spectrum and the regions around (b) Hg, (c) Hb, and (d) Ha. The data and the total model is in black and red, respectively. The broad emission lines are shown by colored curves. The residuals between the data and model are shown in the lower panels. Only the fitted regions are displayed in panel (a), while the lower three panels zoom into the three Balmer line regions.  The absorption lines are only applied to the total model. The fitting results of the other spectra are provided in the online figure set.}
\figsetgrpend

\figsetgrpstart
\figsetgrpnum{2.7}
\figsetgrptitle{Spectral decomposition of J1535-2  }
\figsetplot{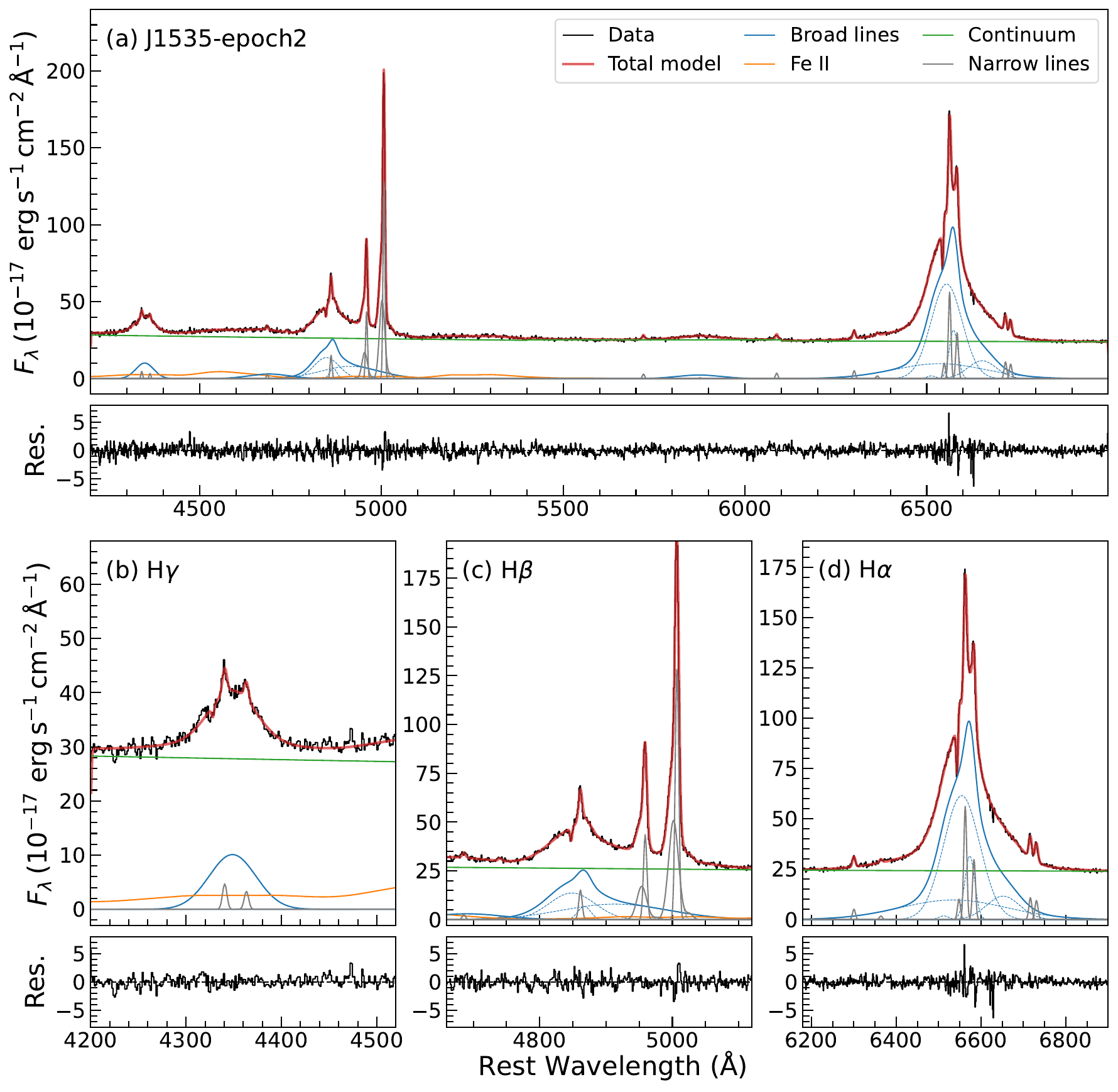}
\figsetgrpnote{(a) the full spectrum and the regions around (b) Hg, (c) Hb, and (d) Ha. The data and the total model is in black and red, respectively. The broad emission lines are shown by colored curves. The residuals between the data and model are shown in the lower panels. Only the fitted regions are displayed in panel (a), while the lower three panels zoom into the three Balmer line regions.  The absorption lines are only applied to the total model. The fitting results of the other spectra are provided in the online figure set.}
\figsetgrpend

\figsetgrpstart
\figsetgrpnum{2.8}
\figsetgrptitle{Spectral decomposition of J1545-0}
\figsetplot{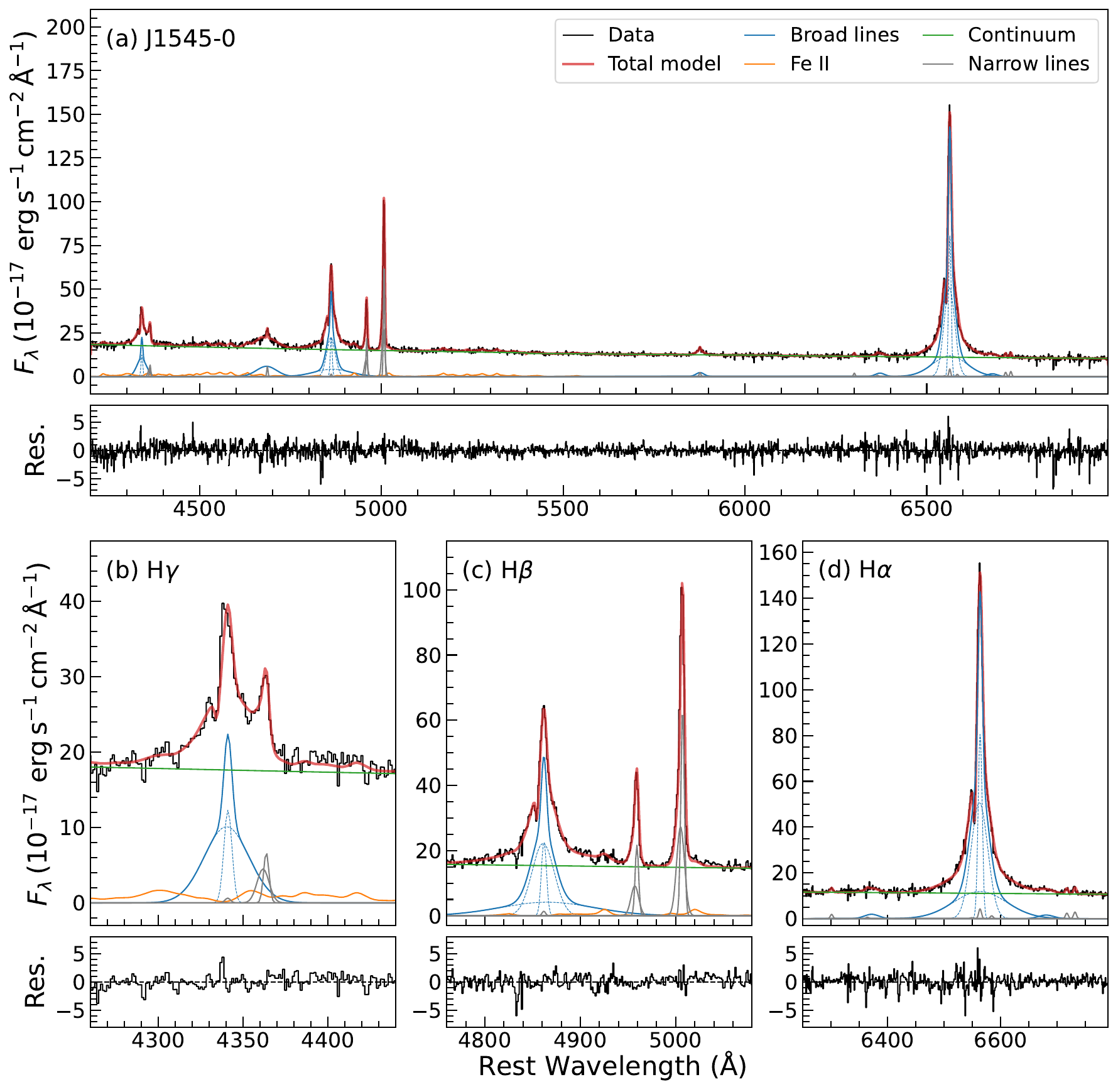}
\figsetgrpnote{(a) the full spectrum and the regions around (b) Hg, (c) Hb, and (d) Ha. The data and the total model is in black and red, respectively. The broad emission lines are shown by colored curves. The residuals between the data and model are shown in the lower panels. Only the fitted regions are displayed in panel (a), while the lower three panels zoom into the three Balmer line regions.  The absorption lines are only applied to the total model. The fitting results of the other spectra are provided in the online figure set.}
\figsetgrpend

\figsetgrpstart
\figsetgrpnum{2.9}
\figsetgrptitle{Spectral decomposition of J1545-1}
\figsetplot{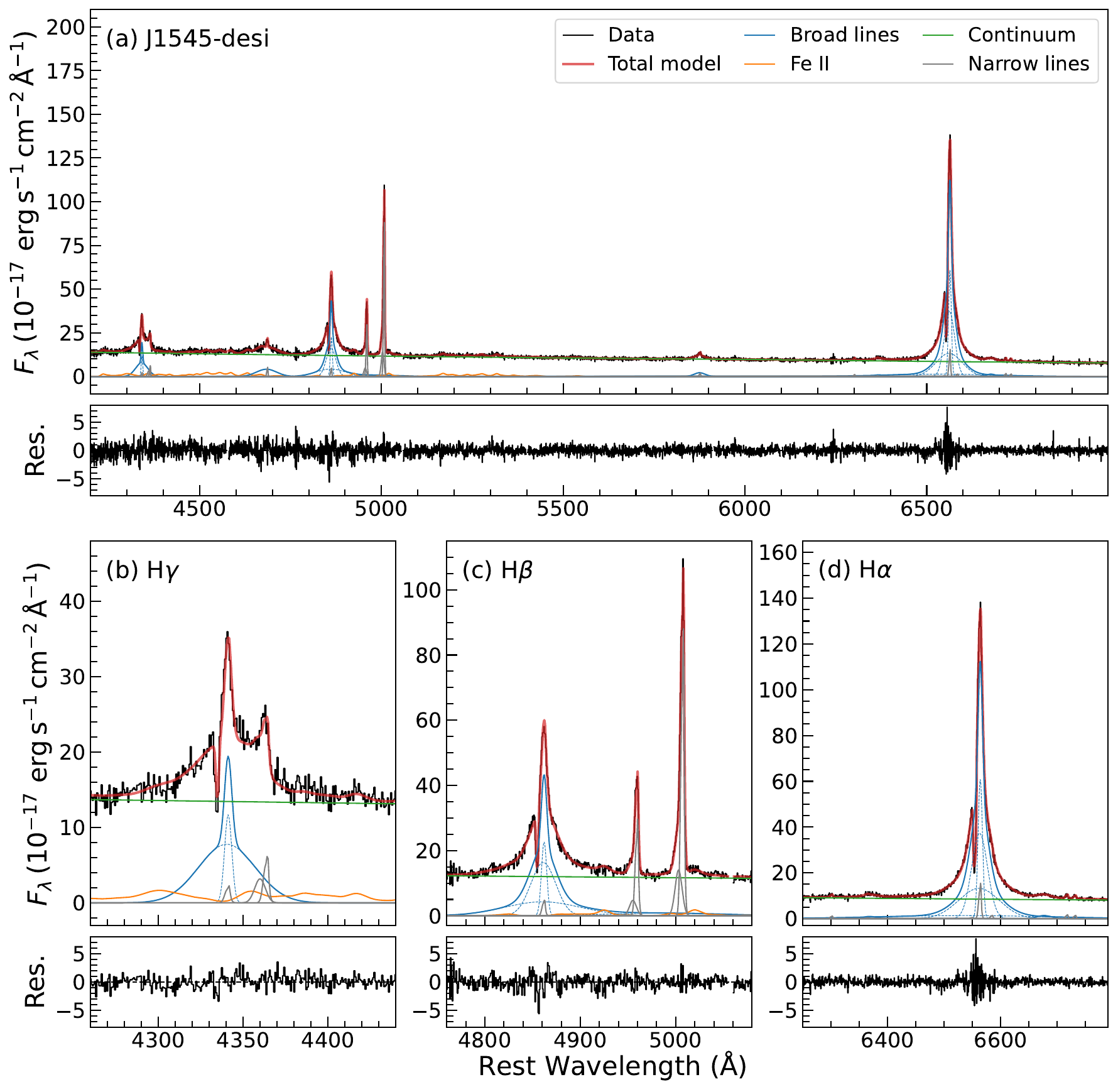}
\figsetgrpnote{(a) the full spectrum and the regions around (b) Hg, (c) Hb, and (d) Ha. The data and the total model is in black and red, respectively. The broad emission lines are shown by colored curves. The residuals between the data and model are shown in the lower panels. Only the fitted regions are displayed in panel (a), while the lower three panels zoom into the three Balmer line regions.  The absorption lines are only applied to the total model. The fitting results of the other spectra are provided in the online figure set.}
\figsetgrpend

\figsetgrpstart
\figsetgrpnum{2.10}
\figsetgrptitle{Spectral decomposition of J2220-0}
\figsetplot{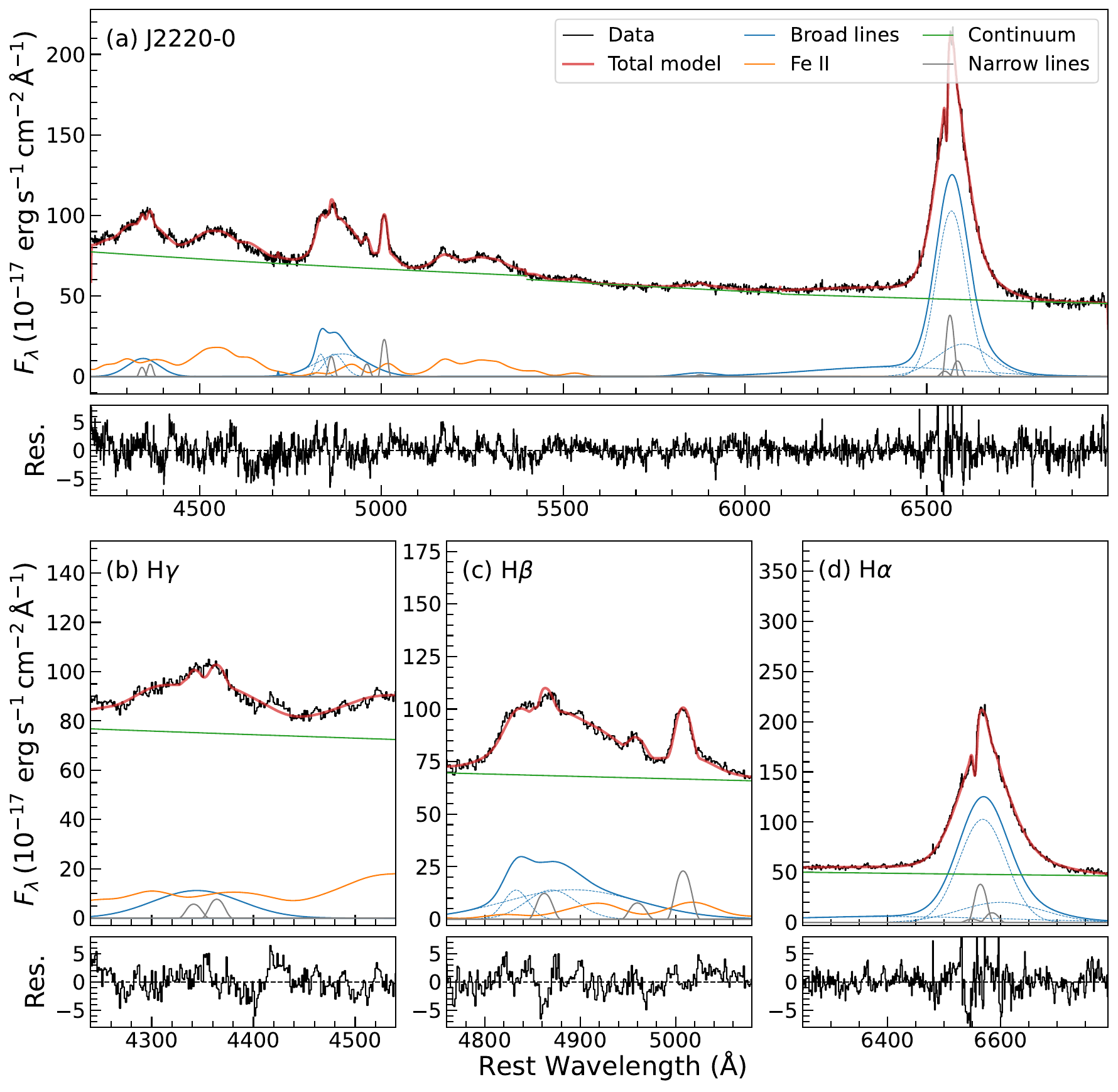}
\figsetgrpnote{(a) the full spectrum and the regions around (b) Hg, (c) Hb, and (d) Ha. The data and the total model is in black and red, respectively. The broad emission lines are shown by colored curves. The residuals between the data and model are shown in the lower panels. Only the fitted regions are displayed in panel (a), while the lower three panels zoom into the three Balmer line regions.  The absorption lines are only applied to the total model. The fitting results of the other spectra are provided in the online figure set.}
\figsetgrpend

\figsetgrpstart
\figsetgrpnum{2.11}
\figsetgrptitle{Spectral decomposition of J2220-1}
\figsetplot{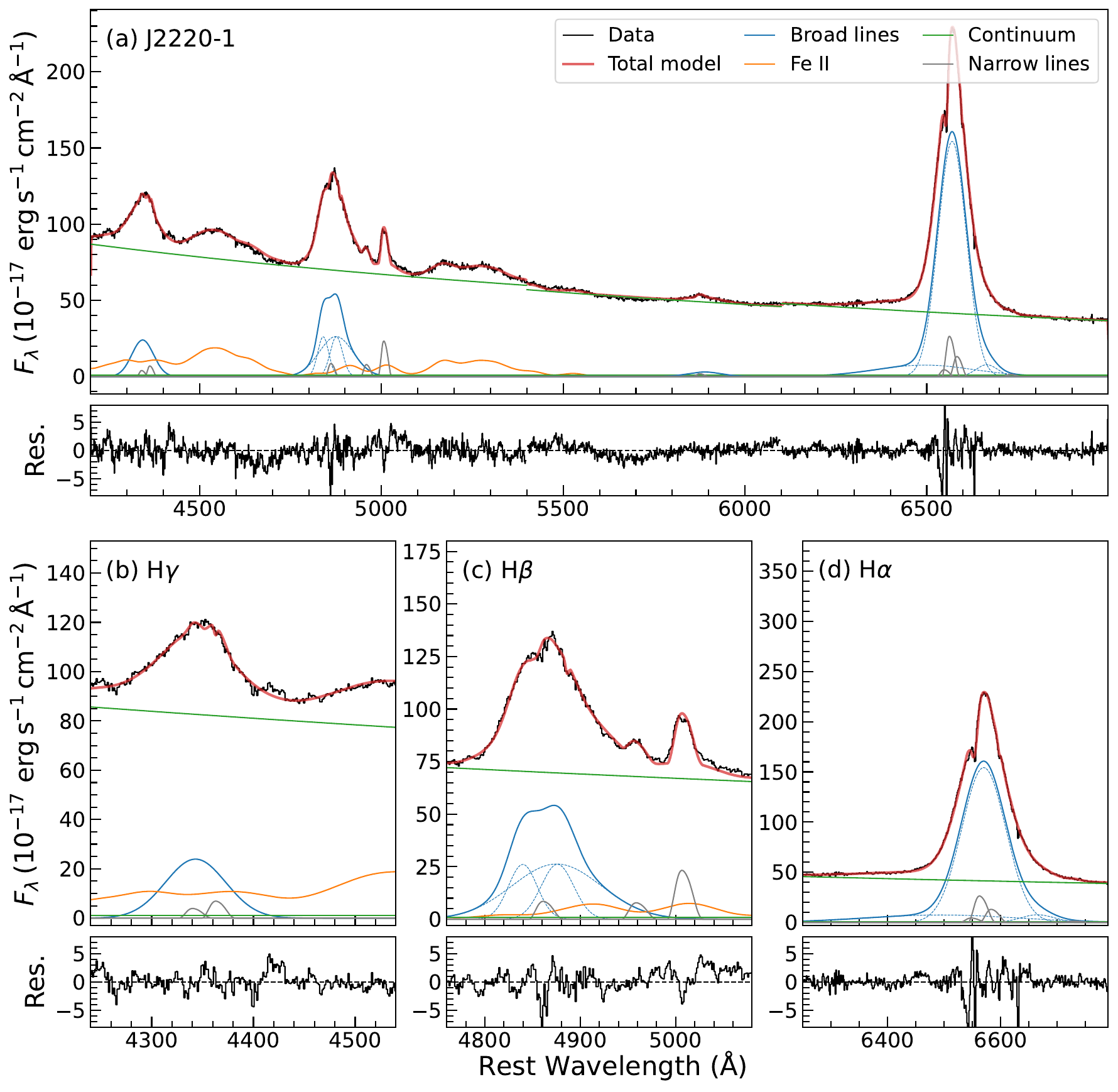}
\figsetgrpnote{(a) the full spectrum and the regions around (b) Hg, (c) Hb, and (d) Ha. The data and the total model is in black and red, respectively. The broad emission lines are shown by colored curves. The residuals between the data and model are shown in the lower panels. Only the fitted regions are displayed in panel (a), while the lower three panels zoom into the three Balmer line regions.  The absorption lines are only applied to the total model. The fitting results of the other spectra are provided in the online figure set.}
\figsetgrpend

\figsetgrpstart
\figsetgrpnum{2.12}
\figsetgrptitle{Spectral decomposition of J2220-2}
\figsetplot{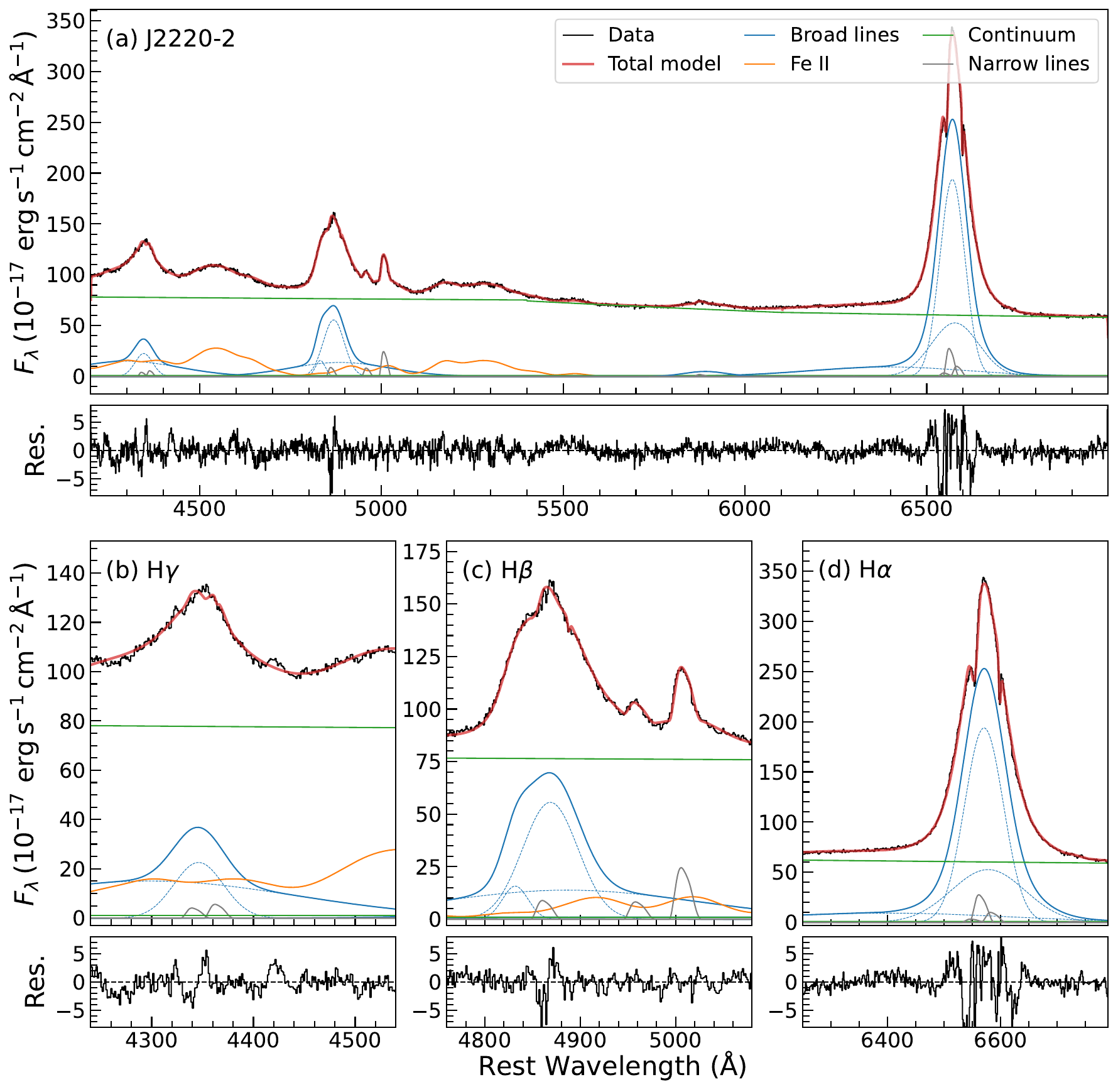}
\figsetgrpnote{(a) the full spectrum and the regions around (b) Hg, (c) Hb, and (d) Ha. The data and the total model is in black and red, respectively. The broad emission lines are shown by colored curves. The residuals between the data and model are shown in the lower panels. Only the fitted regions are displayed in panel (a), while the lower three panels zoom into the three Balmer line regions.  The absorption lines are only applied to the total model. The fitting results of the other spectra are provided in the online figure set.}
\figsetgrpend

\figsetend

\section{Spectral Decomposition} 
\label{sec:fit}

We first summarize our spectral-fitting procedure before discussing individual sources. After correcting for Galactic foreground extinction, we shift each spectrum to the rest frame. We adopt the SDSS pipeline redshift \citep{Ahumada2020} as an initial guess and then refine it using our spectral fits. All modeling is performed with our custom package \galspec,\footnote{\url{https://github.com/jyshangguan/GalSpec}; The code was originally referred to as SAGAN in \citet{Santos2025}.} which provides multi-Gaussian emission-line models and continuum components compatible with the \texttt{Astropy} modeling framework \citep{Astropy2022,Kuhn2024,Santos2025}. Our line models are parameterized in velocity space (line width and centroid shift), which facilitates tying kinematic parameters among different transitions. A description of the adopted components is provided in Appendix~\ref{apx:galspec}. The \galspec\ code also includes an MCMC fitting interface \citep[see also][]{Li2026} based on \texttt{emcee} \citep{Foreman-Mackey2013}. In practice, we first obtain a physically plausible solution using a Levenberg--Marquardt least-squares fit and then initialize the MCMC chains around the best-fit parameters to estimate uncertainties. We use uniform weights for the least-squares stage, as the noise is approximately constant across the fitted wavelength windows; for the MCMC stage we use the SDSS/DESI inverse variance spectra. While the best-fit emission-line parameters from the two methods agree within the quoted uncertainties, MCMC sampling provides a more robust way to explore the weak emission and absorption line parameters and to quantify the associated model uncertainties. Figure~\ref{fig:fit0925} shows an example fit to J0925; analogous fits for the remaining sources are provided in the online figure set.

\subsection{Continuum Fitting} \label{ssec:cont}

As in previous works \citep[e.g.,][]{Greene2005,Ho2009}, we first fit the continuum emission by combining three model components: the galaxy stellar emission, the AGN accretion disk, and the \feii\ pseudo-continuum. Following \cite{Kong2018}, we linearly combine the spectra of red giant stars (F, G, K, and M; luminosity class~III) and an A-type dwarf star (luminosity class~V) to model the galaxy stellar emission. We adopt a power-law model for the accretion disk and a newly developed \feii\ template by \cite{Park2022} for the pseudo-continuum. The fit is performed over spectral windows (4200--4300, 4430--4560, 5060--5400, 5600--5700, 6180--6230, 6800--7000, and 7500--8000~\AA) that avoid strong emission lines but capture the three strongest Balmer lines of interest for this study. Only J0925, J1039, and J1126 show a moderately significant level of stellar emission, as judged by the least-squared fitting and the presence of detectable stellar absorption features (e.g., Ca\,{\sc ii} H+K and Mg\,Ib), whereas the best-fit contribution from stars is negligible for the other four targets. Therefore, the stellar component is included only in the fits for J0925, J1039, and J1126. We do not remove the nonstellar continuum emission from the total spectrum in this step because the Balmer absorption also applies to the continuum \citep[e.g.,][]{Brazzini2025}. Following \cite{Lin2026a}, we also add a blackbody component to account for non-stellar continuum emission rising at $\gtrsim 4000$~\AA\ when fitting the spectrum of the local LRD analog J1025.

\subsection{Emission-line Fitting} \label{ssec:eline}

We focus on the wavelength range 4200--7000~\AA\ for the emission-line fitting (Figure~\ref{fig:fit0925}a). It is necessary to build a narrow-line model using isolated forbidden lines (e.g., the \OIIIc\ or \SIIc\ doublets) in order to decompose the often severely blended narrow and broad components of the Balmer lines \citep{Ho1997}. The \SII\ doublet is usually preferred because \OIII\ often shows a stronger asymmetric, blueshifted profile, likely associated with radial motions in the narrow-line region. In practice, however, the weakness of \SII\ leaves little alternative but \OIII\ as the only viable empirical narrow-line template. 

Both J0925 and J1126 show prominent \SII\ emission. We model the \SII\ doublet by fitting each line with a double-Gaussian profile, allowing the relative amplitudes of the two Gaussian components to vary and simultaneously fitting the local continuum over 6680--6780~\AA. In contrast, the \SII\ lines are only marginally detected in J1025, J1039, J1545, and J2220, and in J1535 they are strongly blended with its exceptionally broad \ha\ line. Therefore, for these five objects, we use the \OIII\ doublet, fit over 4900--5050~\AA, to construct the narrow-line template. Each \OIII\ line is modeled with either a multi-Gaussian profile (2--3 components) or a Gauss--Hermite function, with constant relative amplitude of 2.98 \citep{Osterbrock1989}. We adopt either the full \OIII\ profile (J1025, J1039, and J2220) or only the core component centered at $0~\kms$ (J1535 and J1545) as the narrow-line template. When modeling the (forbidden and permitted) narrow emission lines, we tie the centroid velocity shift across all narrow lines and allow only their amplitudes to vary. In addition to the \OIII\ constraint, we fix the peak-flux ratios of \NIIb/\NIIa\ to 2.96 and \OIb/\OIa\ to 3 \citep{Osterbrock1989}. For J0925, J1126, J1535, and J1545, we include an additional blueshifted ``wind'' component for both \OIIIc\ and \OIIId, and we tie the profile of the \OIIId\ wind component to that of \OIIIc.

Strictly speaking, narrow-line profiles depend on the local physical conditions and can vary in the presence of density stratification, since the gas density is often centrally concentrated \citep{Ho1996}. For example, the \SIIc\ doublet may show different profiles because the two lines arise from different upper levels. Likewise, \OIIId\ can differ from the \OIIIc\ doublet, whereas the two \OIII\ components originate from the same upper level and therefore should share the same intrinsic profile. In practice, however, such differences are typically detectable only in spectra with exceptionally high signal-to-noise ratios \citep{Filippenko1984,Filippenko1985}. We therefore adopt the common and practical approximation that all narrow forbidden lines share the same profile and, when needed, all the wind components share the same profile.

To decompose the spectra, we divide the full wavelength coverage into three segments: 4200--5400~\AA, 5400--6100~\AA, and 6100--7000~\AA. This piecewise approach is necessary because the wavelength range is broad and the continuum generally cannot be represented by a single power law. We begin with the \ha\ region (6100--7000~\AA), since \ha\ is typically the strongest line and exhibits the most complex broad-line profile. We fit the broad \ha\ component with a sum of Gaussian models, starting with a single component and adding additional components until the fit no longer improves, as judged by the residuals becoming flat. In most cases, 2--3 Gaussians suffice, whereas up to five components are required to reproduce the exceptionally broad \ha\ profile of J1535. After obtaining an acceptable fit in the \ha\ region, we extend the fit to include the \hb\ and \hg\ regions (4200--5400~\AA). Owing to their lower signal-to-noise ratios, the broad \hb\ and \hg\ lines generally require fewer Gaussian components than \ha. Whenever possible, we tie the velocity shifts of the \hb\ and \hg\ components to those of the 2--3 highest-peak \ha\ components, while allowing their widths and amplitudes to vary freely. Finally, we include the 5400--6100~\AA\ segment for completeness. In addition to the broad Balmer lines, we also model broad \heia\ and \heiia\ emission. We fit the continuum in each segment with a power-law model truncated at the wavelength limits. As discussed in Section~\ref{ssec:cont}, we add a blackbody component to model the 4200--5400~\AA\ continuum of J1025. Given the limited wavelength range, we fix the temperature at 3800~K to mitigate degeneracy between the power-law and blackbody components.

\subsection{Absorption-line Fitting} 
\label{ssec:aline}

We assume that the absorption medium is located outside the continuum and the broad-line region (BLR) but inside the narrow-line region and the host galaxy \citep{Ji2025b,Lin2026b}. The absorption component, applied as a multiplicative factor to the sum of the broad Balmer emission and the local continuum, is given by
 
\begin{align}\label{eq:abs}
F_\mathrm{abs}(\lambda) &= 1 - \cf + \cf \, e^{-\tau(\lambda)} \\
\tau(\lambda) &= \tau_0 \exp\left[-\frac{(v(\lambda) - v_0)^2}{2\sigma^2}\right],
\end{align}

\noindent
where \cf\ is the covering factor, $\tau_0$ the optical depth, $\sigma$ the velocity dispersion, and $v_0$ the velocity shift of the line center.  It is important to account for the line-spread function (LSF) when fitting the absorption lines, which are observed to be narrower than the spectral resolution of SDSS. We convolve the intrinsic model with the instrumental LSF, $\left[\left(F_\mathrm{cont}(\lambda) + F_\mathrm{broad}(\lambda)\right)\,F_\mathrm{abs}(\lambda)\right] \otimes G_\mathrm{LSF}(\lambda)$, where $F_\mathrm{cont}(\lambda)$ and $F_\mathrm{broad}(\lambda)$ are the continuum and broad-line models, and $G_\mathrm{LSF}(\lambda)$ is the LSF, approximated by a Gaussian function. For stability, we do not convolve the narrow-line components, which are fitted separately from the broad-line complex.

Because the Balmer absorption lines share the same lower energy level ($n=2$), their line-center optical depths are fixed by atomic physics \citep{Draine2011}, provided that they arise from a common line-of-sight absorber that screens the same background (continuum plus broad-line) emission. For a given column density in the lower level and a given line broadening, the central optical depth scales with the oscillator strength and wavelength,\footnote{For a Doppler-broadened transition, the line-center optical depth can be written as $\tau_0=\frac{\sqrt{\pi}e^2}{m_e c}\,\frac{N_\mathrm{H}(n=2)\,f\,\lambda}{\sqrt{2}\sigma}$, where $e$ is the electron charge, $m_e$ the electron mass, and $c$ the speed of light. Here $N_\mathrm{H}(n=2)$ is the column density of absorbers in the lower level of the transition (for Balmer absorption, $n=2$), $f$ is the absorption oscillator strength, $\lambda$ is the rest-frame wavelength, and $\sigma$ is the one-dimensional Gaussian velocity dispersion of the absorbing gas. In the optically thin limit, the equivalent width (EW) is linear in $N_\mathrm{H}(n=2)$; expressed in wavelength units, $\mathrm{EW} \simeq \frac{\pi e^2}{m_e c^2}\,N_\mathrm{H}(n=2) f \lambda^2 C_f$.} such that $\taua:\taub:\taug = 7.13:1:0.34$.\footnote{The Einstein $A$ coefficients for the transitions are taken from the NIST Atomic Spectra Database: \url{https://www.nist.gov/pml/atomic-spectra-database}.} These $\tau_0$ ratios are robust unless the relative $2s/2p$ populations depart substantially from their statistical-weight ratio. Ly$\alpha$ pumping, a key mechanism for populating the $n=2$ level required for Balmer absorption \citep{Hall2007}, changes the ratios by only $\lesssim 10\%$. When fitting a single \ha\ trough alone, $\tau_0$ and \cf\ are strongly degenerate in our data because of the moderate spectral resolution and because \ha\ is often highly saturated. Tying the optical depths of \ha, \hb, and \hg\ according to the ratios above allows us to break this degeneracy partially: for moderate \cf, \ha\ can be saturated while \hb\ and/or \hg\ remain unsaturated, causing their EW ratios to deviate from the optically thin expectation and thereby providing additional leverage on $\tau_0$ and \cf\ \citep{Wang2015}. We further tie the velocity dispersions and centroid shifts of the absorption lines. With these constraints, we typically measure moderate \cf\ and $\tau_{0,\mathrm{H}\alpha}\lesssim 10$; the exceptions are J0925, J1039, and J1126, which require very large optical depths for which we can only place lower limits. With sufficiently high signal-to-noise in the \hg\ absorption profile, we can constrain $\tau_{0,\mathrm{H}\alpha}$ up to values of $\sim 21$ or slightly higher. At significantly larger values, however, the \hg\ line also becomes optically thick, rendering the absorption profile insensitive to further increases in optical depth. Detailed notes for individual sources are provided in Section~\ref{ssec:indi}. We caution that the emission--absorption configuration can be more complex than assumed here, which may also drive EW ratios away from the ideal optically thin values. For example, absorption troughs can be partially filled by unabsorbed or scattered emission, and the absorber may cover the continuum source and the BLR with different covering factors \citep[e.g.,][]{Ogle1999,Hall2002,Hall2007}. Nonetheless, we find that a single-absorber partial-covering model provides an adequate and self-consistent description of the Balmer absorption in almost all objects in our current sample, except for J1025 (redshifted \hb\ absorption) and J2220, which we will discuss in detail in Sections~\ref{sssec:1025} and \ref{sssec:2220}.

\subsection{MCMC Refinement and Uncertainty Estimation}
\label{ssec:mcmc}

We use Markov chain Monte Carlo (MCMC) sampling to refine the best-fit model and to estimate uncertainties for the continuum, emission-line, and absorption-line parameters. We initialize each MCMC run with the best-fit parameters from the least-squares optimization. Even when the least-squares fit does not require an \feii\ component or some narrow-line components, we include them in the MCMC model to obtain a consistent posterior constraint. We restrict the \feii\ velocity shift to $\pm1500$~\kms\ and its line width to within 50\% of the \ha\ FWHM. For the Balmer absorption, we sample the line-center optical depth in logarithmic space with a uniform prior over 0.01--100. Our mock tests (Section~\ref{ssec:rob}) show that the fits become unreliable for $\tau_0 \gtrsim 100$, motivating this upper bound. Because the absorption features are intrinsically narrow, we adopt a uniform prior of 5--300~\kms\ for the velocity dispersion, and a uniform prior of 0--1 for the covering factor. For all other parameters, we adopt sufficiently broad uniform priors so that the posterior constraints are driven by the data instead of by the prior bounds.

We employ the affine-invariant ensemble sampler implemented in \texttt{emcee} \citep{Foreman-Mackey2013} with 200 walkers, which exceeds 3 times the number of free parameters in our most complex model (J1535) and provides adequate exploration of parameter space. We run each chain for 100{,}000 steps and discard the first 50{,}000 steps as burn-in. We verify that the inferred parameters are stable when the total chain length is increased by a factor of 2.

For each derived quantity, we report the posterior median and standard deviation. For some parameters (e.g., the narrow Balmer line fluxes in J1545-0 and \rfeii\ in J1025), the posterior is consistent with a non-detection and should be interpreted as an upper limit. Conversely, for the absorption-line parameters, \taua\ or \cf\ can be consistent with a lower limit. We classify a measurement as an upper (lower) limit when the posterior median lies within $3\,\sigma$ of the lower (upper) boundary of its prior range. In such cases, we quote the 0.3\% and 99.7\% percentiles of the posterior as the corresponding $3\,\sigma$ lower and upper limits.

\subsection{Robustness of the Absorption-line Fitting}
\label{ssec:rob}

\begin{figure*}
\centering
\includegraphics[width=\textwidth]{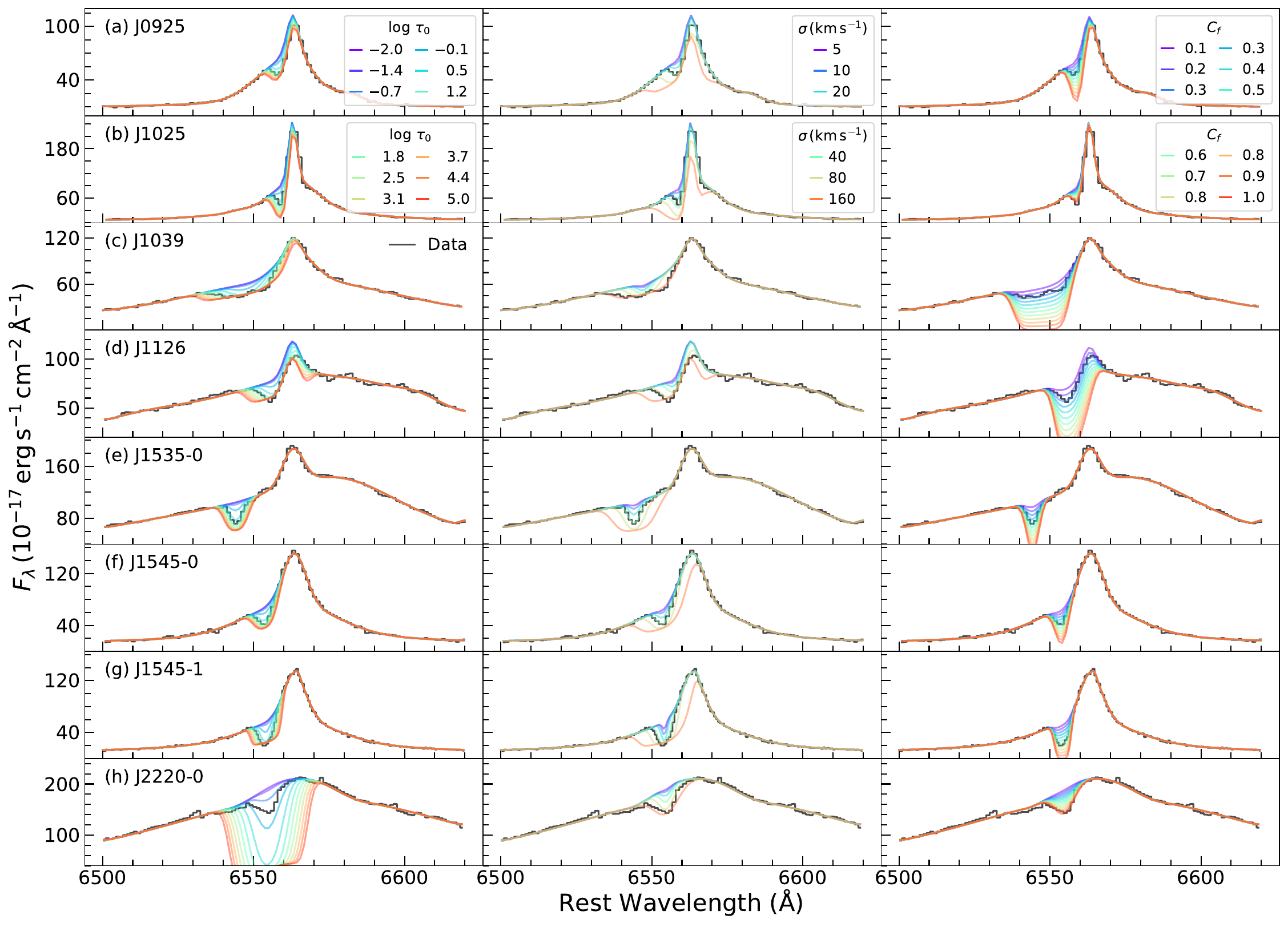}
\caption{Best-fit models with varying absorption parameters $\tau_0$, $\sigma$, and $C_f$, compared with the observed broad \ha\ emission line with the \NII\ doublet, for (a) J0925, (b) J1025, (c) J1039, (d) J1126, (e) J1535-0, (f) J1545-0, (g) J1545-1, and (h) J2220-0. We only show one epoch for J1535 (J1535-0) and J2220 (J2220-0) because the data quality of the other two epochs is largely consistent. Both spectra are shown for J1545 because the SDSS and DESI observations have different spectral resolution.}
\label{fig:mv}
\end{figure*}

We use Figure~\ref{fig:mv} to illustrate how the absorption-line optical depth, velocity dispersion, and covering factor shape the observed \ha\ profiles. For each source, we zoom in on the \ha\ region and generate mock spectra from the best-fit model described in Section~\ref{ssec:indi}. Before fitting the mocks and assessing how well the input parameters are recovered, we first vary one parameter at a time and display the resulting profiles as the colored curves. To construct each mock spectrum, we (1) combine the broad Balmer emission model with the continuum, (2) convolve the intrinsic spectrum with a Gaussian LSF appropriate for the instrumental resolution, and (3) add the narrow-line model to form the full \ha\ profile. The three SDSS epochs of J1535 have comparable data quality and show no significant variability, and we use only the first-epoch spectrum. Both epochs of J1545 are presented because the SDSS and DESI spectra have different resolutions and the absorption varies significantly. We only include the first epoch of J2220 for completeness while caution that its spectral decomposition may not be reliable, as discussed in Section~\ref{sssec:2220}.

As discussed above, $\tau_0$ and \cf\ are highly degenerate when modeling a single absorption profile. We mitigate this degeneracy by fitting \ha, \hb, and \hg\ simultaneously and fixing their optical-depth ratios to the theoretical values. With this in mind, Figure~\ref{fig:mv} is intended as an intuitive guide to the sensitivity of the data to each parameter. We find that the absorption in J0925, J1039, and J1545-1 is consistent with the highest optical-depth mocks ($\log\tau_0 \gtrsim 2$), which are difficult to discriminate from one another. Varying the line width further shows that the absorption troughs for the SDSS spectra are largely washed out when $\sigma \lesssim 10$~\kms, whereas the DESI spectrum retains a detectable signal even for $\sigma=5$~\kms. Finally, the best-fit profiles generally favor moderate covering factors, except for J1025, whose absorption is most consistent with the \cf\ = 1 mock model.

\begin{figure*}
\centering
\includegraphics[width=0.95\textwidth]{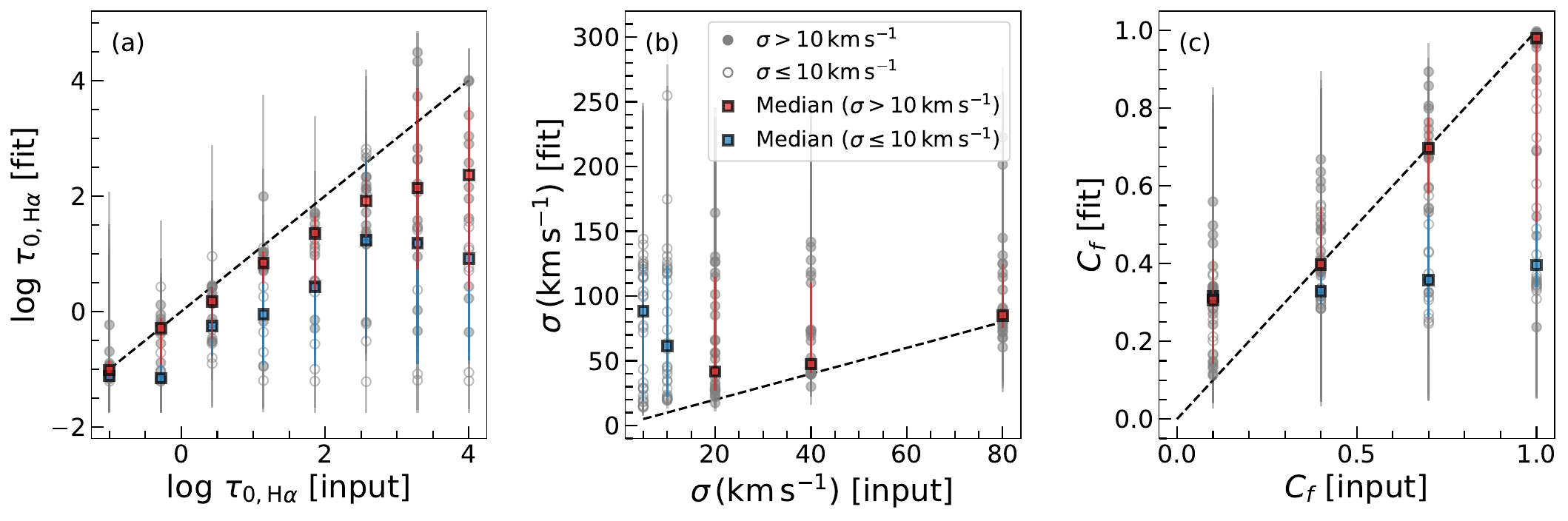}
\caption{Comparison between the input values of the mock spectra and the best-fit absorption-line parameters: (a) optical depth, (b) velocity dispersion, and (c) covering factor. The gray points show the individual fitting results. Open circles mark mock spectra with velocity dispersions below 10~\kms, in which the absorption lines are barely visible. The red and blue boxes represent the median values in each input bin for the $\sigma>10\,\kms$ and $\sigma\leq10\,\kms$ subsamples, respectively.}
\label{fig:mf}
\end{figure*}

To assess whether the absorption-line parameters can be recovered robustly, we performed a suite of mock tests based on the best-fit models of our targets. The results for J0925 are shown in Figure~\ref{fig:mf}, while those for the other sources are provided in the online figure set. In brief, we generate mock spectra using the best-fit \ha, \hb, and \hg\ models, including only the Balmer broad and narrow emission components and the Balmer absorption components. We then vary the input absorption parameters on a grid: \taua\ (10 values uniformly logarithmically spaced from 1 to $10^5$), $\sigma$ (5, 10, 20, 40, and 80~\kms), and \cf\ (10 values uniformly spaced from 0.1 to 1.0), yielding 500 mock spectra in total. For each realization, we initialize the MCMC walkers by perturbing the true input parameters by 30\%. We run 100{,}000 steps with an ensemble size equal to 3 times of the number of free parameters.

The mock tests lead to three main conclusions. First, \taua\ is recovered reliably when $\taua \lesssim 100$ but becomes poorly constrained at larger values because the absorption-trough depth is then primarily controlled by \cf\ instead of by further increases in optical depth. We therefore adopt 100 as the upper bound of the \taua\ prior. Second, the absorption-line width is recovered robustly only when $\sigma \gtrsim 10$~\kms, as intrinsically narrower troughs are strongly smoothed by instrumental broadening; accordingly, when summarizing the mock results we focus on realizations with $\sigma>10$~\kms. Third, \cf\ is generally well constrained, although the uncertainties increase as \cf\ approaches zero, when the absorption signal becomes weak. Overall, these tests support the robustness of our absorption-line measurements over the parameter ranges relevant to our data. The ability to constrain \taua\ up to $\sim 100$ is expected because we fit three Balmer lines simultaneously: even when $\taua=100$, the \hg\ optical depth remains only moderate given $\taua/\taug \approx 21$, providing additional leverage on the optical-depth regime.

\figsetstart
\figsetnum{4}
\figsettitle{Mock test of the robustness of the model parameters of the absorption line model}

\figsetgrpstart
\figsetgrpnum{4.1}
\figsetgrptitle{Comparison for J0925}
\figsetplot{fig4_0.pdf}
\figsetgrpnote{Comparison between the input values of the mock spectra and the best-fit absorption-line parameters.}
\figsetgrpend

\figsetgrpstart
\figsetgrpnum{4.2}
\figsetgrptitle{Comparison for J1025}
\figsetplot{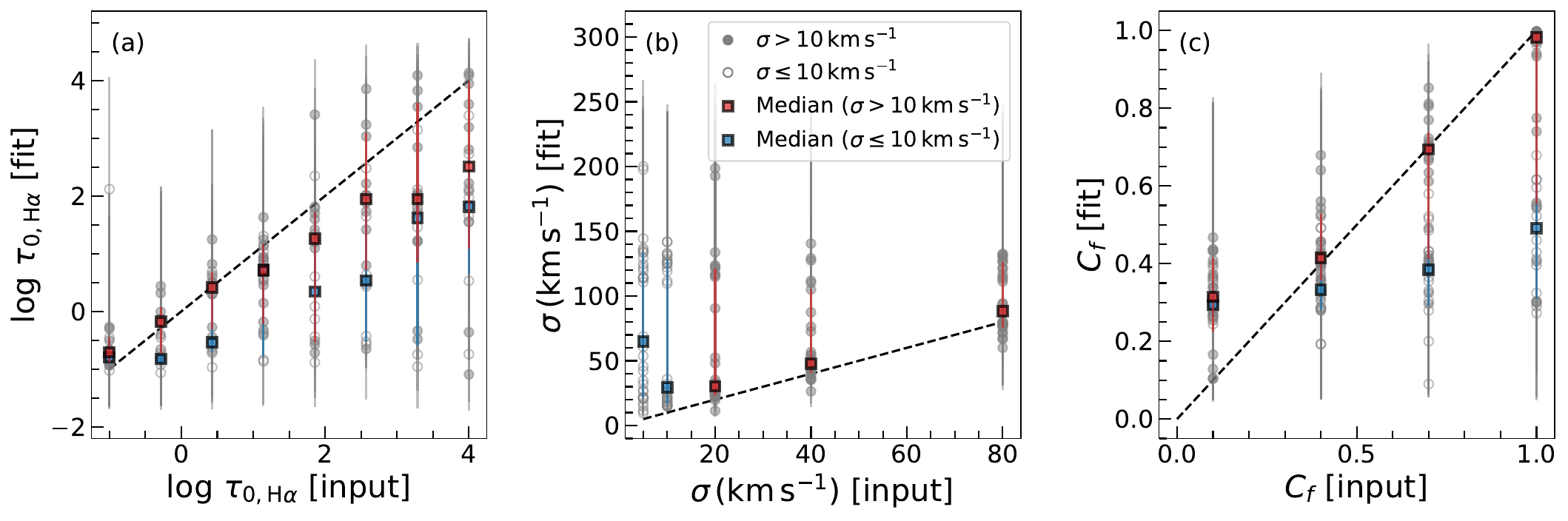}
\figsetgrpnote{Comparison between the input values of the mock spectra and the best-fit absorption-line parameters.}
\figsetgrpend

\figsetgrpstart
\figsetgrpnum{4.3}
\figsetgrptitle{Comparison for J1039}
\figsetplot{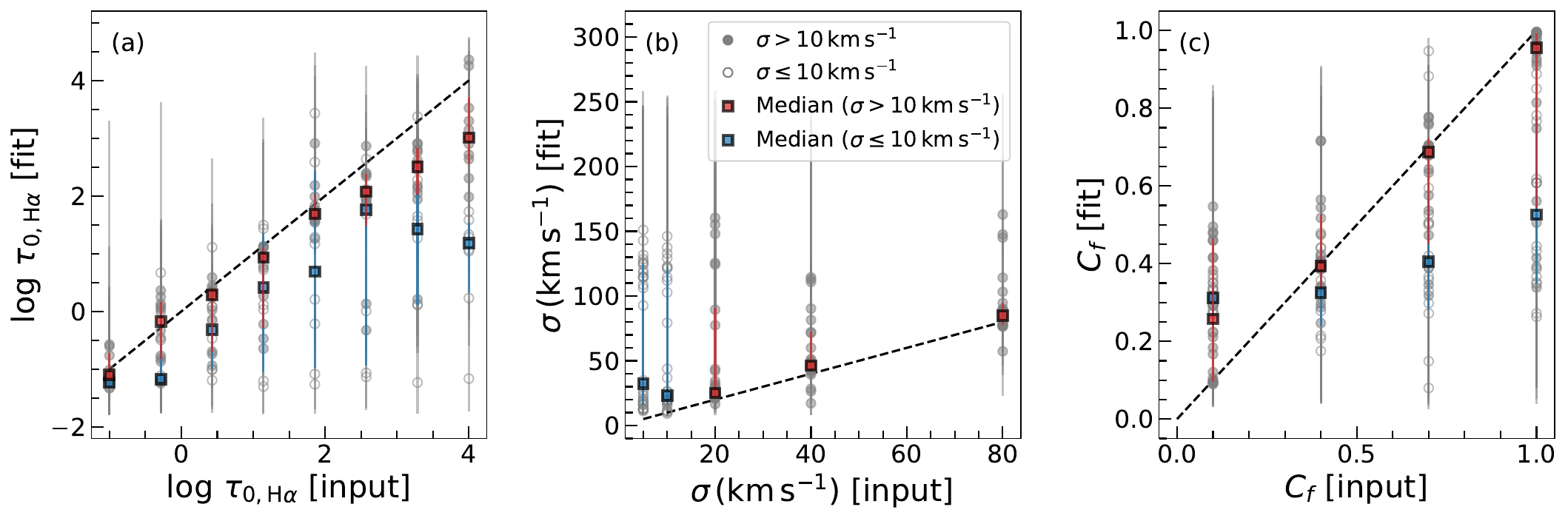}
\figsetgrpnote{Comparison between the input values of the mock spectra and the best-fit absorption-line parameters.}
\figsetgrpend

\figsetgrpstart
\figsetgrpnum{4.4}
\figsetgrptitle{Comparison for J1126}
\figsetplot{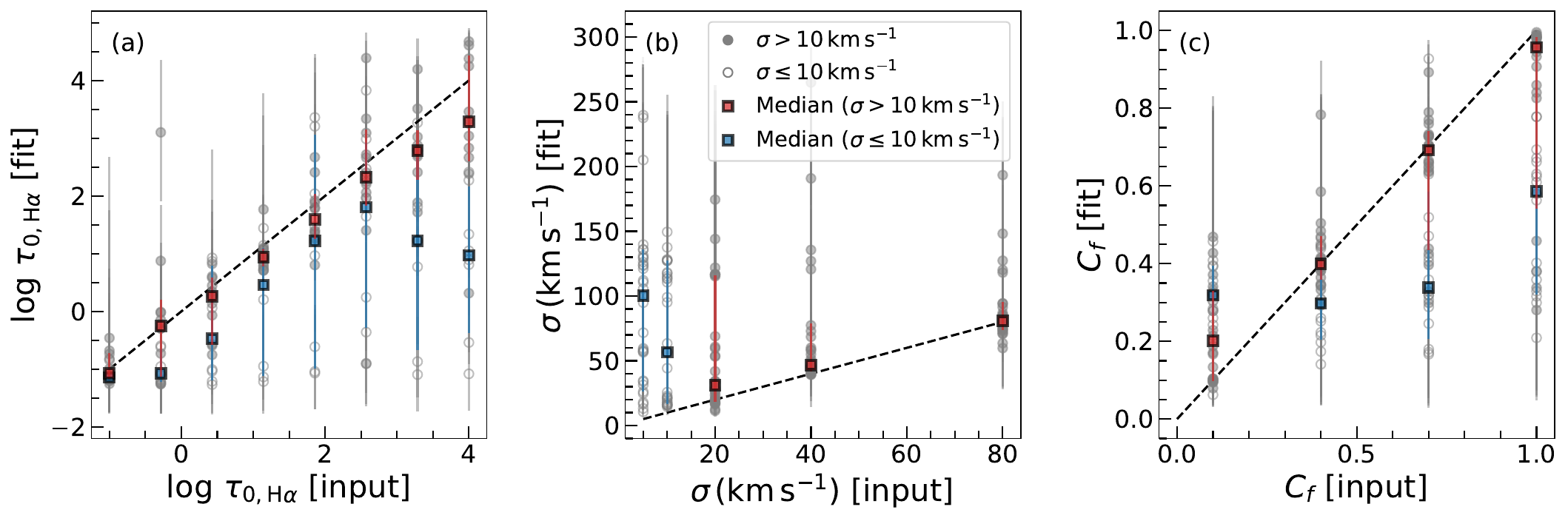}
\figsetgrpnote{Comparison between the input values of the mock spectra and the best-fit absorption-line parameters.}
\figsetgrpend

\figsetgrpstart
\figsetgrpnum{4.5}
\figsetgrptitle{Comparison for J1535-0}
\figsetplot{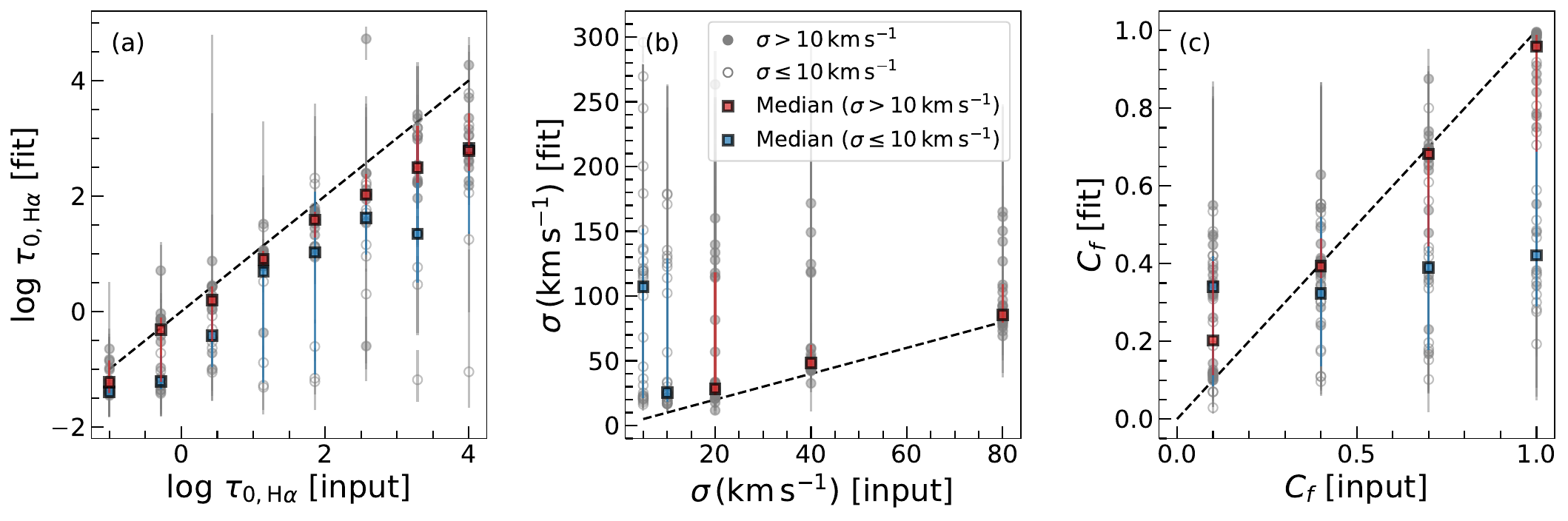}
\figsetgrpnote{Comparison between the input values of the mock spectra and the best-fit absorption-line parameters.}
\figsetgrpend

\figsetgrpstart
\figsetgrpnum{4.6}
\figsetgrptitle{Comparison for J1545-0}
\figsetplot{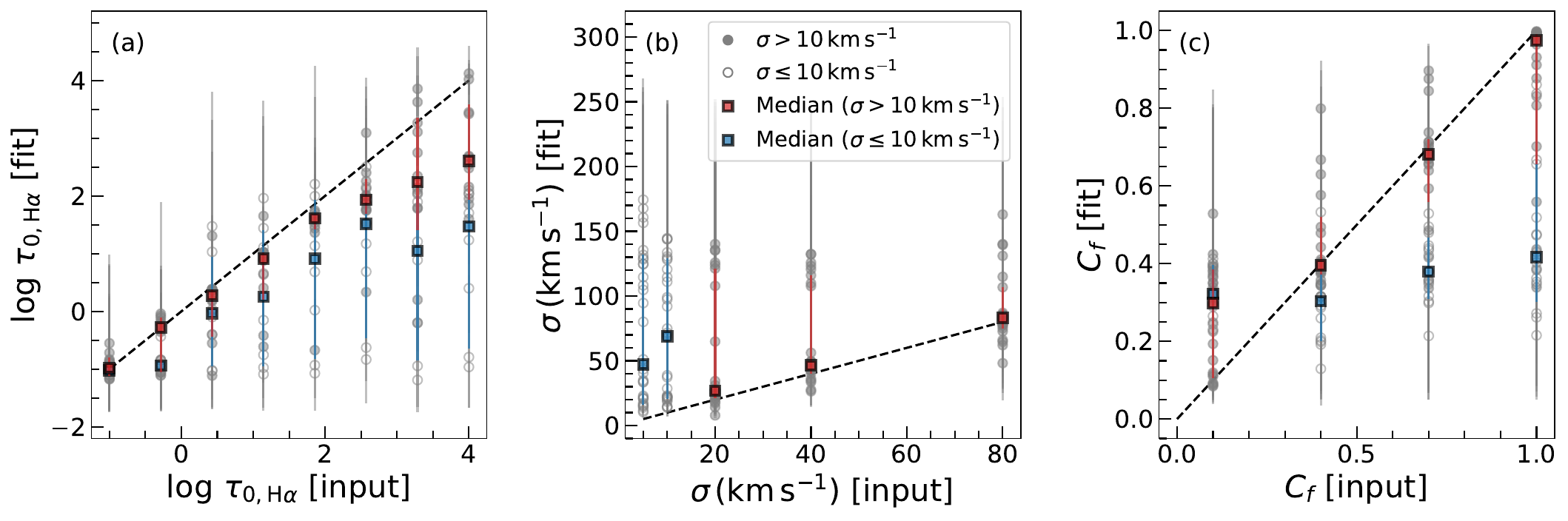}
\figsetgrpnote{Comparison between the input values of the mock spectra and the best-fit absorption-line parameters.}
\figsetgrpend

\figsetgrpstart
\figsetgrpnum{4.7}
\figsetgrptitle{Comparison for J1545-1}
\figsetplot{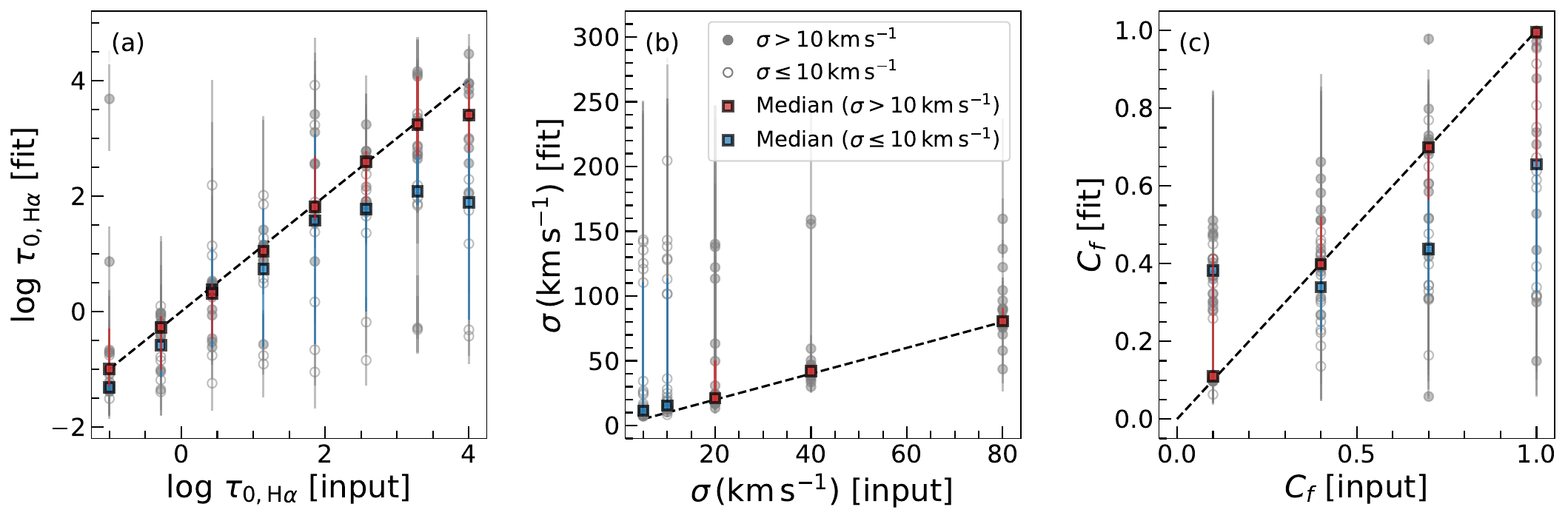}
\figsetgrpnote{Comparison between the input values of the mock spectra and the best-fit absorption-line parameters.}
\figsetgrpend

\figsetgrpstart
\figsetgrpnum{4.8}
\figsetgrptitle{Comparison for J2220-0}
\figsetplot{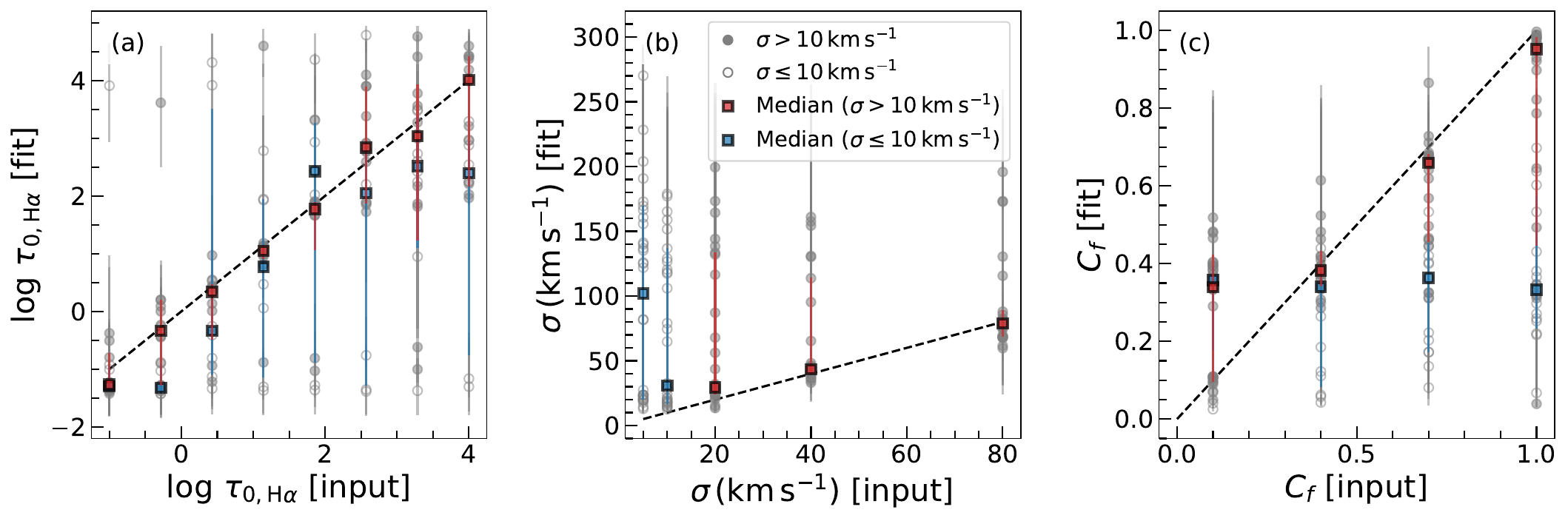}
\figsetgrpnote{Comparison between the input values of the mock spectra and the best-fit absorption-line parameters.}
\figsetgrpend

\figsetgrpstart
\figsetgrpnum{4.9}
\figsetgrptitle{Comparison for J2220-2}
\figsetplot{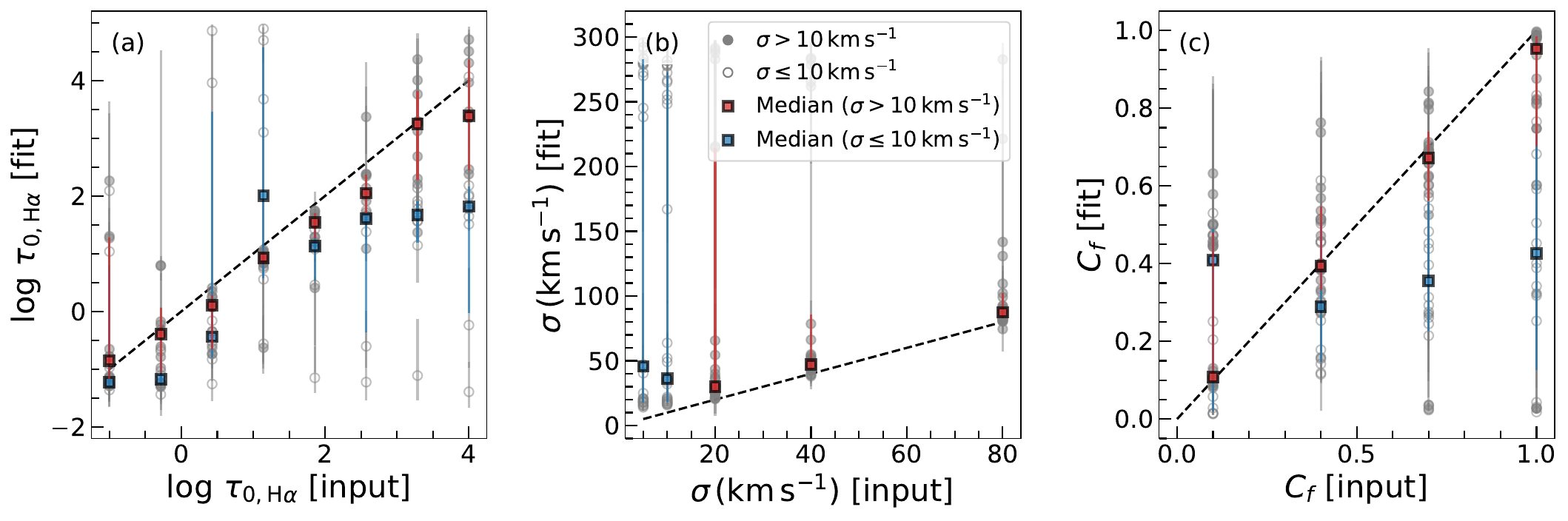}
\figsetgrpnote{Comparison between the input values of the mock spectra and the best-fit absorption-line parameters.}
\figsetgrpend

\figsetend


\begin{deluxetable*}{ccccccccc}
\tabletypesize{\footnotesize}
\tablecaption{Fluxes of Narrow Emission Lines \label{tab:narrowlines}}
\tablecolumns{9}
\tablewidth{0pt}
\tablehead{
\colhead{Name} &
\colhead{H$\alpha$} &
\colhead{H$\beta$} &
\colhead{H$\gamma$} &
\colhead{[O\,{\sc iii}] $\lambda$5007} &
\colhead{[N\,{\sc ii}] $\lambda$6584} &
\colhead{[S\,{\sc ii}] $\lambda\lambda$6716, 6731} &
\colhead{[O\,{\sc i}] $\lambda$6300} &
\colhead{EW$_{\mathrm{[O\,{\sc III}]}}$} 
}
\colnumbers
\startdata
J0925   & $3.34 \pm 0.19$ & $0.38 \pm 0.09$ & $0.25 \pm 0.05$ & "$9.17 \pm 0.07$ & $0.58 \pm 0.09$ & $1.15 \pm 0.03$ & $0.40 \pm 0.02$ & "$74.19 \pm 0.72$ \\
J1025   & $5.45 \pm 0.09$ & $1.97 \pm 0.07$ & $0.90 \pm 0.03$ & $11.56 \pm 0.08$ & $0.16 \pm 0.03$ & $0.21 \pm 0.02$ & $0.15 \pm 0.01$ & $251.41 \pm 2.46$ \\
J1039   & $1.90 \pm 0.33$ & $0.79 \pm 0.05$ & $0.17 \pm 0.05$ & "$3.43 \pm 0.05$ & $0.41 \pm 0.07$ & $0.20 \pm 0.04$ & $0.14 \pm 0.03$ & "$20.50 \pm 0.32$ \\
J1126   & $2.80 \pm 0.15$ & $0.61 \pm 0.06$ & $0.20 \pm 0.05$ & $14.75 \pm 0.13$ & $0.92 \pm 0.10$ & $1.15 \pm 0.05$ & $0.46 \pm 0.03$ & "$76.69 \pm 0.85$ \\
J1535-0 & $4.57 \pm 0.14$ & $1.26 \pm 0.08$ & $0.27 \pm 0.05$ & $18.54 \pm 0.15$ & $2.70 \pm 0.12$ & $2.19 \pm 0.07$ & $0.41 \pm 0.04$ & "$54.23 \pm 0.43$ \\
J1535-1 & $4.63 \pm 0.26$ & $1.20 \pm 0.06$ & $0.22 \pm 0.03$ & $17.13 \pm 0.16$ & $2.03 \pm 0.18$ & $1.79 \pm 0.05$ & $0.45 \pm 0.03$ & "$62.40 \pm 0.58$ \\
J1535-2 & $4.75 \pm 0.14$ & $1.01 \pm 0.07$ & $0.28 \pm 0.03$ & $18.55 \pm 0.14$ & $2.56 \pm 0.11$ & $1.82 \pm 0.06$ & $0.43 \pm 0.04$ & "$71.56 \pm 0.59$ \\
J1545-0 & $<1.58$         & $<0.38$         & $<0.15$         & "$5.81 \pm 0.06$ & $<0.21$         & $0.33 \pm 0.04$ & $0.10 \pm 0.02$ & "$39.05 \pm 0.40$ \\
J1545-1 & $0.87 \pm 0.14$ & $0.21 \pm 0.03$ & $0.09 \pm 0.01$ & "$4.79 \pm 0.04$ & $0.09 \pm 0.02$ & $0.17 \pm 0.02$ & $0.07 \pm 0.01$ & "$41.02 \pm 0.39$ \\
J2220-0 & $9.06 \pm 0.36$ & $2.17 \pm 0.08$ & $0.89 \pm 0.03$ & "$4.32 \pm 0.11$ & $2.49 \pm 0.36$ & $<0.14$         & $<0.36$         & ""$6.46 \pm 0.16$ \\
J2220-1 & $7.01 \pm 0.36$ & $1.67 \pm 0.09$ & $0.69 \pm 0.04$ & "$4.56 \pm 0.07$ & $3.48 \pm 0.35$ & $0.47 \pm 0.07$ & $<0.24$         & ""$6.83 \pm 0.11$ \\
J2220-2 & $6.55 \pm 0.59$ & $1.57 \pm 0.14$ & $0.64 \pm 0.06$ & "$4.40 \pm 0.08$ & $2.23 \pm 0.55$ & $0.29 \pm 0.08$ & $<0.12$         & ""$5.79 \pm 0.12$ \\
\enddata
\tablecomments{
Col. (1): Object name.  
Cols. (2)--(8): Fluxes $5\,\sigma$ upper limits of narrow emission lines in units of $10^{-15}\,\mathrm{erg\,s^{-1}\,cm^{-2}}$.
Col. (9): Equivalent width of [O\,{\sc iii}] $\lambda$5007, in units of \AA.\\
}
\end{deluxetable*}


%

\begin{deluxetable*}{ccccccc}
\tabletypesize{\footnotesize}
\tablecaption{Fluxes and Widths of Broad Emission Lines \label{tab:broadlines}}
\tablecolumns{7}
\tablewidth{0pt}
\tablehead{
\colhead{Name} &
\colhead{H$\alpha$} &
\colhead{EW$_\mathrm{H\alpha}$} &
\colhead{H$\beta$} &
\colhead{$\mathrm{FWHM}_{\mathrm{H\beta}}$} &
\colhead{H$\gamma$} &
\colhead{$\mathrm{FWHM}_{\mathrm{H\gamma}}$}
\\
\colhead{} &
\colhead{($10^{-15}\,\mathrm{erg\,s^{-1}\,cm^{-2}}$)} &
\colhead{(\AA)} &
\colhead{($10^{-15}\,\mathrm{erg\,s^{-1}\,cm^{-2}}$)} &
\colhead{(\kms)} &
\colhead{($10^{-15}\,\mathrm{erg\,s^{-1}\,cm^{-2}}$)} &
\colhead{(\kms)} 
}
\colnumbers
\startdata
J0925   & "$16.63 \pm 0.27$ & $185.52 \pm 3.15$ & "$4.13 \pm 0.13$ & "$991 \pm 84$" & "$1.22 \pm 0.10$ & $1203 \pm 118$ \\
J1025   & "$26.82 \pm 0.17$ & $381.85 \pm 2.87$ & "$1.06 \pm 0.09$ & "$632 \pm 127$ & "$0.18 \pm 0.05$ & "$619 \pm 127$ \\
J1039   & "$55.70 \pm 0.51$ & $413.75 \pm 4.32$ & $13.06 \pm 0.20$ & $4407 \pm 82$" & "$2.94 \pm 0.58$ & $4099 \pm 812$ \\
J1126   & "$76.16 \pm 0.50$ & $497.29 \pm 4.75$ & $13.68 \pm 0.23$ & $4726 \pm 96$" & "$4.21 \pm 0.21$ & $4453 \pm 234$ \\
J1535-0 & $129.89 \pm 0.70$ & $436.14 \pm 3.25$ & $37.13 \pm 0.59$ & $4032 \pm 102$ & "$7.82 \pm 0.19$ & $3548 \pm 101$ \\
J1535-1 & $119.98 \pm 0.46$ & $472.18 \pm 2.42$ & $33.02 \pm 0.56$ & $5016 \pm 101$ & "$8.08 \pm 0.16$ & $4193 \pm 87$" \\
J1535-2 & $117.44 \pm 0.51$ & $485.49 \pm 2.98$ & $29.43 \pm 0.60$ & $4956 \pm 152$ & "$6.30 \pm 0.16$ & $4037 \pm 106$ \\
J1545-0 & "$39.57 \pm 0.38$ & $358.03 \pm 3.89$ & $12.84 \pm 0.26$ & "$854 \pm 42$" & "$3.94 \pm 0.13$ & "$717 \pm 65$" \\
J1545-1 & "$35.90 \pm 0.46$ & $422.84 \pm 7.52$ & $12.14 \pm 0.22$ & "$729 \pm 40$" & "$3.94 \pm 0.11$ & "$544 \pm 57$" \\
J2220-0 & $176.17 \pm 1.91$ & $367.47 \pm 5.15$ & $37.56 \pm 0.78$ & $6527 \pm 119$ & $11.82 \pm 0.73$ & $6507 \pm 512$ \\
J2220-1 & $181.10 \pm 0.87$ & $436.48 \pm 2.59$ & $51.44 \pm 0.32$ & $5089 \pm 62$" & $18.04 \pm 0.34$ & $4777 \pm 90$" \\
J2220-2 & $304.52 \pm 2.14$ & $505.17 \pm 4.57$ & $92.86 \pm 2.95$ & $5459 \pm 64$" & $68.01 \pm 3.91$ & $5779 \pm 241$ \\
\enddata
\tablecomments{
Col. (1): Target name.  
Cols. (2), (4), and (6): Flux of broad H$\alpha$, H$\beta$, and H$\gamma$, respectively. 
Col. (3): EW of broad H$\alpha$. 
Cols. (5), and (7): FWHM of broad H$\beta$ and H$\gamma$, respectively. 
}
\end{deluxetable*}


%

\begin{deluxetable*}{cccccccc}
\tabletypesize{\footnotesize}
\tablecaption{Measurements for the Absorption Lines \label{tab:absorption}}
\tablecolumns{8}
\tablewidth{0pt}
\tablehead{
\colhead{Name} &
\colhead{EW$_{\mathrm{abs,H\alpha}}$} &
\colhead{EW$_{\mathrm{abs,H\beta}}$} &
\colhead{EW$_{\mathrm{abs,H\gamma}}$} &
\colhead{$\log\,\tau_{\mathrm{0,H\alpha}}$} &
\colhead{$\Delta V_\mathrm{abs}$} &
\colhead{$\sigma_\mathrm{abs}$} &
\colhead{$C_f$} \\
\colhead{} &
\colhead{(\AA)} &
\colhead{(\AA)} &
\colhead{(\AA)} &
\colhead{} &
\colhead{(\kms)} &
\colhead{(\kms)} &
\colhead{} 
}
\colnumbers
\startdata
J0925   & $2.29 \pm 0.26$ & $1.27 \pm 0.16$ & $0.80 \pm 0.13$ & $>1.36$               & "$-145.6 \pm 13.2$        & "$43.0 \pm 5.3$" & $0.40 \pm 0.02$ \\
J1025   & $2.09 \pm 0.06$ & $0.47 \pm 0.09$ & $<0.48$         & $\ph{-}0.67 \pm 0.18$ & "$-157.9 \pm 4.0$"        & "$24.9 \pm 3.8$" & $>0.79$         \\
J1039   & $4.57 \pm 0.23$ & $2.48 \pm 0.13$ & $1.57 \pm 0.20$ & $>1.26$               & "$-705.2 \pm 11.7$        & $149.6 \pm 9.6$" & $0.23 \pm 0.01$ \\
J1126   & $3.05 \pm 0.17$ & $1.71 \pm 0.10$ & $1.17 \pm 0.10$ & $>1.47$               & "$-223.3 \pm 11.4$        & "$91.8 \pm 4.6$" & $0.24 \pm 0.01$ \\
J1535-0 & $1.54 \pm 0.07$ & $0.48 \pm 0.06$ & $0.18 \pm 0.04$ & $\ph{-}0.88 \pm 0.15$ & "$-845.2 \pm 3.9$"        & "$36.5 \pm 5.6$" & $0.44 \pm 0.05$ \\
J1535-1 & $2.04 \pm 0.08$ & $0.60 \pm 0.04$ & $0.22 \pm 0.02$ & $\ph{-}0.81 \pm 0.07$ & "$-842.2 \pm 3.7$"        & "$53.8 \pm 3.8$" & $0.41 \pm 0.02$ \\
J1535-2 & $1.66 \pm 0.07$ & $0.55 \pm 0.05$ & $0.21 \pm 0.03$ & $\ph{-}0.92 \pm 0.09$ & "$-852.1 \pm 3.8$"        & "$48.6 \pm 4.0$" & $0.35 \pm 0.02$ \\
J1545-0 & $2.16 \pm 0.12$ & $0.89 \pm 0.08$ & $0.39 \pm 0.06$ & $\ph{-}1.15 \pm 0.11$ & "$-406.8 \pm 4.9$"        & "$42.6 \pm 4.3$" & $0.47 \pm 0.03$ \\
J1545-1 & $3.23 \pm 0.04$ & $1.54 \pm 0.04$ & $0.76 \pm 0.04$ & $\ph{-}1.36 \pm 0.05$ & "$-394.3 \pm 1.2$"        & "$49.1 \pm 0.9$" & $0.56 \pm 0.01$ \\
J2220-0 & $2.52 \pm 0.13$ & $0.31 \pm 0.03$ & $0.10 \pm 0.01$ & $-0.24 \pm 0.15$      & "$-373.9 \pm 6.3$"        & $125.9 \pm 9.1$" & $>0.35$         \\
J2220-1 & $3.04 \pm 0.20$ & $0.36 \pm 0.02$ & $0.11 \pm 0.01$ & $-0.50 \pm 0.09$      & "$-385.9 \pm 9.5$"        & $216.6 \pm 14.6$ & $>0.57$         \\
        & $0.29 \pm 0.05$ & $0.19 \pm 0.03$ & $0.14 \pm 0.04$ & $>1.34$               & $\ph{-}1548.0 \pm 16.7$   & "$33.4 \pm 10.1$ & $0.05 \pm 0.01$ \\
J2220-2 & $3.02 \pm 0.13$ & $0.35 \pm 0.02$ & $0.11 \pm 0.01$ & $-0.45 \pm 0.08$      & "$-374.1 \pm 6.2$"        & $206.7 \pm 9.1$" & $>0.53$         \\
        & $0.69 \pm 0.03$ & $0.11 \pm 0.02$ & $0.04 \pm 0.01$ & $\ph{-}0.14 \pm 0.28$ & $\ph{-}1626.6 \pm 3.7$"   & "$23.8 \pm 5.0$" & $>0.28$         \\
\enddata
\tablecomments{
Col. (1): Target name.  
Cols. (2)--(4): Equivalent width of H$\alpha$, H$\beta$, and H$\gamma$, respectively.
Col. (5): Optical depth of H$\alpha$ absorption.  
Col. (6): Velocity offset of H$\alpha$ absorption from the systemic redshift.
Col. (7): Width of H$\alpha$ absorption.
Col. (8): Covering factor of the absorbing gas.
For J2220-1 and J2220-2, the first and second rows list the measurements for the blueshifted and redshifted absorption components, respectively.
}
\end{deluxetable*}

\section{Results} 
\label{sec:res}

\subsection{Spectral Decomposition of Individual Sources}
\label{ssec:indi}

The section presents the spectral decomposition results for each source, emphasizing any source-specific modeling choices. For each target, we first describe the continuum decomposition and motivate the adopted narrow-line template. We then summarize the emission-line modeling for both the broad and narrow components, and describe the absorption-line model used to reproduce the Balmer series troughs. We also present the narrow-line decomposition of the Balmer lines as well as the absorption-line components. The measurements of the narrow and broad emission lines, together with those of the absorption-line components, are reported in Tables~\ref{tab:narrowlines}, \ref{tab:broadlines}, and \ref{tab:absorption}, respectively.

\subsubsection{J0925+6409} 
\label{sssec:0925}

J0925 shows a relatively simple continuum and is well described by an AGN power-law component plus a moderate contribution from host galaxy starlight. The best-fit stellar continuum is dominated by K- and M-type templates. Because the \SII\ doublet is reasonably strong, we use its profile to construct the narrow-line template. We fit each \SII\ line with two Gaussian components and apply the resulting template to the other forbidden lines and the narrow component of \ha. The full spectral decomposition, after subtracting the host contribution, is shown in Figure~\ref{fig:fit0925}.

\begin{figure}
\centering
\includegraphics[width=\linewidth]{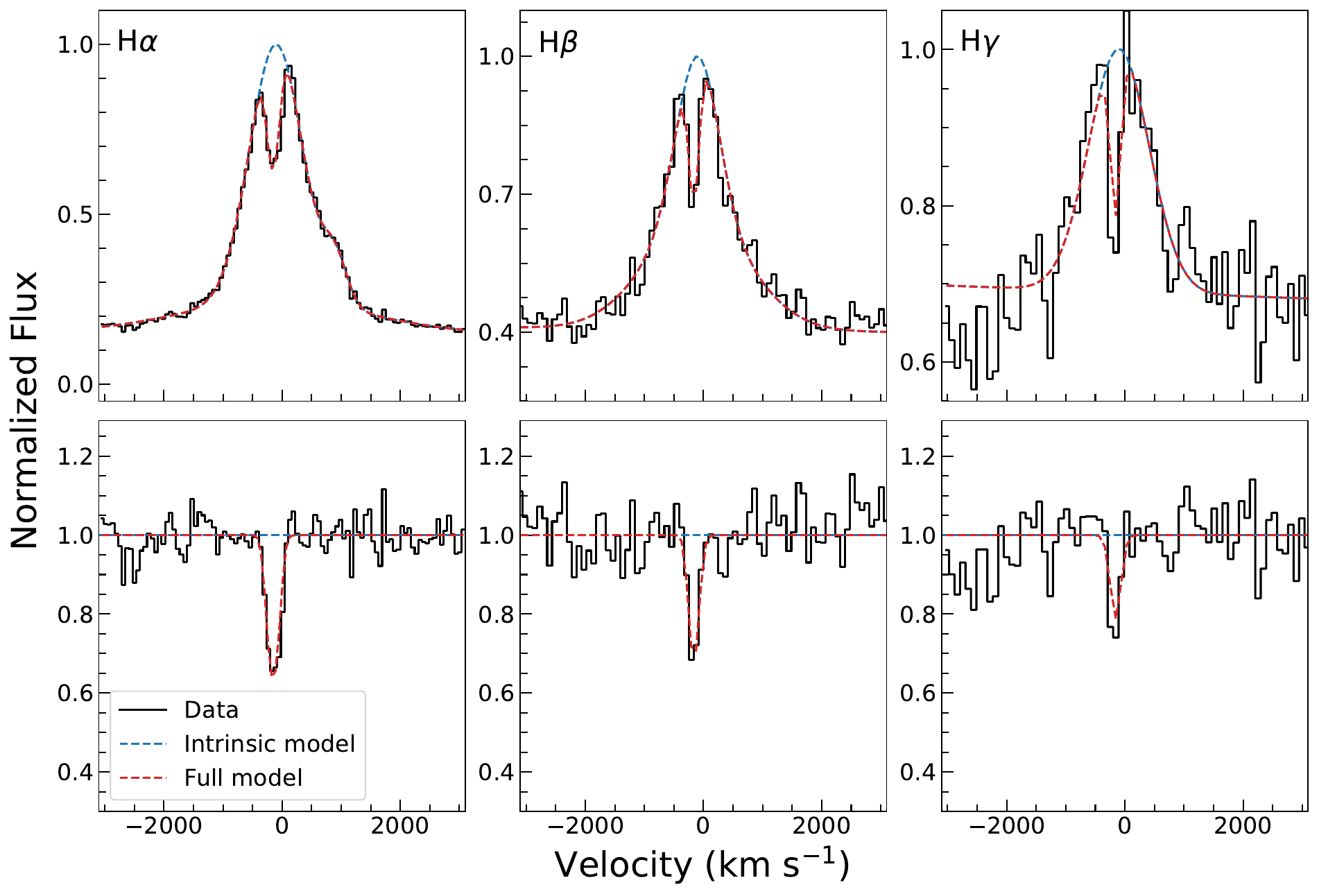}
\caption{Absorption features in the decomposed broad \ha, \hb, and \hg\ lines of J0925. Upper panels display the broad components with absorption, and the lower panels show the absorption profiles normalized by the best-fit emission-line model. The black dashed curves indicate the intrinsic emission without the absorption, while the red dashed curves are the full model.}
\label{fig:bal0925}
\end{figure}

The broad \ha\ and \hb\ profiles are each well reproduced with two Gaussian components. Given the limited signal-to-noise ratio, the broad \hg\ line requires only a single Gaussian component. We do not detect significant \feii\ emission in the least-squares fit; nevertheless, we retain an \feii\ component in the MCMC sampling, which yields a very weak contribution. Figure~\ref{fig:bal0925} shows the broad Balmer profiles after subtracting the narrow-line component. Notably, the \hg\ absorption trough is prominent in addition to those in \ha\ and \hb. In the fully covered, optically thin limit, the expected absorption EW ratios are $\mathrm{EW}_{\ha}:\mathrm{EW}_{\hb}:\mathrm{EW}_{\hg}\approx 9.8:1:0.3$. In contrast, Figure~\ref{fig:bal0925} shows that the \hb\ and \hg\ absorption troughs are comparable in depth to that of \ha. This behavior requires a highly optically thick absorber with partial covering. Consistent with this interpretation, our fit favors a very large \ha\ optical depth, for which we can only place a lower limit.

\subsubsection{J1025+1402}
\label{sssec:1025}

With a characteristic ``V-shaped'' optical SED with a turnover at $\sim 4000-5000$~\AA, J1025 was identified as a low-$z$ analog of LRDs by \cite{Lin2026a}. Following these authors, we model the continuum with a power law plus a blackbody component. We also tried to include an additional stellar component to account for host galaxy starlight, but the least-squares fit does not require it and we omit it. The spectrum exhibits exceptionally strong narrow emission lines (e.g., \OIII\ and the Balmer lines), whereas the \SII\ doublet is barely visible. We thus construct the narrow-line template from \OIII\ and verify that the overall fit is insensitive to whether the template is based on the full \OIII\ profile or only its narrow core component alone. For the \ha\ region, a simple local power law is sufficient to describe the continuum. For the \hb\ and \hg\ region, we retain the power law plus blackbody continuum because the SED turnover lies within this wavelength range.

The broad \ha\ line requires three Gaussian components, while broad \hb\ is adequately described by a single Gaussian. A broad component is not required by the least-squares fit of \hg, but we still include one in the MCMC sampling to place a consistent constraint. The MCMC model ties the velocity offsets of broad \hb\ and \hg\ to that of the central component of broad \ha, all the while enforcing a common width for broad \hb\ and \hg. The posterior distribution yields only a marginally detected broad \hg\ flux, slightly above the $3\,\sigma$ level (Table~\ref{tab:broadlines}).

We detect a blueshifted absorption feature at $v \approx -180~\kms$ superposed on the broad \ha\ profile. In the SDSS spectrum, an absorption counterpart at the same velocity is not apparent in \hb, likely because the \hb\ absorption is intrinsically weaker and is strongly blended with the narrow \hb\ emission component. Spectroscopy of higher resolution and higher signal-to-noise ratio presented by \cite{Lin2026a} confirms that \hb\ absorption is present at the same velocity as the \ha\ trough. Motivated by this, we include a single blueshifted absorber applied to \ha, \hb, and \hg, tying their line-center optical depths to the theoretical ratios and enforcing common kinematics (velocity shift and line width). Because broad \hg\ emission is only marginally detected, the MCMC sampling yields only an upper limit on the \hg\ absorption EW (Table~\ref{tab:absorption}).

J1025 also exhibits a distinct \hb\ absorption feature at $\Delta V \approx +200~\kms$, while a corresponding \ha\ trough is not evident. Since any redshifted \ha\ absorption could in principle be masked by the narrow \ha\ emission, we attempted to include a redshifted \ha\ absorption component and tie its optical depth to that of the redshifted \hb\ absorption. We were unable to obtain a physically plausible fit under these constraints, implying that simple blending is unlikely to explain the apparent absence of redshifted \ha\ absorption. The origin of this deviation from the expected Balmer absorption ratios remains unclear. Similar inconsistencies between Balmer absorption strengths and velocities may also occur in JWST spectra of other LRDs \citep{Ji2025a,DEugenio2025,Chen2026}. \cite{Ji2025b} emphasized the mismatch in J1025 and interpreted it as evidence for multiple absorbing components: the deep \ha\ absorption (and part of Na~D) traces slowly outflowing gas, whereas the redshifted \hb\ absorption suggests an additional component of lower optical depth that may be associated with inflowing material.

\begin{figure}
\centering
\includegraphics[width=0.67\linewidth]{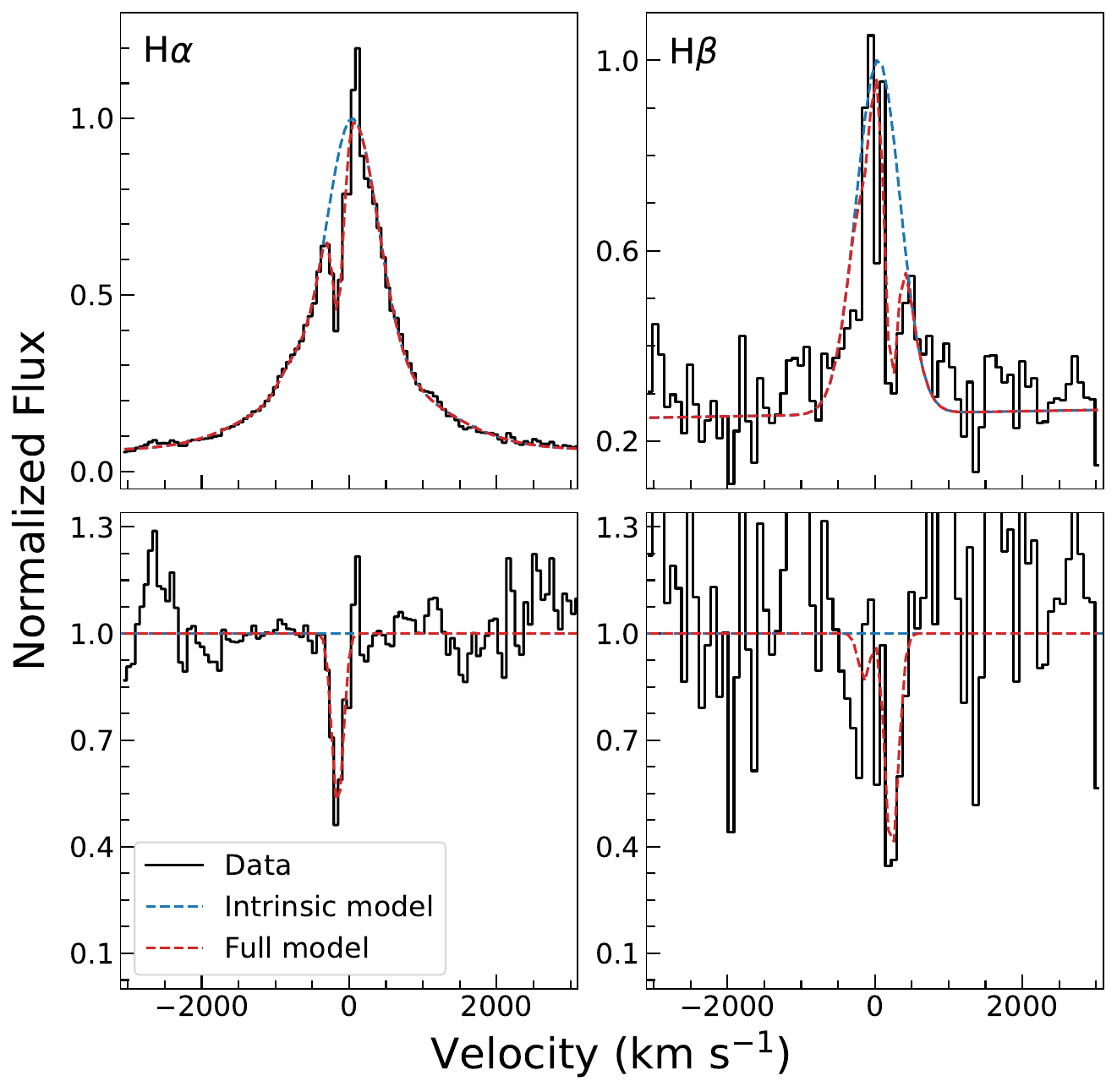}
\caption{Similar to Figure~\ref{fig:bal0925}, but for J1025, for which only broad \ha\ and \hb\ are robustly detected. The \hb\ absorption requires two components: a dominant redshifted trough and a weaker blueshifted trough. In contrast, \ha\ shows only the blueshifted absorption component. We report and interpret the properties of the blueshifted absorber; the origin of the redshifted \hb\ absorption is uncertain.}
\label{fig:bal1025}
\end{figure}

Given this complexity, we do not attempt a physical interpretation of the redshifted \hb\ absorber. We include it phenomenologically as an additional absorption component affecting only \hb. Because this feature lies close to the noise level, leaving all absorber parameters free yields unstable solutions. We therefore fix $\cf=0.8$ and $\sigma=30~\kms$ for this component; this choice does not affect our main conclusions because $\cf$ is largely degenerate with optical depth and the feature is unresolved at SDSS resolution. We show only the decomposed broad \ha\ and \hb\ profiles in Figure~\ref{fig:bal1025}, as the broad \hg\ emission is not securely detected. The blueshifted \hb\ absorption, whose optical depth is tied to that of \ha, is weak and comparable to (or below) the noise. The resulting large absorption-line EW ratio favors a high covering factor, $\cf \approx 1$, consistent with our best-fit value. We caution, however, that the uncertain origin of the redshifted \hb\ component may introduce additional systematic uncertainty in the inferred properties of the blueshifted absorber. Nevertheless, as discussed in Section~\ref{ssec:abs}, our inferred high $\cf$ is broadly consistent with the covering factors measured for LRDs by \cite{Chen2026}. Similarly high covering factors have also been reported by \cite{Lin2026b} for newly discovered low-$z$ LRD analogs identified in DESI spectra.

\subsubsection{J1039+1002}
\label{sssec:1039}

The continuum of J1039 is dominated by a power-law model and a moderate \feii\ pseudo-continuum, with only a modest host galaxy stellar contribution (primarily K-type stars) required in the least-squares fit. The \SII\ doublet is too weak to provide a reliable narrow-line template, and we therefore construct the template from the \OIII\ doublet. We find that the full \OIII\ profile provides a satisfactory template for the other narrow lines. For the broad-line decomposition, we adopt three Gaussian components to model broad \ha, while broad \hb\ and \hg\ are each adequately described by a single Gaussian component. We tie the centroid velocity shifts of broad \hb\ and \hg\ to that of the primary (highest-peak) broad \ha\ component. A moderate \feii\ pseudo-continuum is detected and included in the fit. Based on the measured broad-line width and luminosity, we derive an intermediate BH mass of $M_{\rm BH}=10^{7.29}$~\Msun, consistent with J1039 being among the objects with intermediate luminosity and line width within our Balmer-absorption sample (Table~\ref{tab:sample}).

\begin{figure}
\centering
\includegraphics[width=\linewidth]{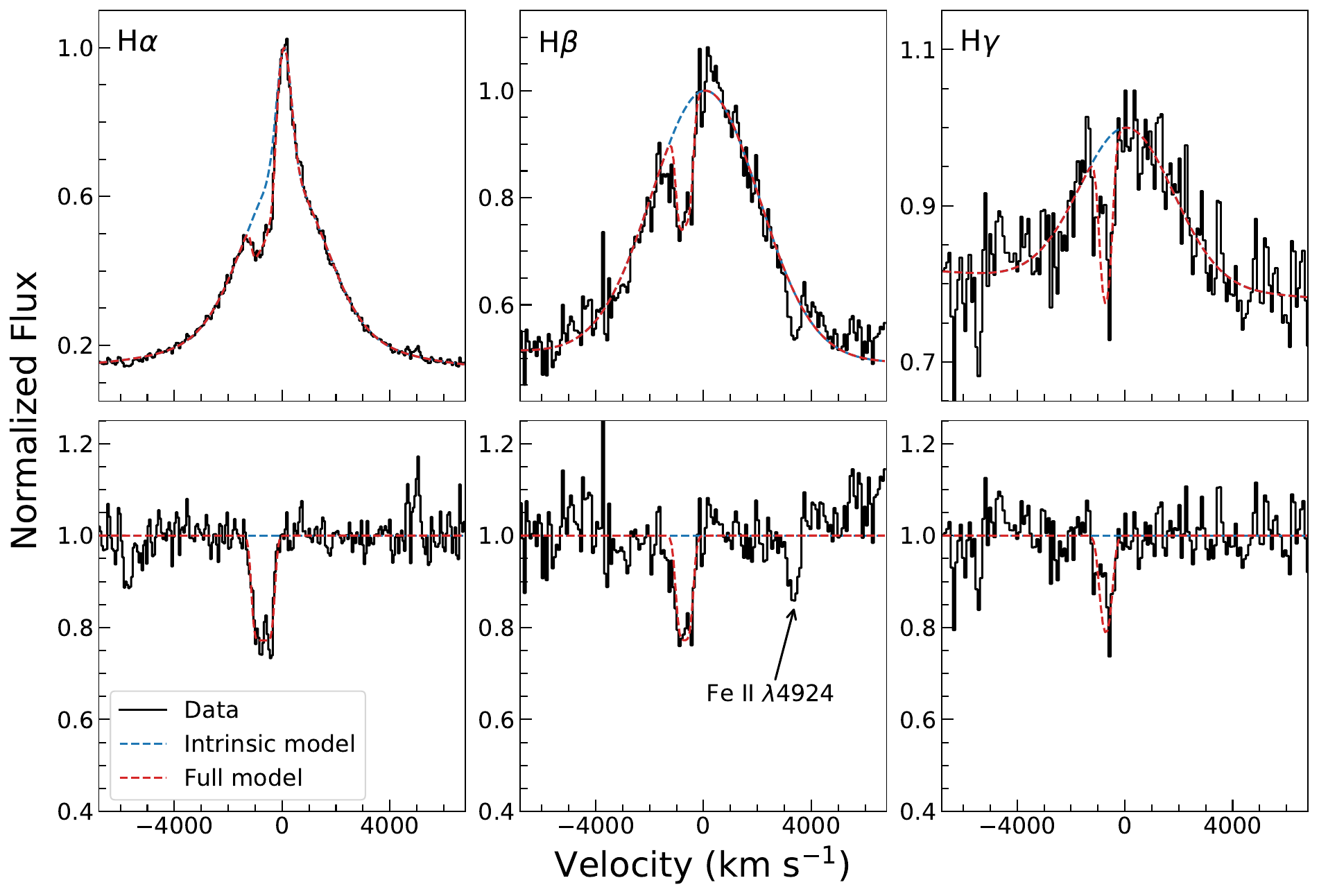}
\caption{Similar to Figure~\ref{fig:bal0925}, but for J1039. The \feii~$\lambda4924$ absorption line at $\sim 3400$~\kms\ from the \hb\ is marked.}
\label{fig:bal1039}
\end{figure}

The decomposed broad Balmer lines and the associated absorption components are shown in Figure~\ref{fig:bal1039}. Our partially covering absorption model provides a good description of the observed Balmer absorption troughs. Similar to J0925, the absorption EWs of \ha, \hb, and \hg\ are broadly comparable, which is difficult to reconcile with an optically thin absorber. Instead, this behavior indicates that the absorber is highly optically thick and only partially covers the emitting region. Consistent with this interpretation, our fit yields a lower limit on the line-center optical depth, $\tau_{\mathrm{H}\alpha}\gtrsim 18$, and a moderate covering factor of $\cf \approx 0.23$.

In addition to the Balmer absorption, J1039 also shows \feii\ absorption features, most notably near \feii~$\lambda4924$ (appearing at about $+4000$~\kms\ relative to the \hb\ absorption line in Figure~\ref{fig:bal1039}) and \feii~$\lambda5169$. We note that similar absorption features are also present in J1126 (Figure~\ref{fig:bal1126}), and are seen more weakly in J0925 and J1535 (Figure~\ref{fig:bal1535}). Because these lines are not the main focus of this paper, we mask them in the spectral decomposition. With much higher-sensitivity spectra, \cite{Lin2026a} revealed numerous \feii\ emission and absorption lines, suggesting the presence of a cool, metal-rich gas near the central accreting BH.

\subsubsection{J1126+4252}
\label{sssec:1126}

J1126 was previously identified as a Balmer-absorption AGN by \citet{Wang2015}. They performed a detailed spectral decomposition by first modeling and subtracting a continuum consisting of host galaxy starlight, an AGN power law, and the \feii\ pseudo-continuum, and then fitting the emission and absorption features using multiple Gaussian components. \citet{Wang2015} reported that the inferred Balmer-absorption EW ratios depend on how the spectrum is normalized. When the absorption is normalized by the continuum alone, the \ha/\hb\ absorption EW ratio is close to the optically thin expectation. In contrast, when the absorption is normalized by the broad-line emission component, the EW ratio becomes close to unity, implying that the absorber is highly optically thick and only partially covers the BLR.

In this work, we model the absorption as a multiplicative factor applied to the sum of the continuum and broad-line emission, as discussed in Section~\ref{ssec:aline}. For J1126, our continuum model is broadly consistent with that of \citet{Wang2015}, consisting of a dominant power-law component, a moderate stellar contribution, and a significant \feii\ pseudo-continuum. We use the \SII\ doublet to construct the narrow-line template, fitting each \SII\ line with two Gaussian components.

For the broad-line decomposition, we adopt four Gaussian components to model the broad \ha\ profile. Broad \hb\ and \hg\ are described adequately by a single Gaussian component each, with their velocity shifts tied to that of the central \ha\ component. Using these measurements, we estimate a BH mass of $M_{\rm BH} = 1.12\times10^8~\Msun$ and an Eddington ratio of $\lambda_\mathrm{Edd} \simeq 0.06$ (Table~\ref{tab:sample}), in good agreement with \citet{Wang2015}. The Balmer absorption in J1126 has a velocity dispersion of $\sim 90$~\kms\ and a centroid offset of approximately $-220$~\kms. These values are consistent with those given in \citet{Wang2015}, and any residual differences likely stem from details of the parameterization and our explicit treatment of the instrumental LSF.

\begin{figure}
\centering
\includegraphics[width=\linewidth]{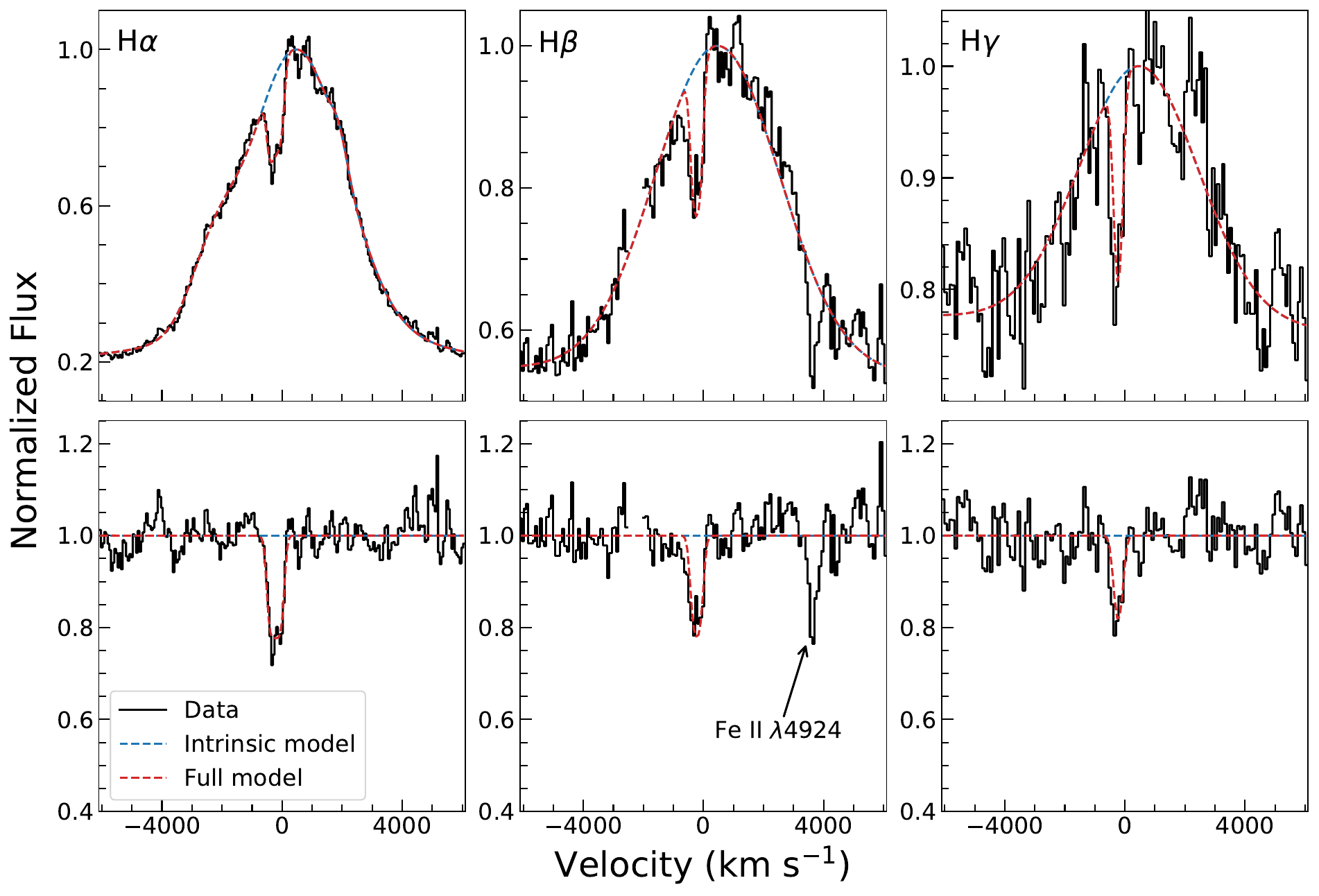}
\caption{Similar to Figure~\ref{fig:bal0925}, but for J1126. The bad pixels at approximately $-2000$~\kms\ are masked in the fitting.  The \feii~$\lambda4924$ absorption line at $\sim 3450$~\kms\ from the \hb\ is marked.}
\label{fig:bal1126}
\end{figure}

Finally, the Balmer absorption in J1126 is well reproduced by a partially covering model in the high-optical depth regime. We can only place a lower limit on the \ha\ optical depth, $\tau_{\mathrm{H}\alpha}\gtrsim 36$, while the \hb\ and \hg\ optical depths are tied to \ha\ by their theoretical ratios. Consistent with this, Figure~\ref{fig:bal1126} shows that the absorption trough depths are similar among \ha, \hb, and \hg, indicating that the observed trough depth is limited primarily by the covering factor instead of the line-center optical depth. This behavior is qualitatively consistent with the interpretation of \citet{Wang2015} when the absorption is considered relative to the BLR emission; however, we emphasize that both the power-law continuum and the broad Balmer emission should be included when defining the normalization and measuring absorption EWs.

\subsubsection{J1535+5644}
\label{sssec:1535}

\begin{figure}
\centering
\includegraphics[width=\linewidth]{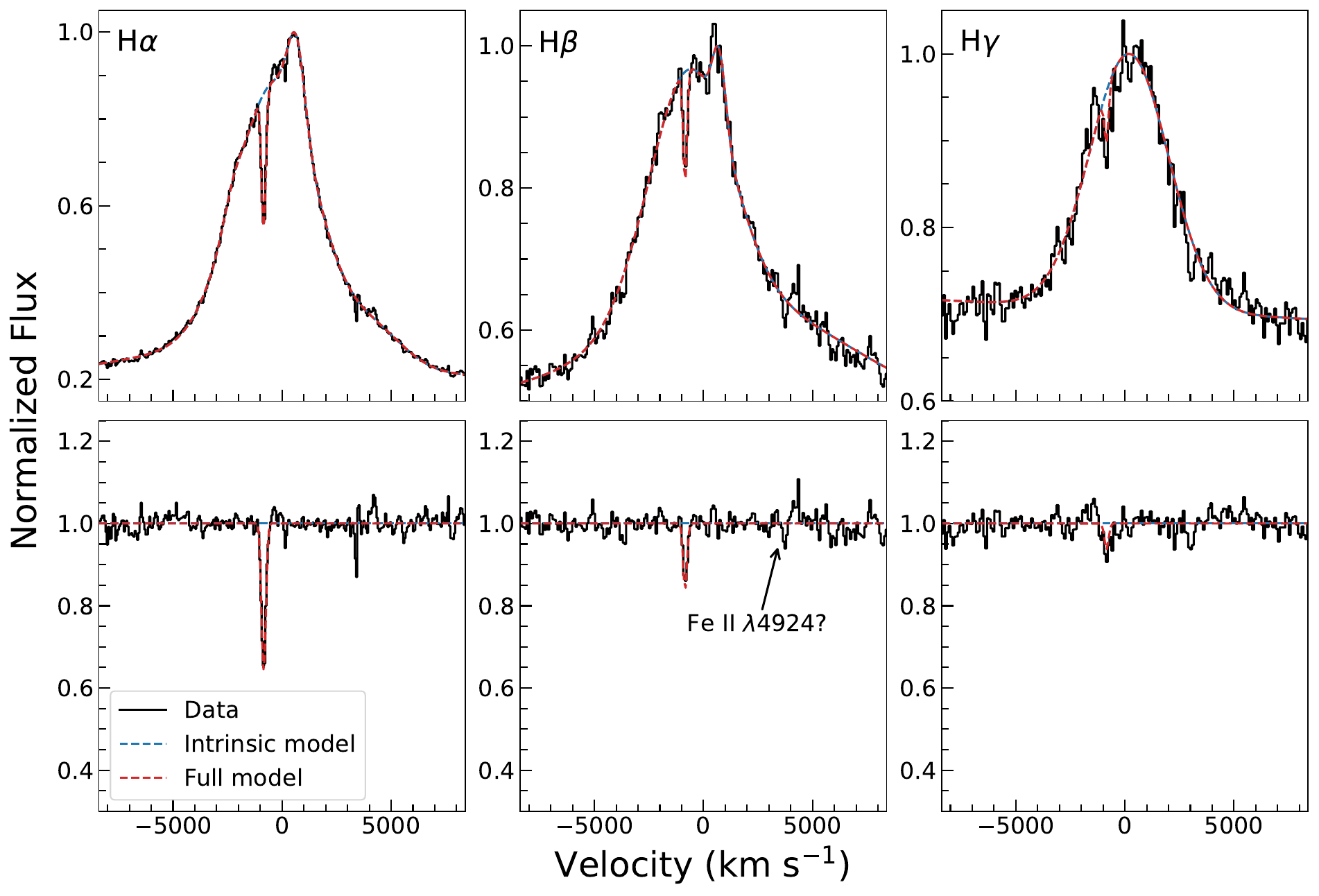}
\caption{Similar to Figure~\ref{fig:bal0925}, but for J1535-1. The \feii~$\lambda4924$ absorption line at $\sim 3450$~\kms\ from the \hb\ is marked.}
\label{fig:bal1535}
\end{figure}

J1535 was observed by SDSS in three epochs between 2001 and 2018. The best-fit models indicate only very moderate variability across these spectra. We therefore show the narrow-line-decomposed Balmer profiles only for the second epoch in Figure~\ref{fig:bal1535}, and provide the corresponding figures for the remaining epochs in the online figure set. We do not detect a measurable host galaxy contribution to the continuum. Unlike the other objects in our sample, J1535 exhibits a clear \feii\ pseudo-continuum and exceptionally broad Balmer emission, with $\mathrm{FWHM} \gtrsim 4000$~\kms. Because the very broad \ha\ profile strongly blends with the \SII\ doublet, we construct the narrow-line template using \OIII. We fit each of the \OIII\ doublets with two Gaussians and adopt only the narrow core component as the template; including the extended \OIII\ wing produces systematically poorer fits to the narrow-line cores. In addition to \OIII, the standard narrow-line species (e.g., \NIIc, \SIIc, and \OIc) are all well detected in J1535.

The broad \ha\ line requires five Gaussian components to achieve a satisfactory fit, while we use three Gaussian components for \hb\ and one for \hg. Since these Gaussian components are purely empirical and do not have a unique physical interpretation, we do not tie the individual components of \hb\ and \hg\ to those of \ha. The large broad-line width and the moderately high luminosity imply a high BH mass, $\mbh \approx 10^{8.2}$~\Msun, more than an order of magnitude larger than that of the other sources in our sample. Although the \mbh\ variations across epochs are not expected to be physically meaningful, we report the measurements for each epoch to reflect the measurement uncertainties. As shown in Figure~\ref{fig:bal1535}, the \ha\ absorption is very strong, \hb\ absorption is moderate but significant against the noise, and \hg\ absorption is close to the noise level. This pattern suggests that the absorber has a moderate optical depth, with the \hb\ and \hg\ absorption being close to the optically thin regime.

\figsetstart
\figsetnum{9}
\figsettitle{Decomposed broad Balmer line profiles of J1535}

\figsetgrpstart
\figsetgrpnum{9.1}
\figsetgrptitle{Absorption features in the decomposed broad H$\alpha$, H$\beta$, and H$\gamma$ lines of J1535-0.}
\figsetplot{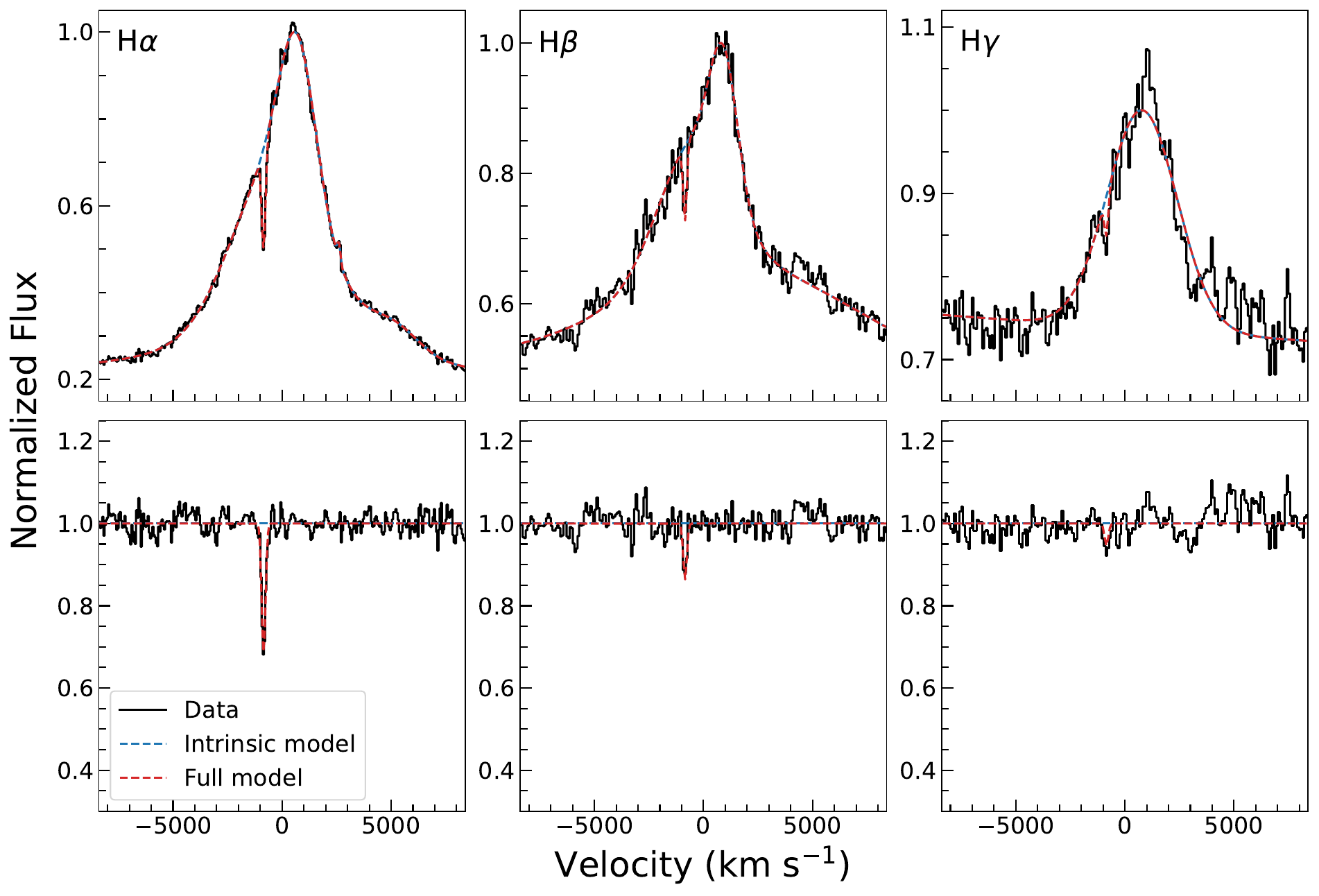}
\figsetgrpnote{Similar to Figure 5, but for J1535.}
\figsetgrpend

\figsetgrpstart
\figsetgrpnum{9.2}
\figsetgrptitle{Absorption features in the decomposed broad H$\alpha$, H$\beta$, and H$\gamma$ lines of J1535-1.}
\figsetplot{fig9_1.pdf}
\figsetgrpnote{Similar to Figure 5, but for J1535.}
\figsetgrpend

\figsetgrpstart
\figsetgrpnum{9.3}
\figsetgrptitle{Absorption features in the decomposed broad H$\alpha$, H$\beta$, and H$\gamma$ lines of J1535-2.}
\figsetplot{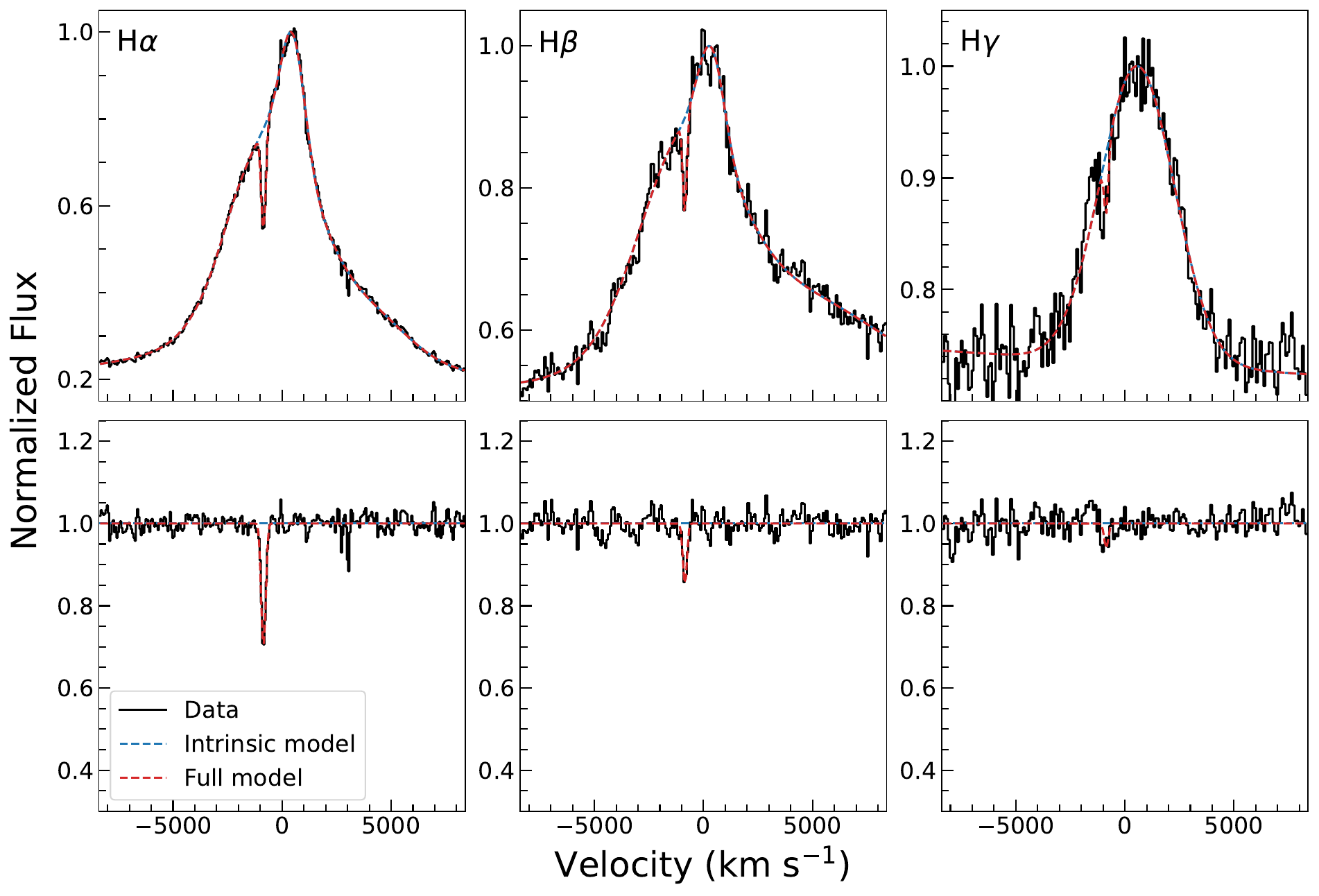}
\figsetgrpnote{Similar to Figure 5, but for J1535.}
\figsetgrpend

\figsetend

\subsubsection{J1545+2238} 
\label{sssec:1545}

Observations of J1545 were taken by SDSS in 2005 and by DESI in 2021. We detect moderate variability in both the continuum and the broad-line emission between the two epochs. When fitting the continuum, we do not detect any host galaxy stellar emission, while we can measure a moderate \feii\ pseudo-continuum. Because \SII\ doublet is very weak, we use \OIII\ to construct the narrow-line template. For the DESI spectrum ($R\approx 4000$ at $\sim 6500$~\AA), we fit each of the \OIII\ lines with three Gaussian components. The other narrow lines in the source are well reproduced if we include only the two narrow components in the template. For the SDSS spectrum, whose resolution is more than a factor of 2 lower, two Gaussian components are sufficient for the \OIII\ fit, and we use only the core component as the narrow-line template.

We fit the broad \ha\ line with three Gaussians. The \NII\ doublet is hardly detected, while \SII\ and \OIb\ are only marginally measured. We nevertheless include \NII\ in the fit to derive an upper limit on \NIIb, which is needed to place J1545 on the optical line-intensity ratio diagnostic diagrams \citep{Baldwin1981} in Section~\ref{sssec:bpt}. The broad Balmer lines are very narrow, with $\mathrm{FWHM}\lesssim 700$~\kms. When the signal-to-noise ratio is low, it is challenging to separate the broad and narrow components for \hb\ and \hg. We therefore fix the narrow-line flux ratios to $\ha:\hb:\hg = 3.1:1:0.47$, based on empirical and theoretical values for AGNs (e.g., \citealp{Gaskell1984,Lu2019}). We use three Gaussian components for \hb\ and two for \hg, tying the velocity shift of each \hb\ component to the corresponding \ha\ component, and the velocity shifts of the two \hg\ components to those of the two strongest \ha\ components. Both the SDSS and DESI spectra show broad \feii~$\lambda6369$ and \hei~$\lambda6678$. From the DESI spectrum, we measure $\mathrm{FWHM}\approx 1200$~\kms\ for both lines, but the low signal-to-noise ratio of the SDSS spectrum prevents a robust line width measurement. We thus fix their velocity dispersion to 600~\kms\ when fitting the SDSS spectrum and confirm that the results are insensitive to this choice. 

\begin{figure}
\centering
\includegraphics[width=\linewidth]{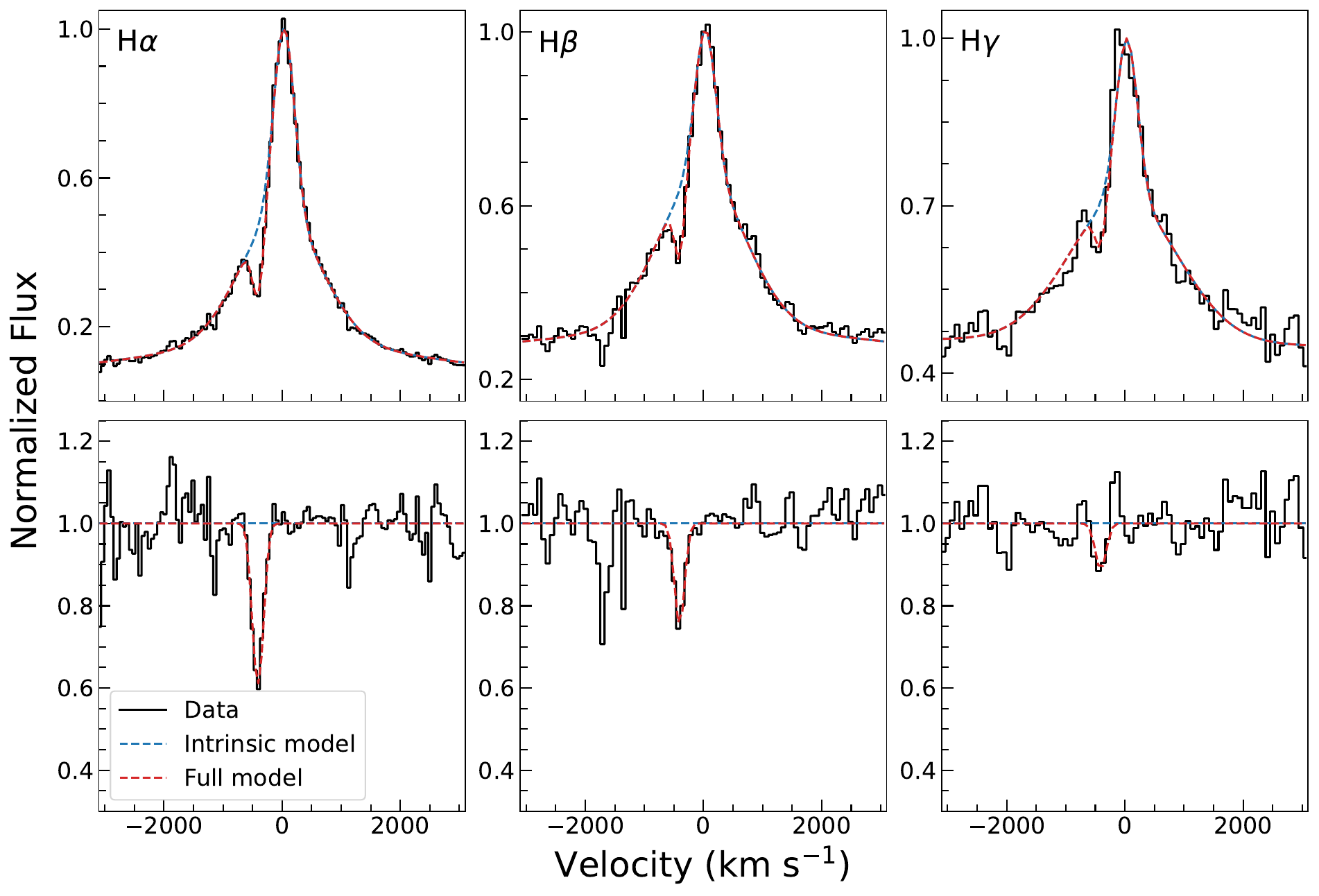}
\caption{Similar to Figure~\ref{fig:bal0925}, but for J1545-0.}
\label{fig:bal15450}
\end{figure}

\begin{figure}
\centering
\includegraphics[width=\linewidth]{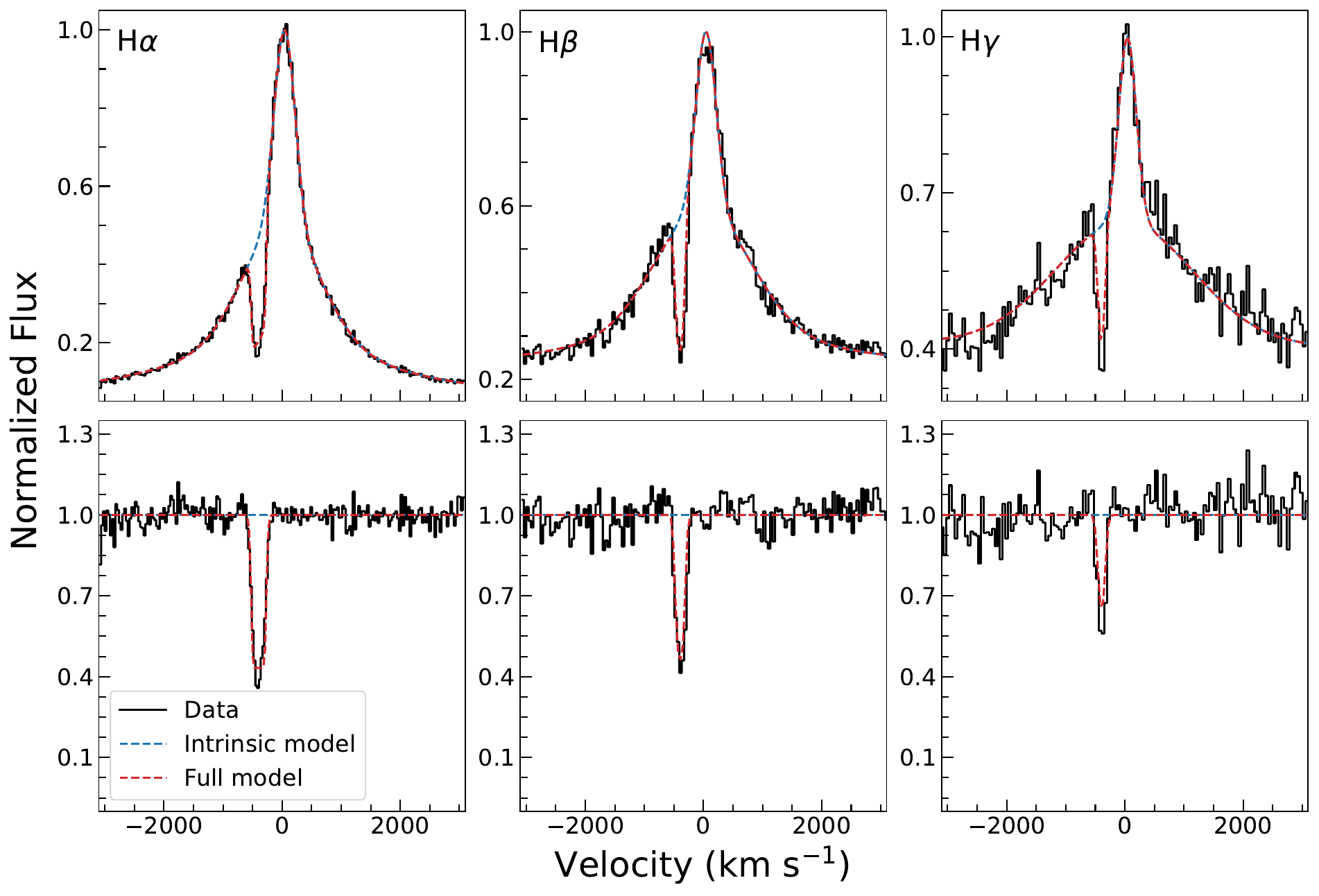}
\caption{Similar to Figure~\ref{fig:bal0925}, but for J1545-1.}
\label{fig:bal15451}
\end{figure}

The decomposed Balmer lines of the SDSS and DESI epochs are shown in Figures~\ref{fig:bal15450} and \ref{fig:bal15451}, respectively. The \hb\ and \hg\ absorption troughs are deeper in the DESI spectrum. This change cannot be attributed solely to the difference in spectral resolution, because the equivalent width does not depend on spectral resolution. We will discuss the variability of J1545 in Section~\ref{ssec:var}.

\subsubsection{J2220+0109}
\label{sssec:2220}

J2220 was observed by SDSS in three epochs: the first in 2001, and the latter two in 2017 separated by only two months. Intriguingly, the first two spectra show moderate variability in both the continuum and the broad Balmer lines, while the optical continuum and broad emission lines brighten rapidly between the last two epochs. The Balmer line profiles also exhibit noticeable changes. In addition, a secondary Balmer absorption component appears in the second-epoch spectrum and becomes more prominent in the third epoch. We fit the three spectra with largely the same model, with only minor modifications to accommodate the observed variability.

The narrow emission lines are all very weak, if detected at all, except for the \OIII\ doublet, which is broad and prominent. We therefore use \OIII\ to construct the narrow-line template. Because a single Gaussian cannot adequately reproduce the peak of the \OIII\ profile, we fit the \OIII\ doublet with a Gauss--Hermite function to determine a reliable systemic redshift and template shape. Similar to J1545, we tie the amplitudes of the narrow Balmer lines to empirical AGN line ratios to stabilize the fitting. We note, however, that this narrow-line template may not accurately represent the profiles of other narrow lines, particularly \SII\ and the narrow Balmer components. In particular, the inferred \SII\ flux in the first epoch is inconsistent with that in the other two epochs, even though narrow-line variability is not expected on a 16 yr timescale. We modeled broad \ha\ and \hb\ with three Gaussian components in all three epochs, while broad \hg\ was fit with one Gaussian in the first two epochs and two Gaussians in the third epoch, reflecting the profile variability. Despite these changes, the BH masses inferred from the three single-epoch spectra remained broadly consistent, $\mbh \approx 10^{8.4}$--$10^{8.5}~\Msun$, within the uncertainty of the single-epoch virial method.

\begin{figure}
\centering
\includegraphics[width=\linewidth]{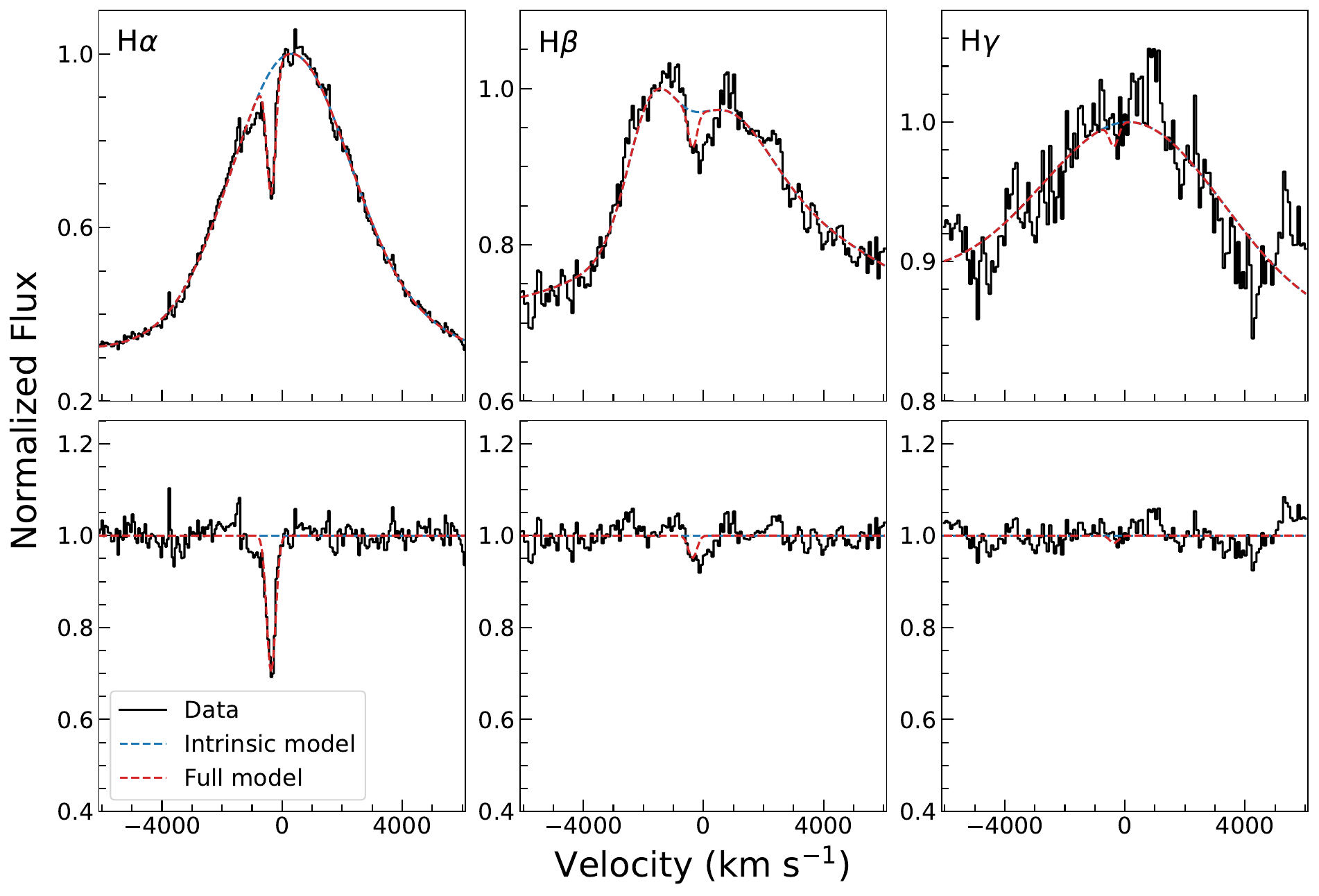}
\caption{Similar to Figure~\ref{fig:bal0925}, but for J2220-0.}
\label{fig:bal22200}
\end{figure}

\begin{figure}
\centering
\includegraphics[width=\linewidth]{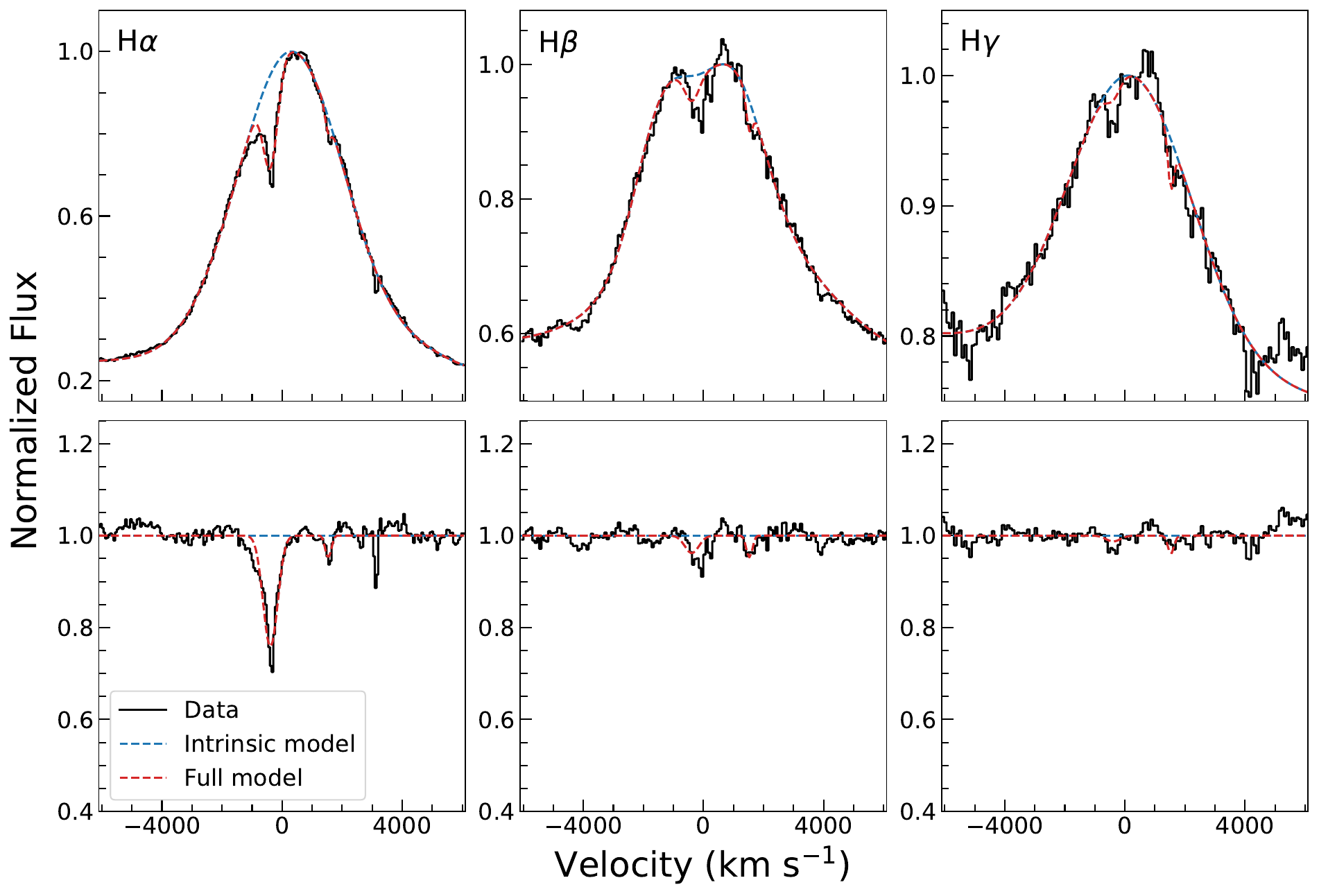}
\caption{Similar to Figure~\ref{fig:bal0925}, but for J2220-1.}
\label{fig:bal22201}
\end{figure}

\begin{figure}
\centering
\includegraphics[width=\linewidth]{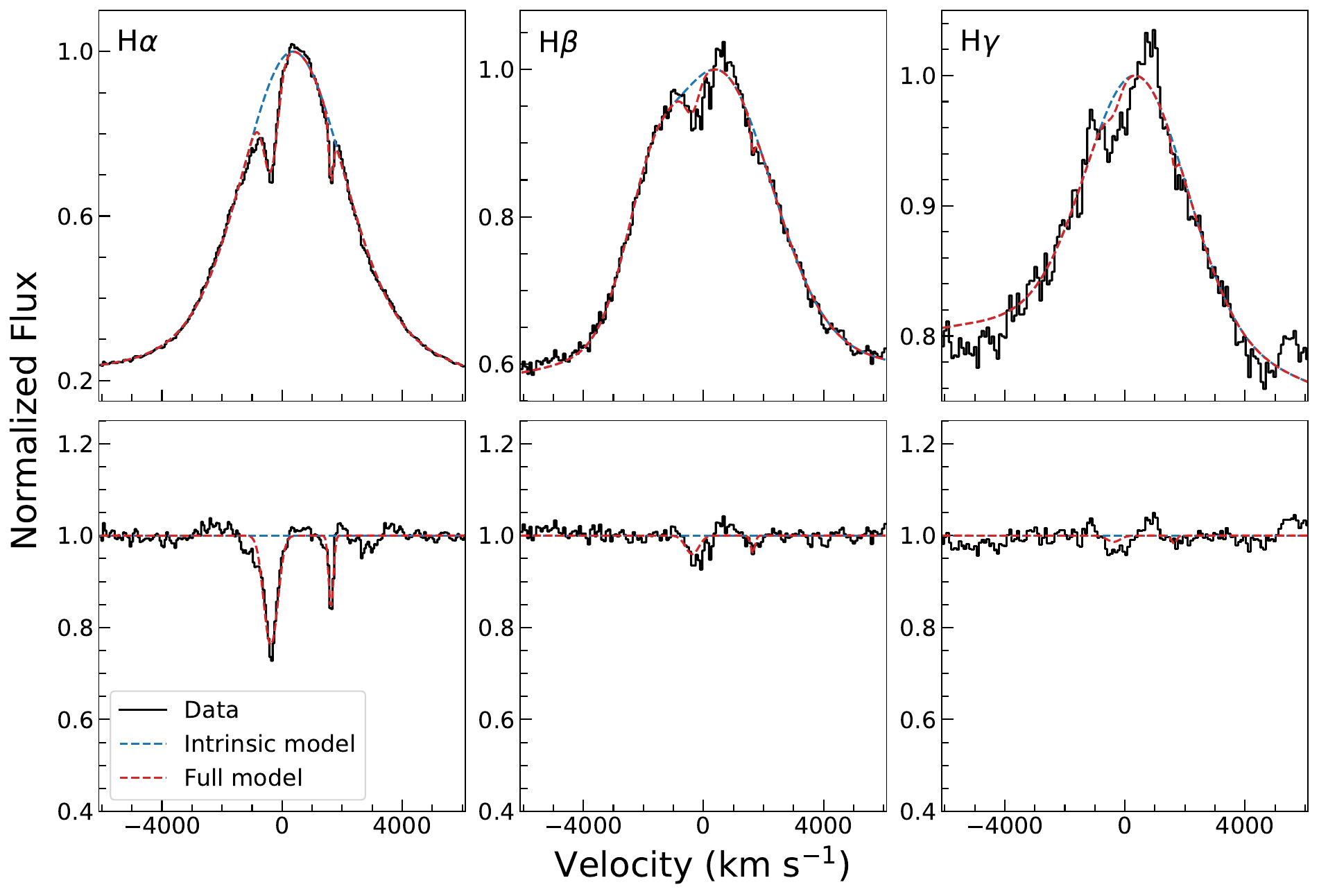}
\caption{Similar to Figure~\ref{fig:bal0925}, but for J2220-2.}
\label{fig:bal22202}
\end{figure}

For the absorption modeling, we include one blueshifted absorber component in the first epoch (Figure~\ref{fig:bal22200}) and an additional redshifted component in the last two epochs (Figures~\ref{fig:bal22201} and \ref{fig:bal22202}). The model reproduce the blueshifted \ha\ trough reasonably well, but the decomposed \hb\ troughs are systematically broader than the best-fit model in all three epochs, and the \hg\ troughs in the last two epochs are also deeper than predicted. These discrepancies are likely caused by imperfections in the narrow-line template, which can significantly bias the inferred absorption profiles of \ha, \hb, and \hg\ through blending with the narrow Balmer lines and, for \ha, the \NII\ doublet. Consequently, although the formal fit implies that J2220 is the only source with $\tau_{\mathrm{H}\alpha}<1$, we consider this result uncertain. The redshifted absorption is less affected because of its larger velocity offset, although redshifted \ha\ still partially overlaps with \NIIb. We therefore report the fitted parameters of both the blueshifted and redshifted absorbers in Table~\ref{tab:absorption}, but exclude J2220 from the physical interpretation of absorber properties in Section~\ref{ssec:abs}. Nevertheless, the rapid variability of the redshifted component over two months strongly indicates short-timescale changes in the absorbing gas (Section~\ref{sssec:v2}).

\subsection{AGN Emission-line and Continuum Properties}
\label{ssec:agn}

We compare our AGN emission-line and continuum measurements with those reported for the parent sample in \cite{Liu2019}, who performed a careful spectral decomposition for the full sample and adopted techniques similar to ours, including constructing a narrow-line template and decomposing the \ha\ and \hb\ regions. Overall, we find good agreement in the continuum luminosities and broad-line fluxes. However, our measured broad Balmer FWHMs for J0925, J1025, J1039, and J1545 tend to be $\gtrsim 30\%$ lower. This difference is likely because these Balmer lines are intrinsically narrow and/or strongly blended with the narrow-line components, making the broad--narrow decomposition sensitive to details of the modeling. In contrast, our broad Balmer FWHM for J1535 is $\sim 30\%$ higher, possibly because we explicitly model the Balmer absorption, which affects the recovered broad-line profile. Finally, the narrow-line measurements and the inferred \rfeii\ values are qualitatively consistent between our work and \cite{Liu2019}, although fitting the large sample suffers more uncertainties.

\subsubsection{Eigenvector 1 Sequence}
\label{sssec:eigen}

\begin{figure*}
\includegraphics[width=0.95\textwidth]{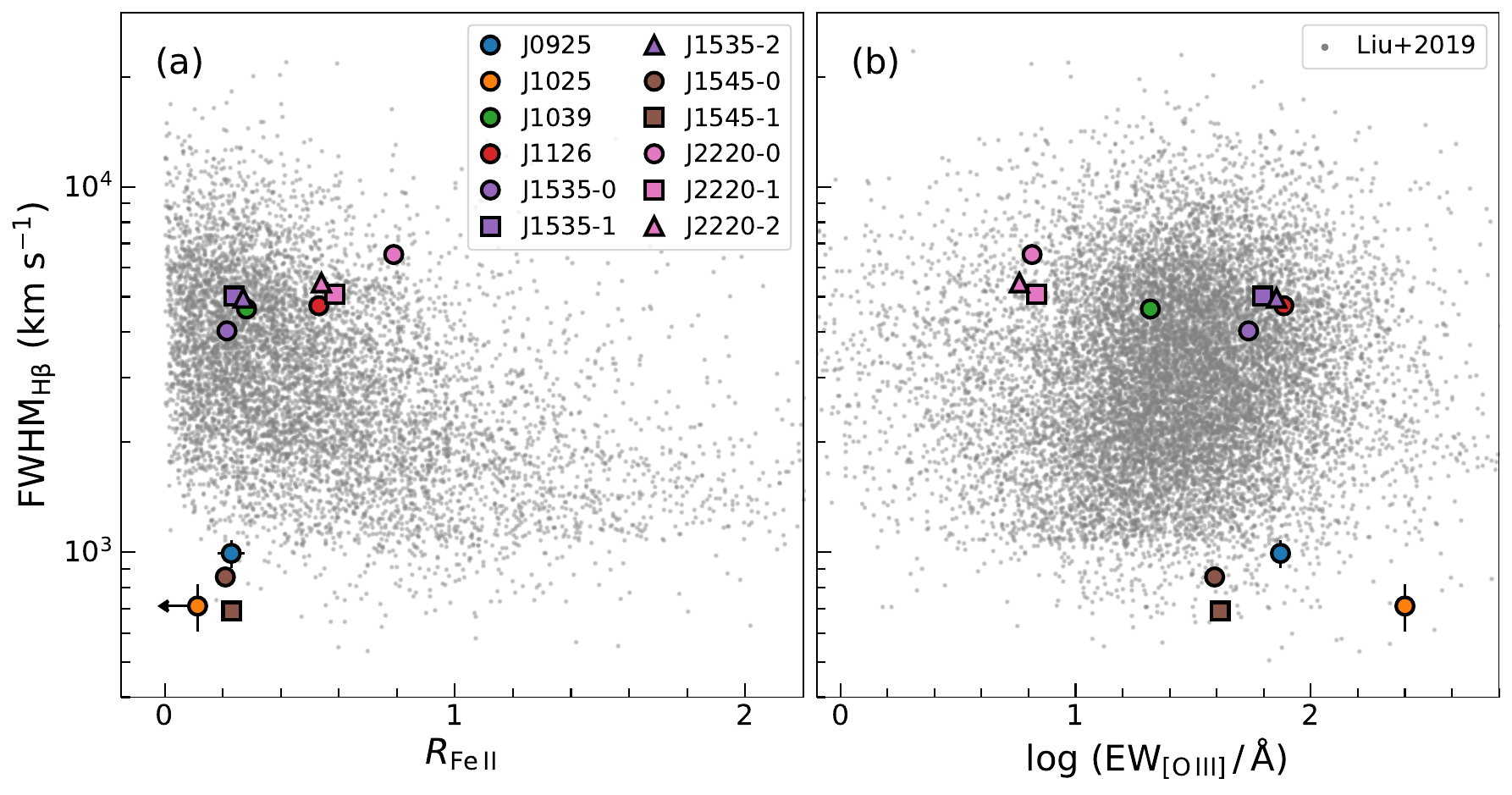}
\caption{Distribution of broad \hb\ line width ($\mathrm{FWHM}_{\mathrm{H}\beta}$) versus (a) the Fe\,{\sc ii} emission strength (\rfeii) and (b) the EW of \OIIIb\ for the parent AGN sample. Sources with Balmer absorption are highlighted in color. Three sources (J0925, J1025, and J1545) stand out in the lower-left corner of panel (a) and the lower-right corner of panel (b), while J1039, J1126, J1535, and J2220 fall within the main locus of typical broad-line AGNs.}
\label{fig:comp1}
\end{figure*}

\begin{figure}
\centering
\includegraphics[width=0.95\linewidth]{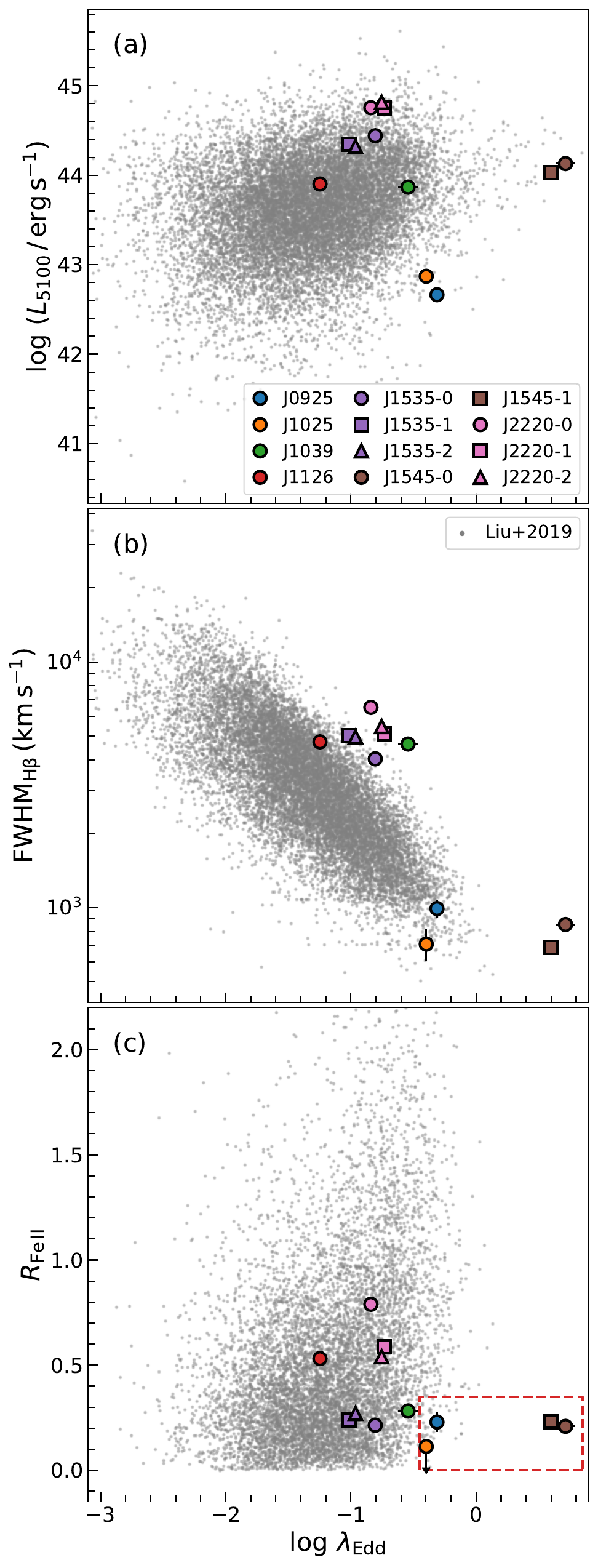}
\caption{Variation with Eddington ratio of (a) $L_{5100}$, (b) $\rm FWHM_{H\beta}$, and (c) \rfeii\ for the parent AGN sample. The sources with Balmer absorption, highlighted in color, show diverse properties. Four sources resemble the parent population, but three sources (dashed red box) occupy the regime of high Eddington ratio in panel (c).}
\label{fig:comp2}
\end{figure}

Figure~\ref{fig:comp1} compares the velocity width of broad \hb\ (\fwb) with \rfeii\ and the EW of \OIIIb\ of our Balmer-absorption-selected AGNs with those of the parent sample. These quantities are widely used to describe the AGN eigenvector~1 sequence \citep{Boroson1992,Shen2014}: \rfeii\ traces the relative strength of optical \feii\ emission, \fwb\ characterizes BLR kinematics (with additional sensitivity to orientation), and \OIII\ EW reflects the strength of the narrow-line region. In the canonical interpretation, higher accretion rates (or Eddington ratios) tend to be associated with larger \rfeii, narrower broad Balmer lines, and weaker \OIII\ emission \citep{Boroson1992,Shen2014}. In apparent tension with this expectation, most of our targets show low \rfeii, and their \OIII\ EWs lie toward the high end of the parent distribution. Some of them also have relatively narrow broad Balmer lines ($\fwb \lesssim 1000$~\kms; J0925, J1025, and J1545). Taken at face value, these eigenvector~1 indicators would not place our objects among the high-accretion rate population. However, Figure~\ref{fig:comp2} shows that at least four of the seven sources (all except J1126, J1535, J2220) have among the highest Eddington ratios in the parent sample. Although J1126, J1535, and J2220 have $\mbh \gtrsim 10^8\,\Msun$, their Eddington ratios are still high compared to parent objects with similar line widths, suggesting that the disky broad-line region has a high inclination angle to the sight line. Notably, the three highest Eddington ratio sources are strong outliers from the overall \rfeii--\lEdd\ trend (Figure~\ref{fig:comp2}c): the subsample highlighted by the red dashed box occupies only 0.4\% of the parent sample among objects with a measured \rfeii. We defer further discussion until Section~\ref{ssec:stat}.

\subsubsection{Optical Emission-line Diagnostics}
\label{sssec:bpt}

\begin{figure*}
\centering
\includegraphics[width=0.95\textwidth]{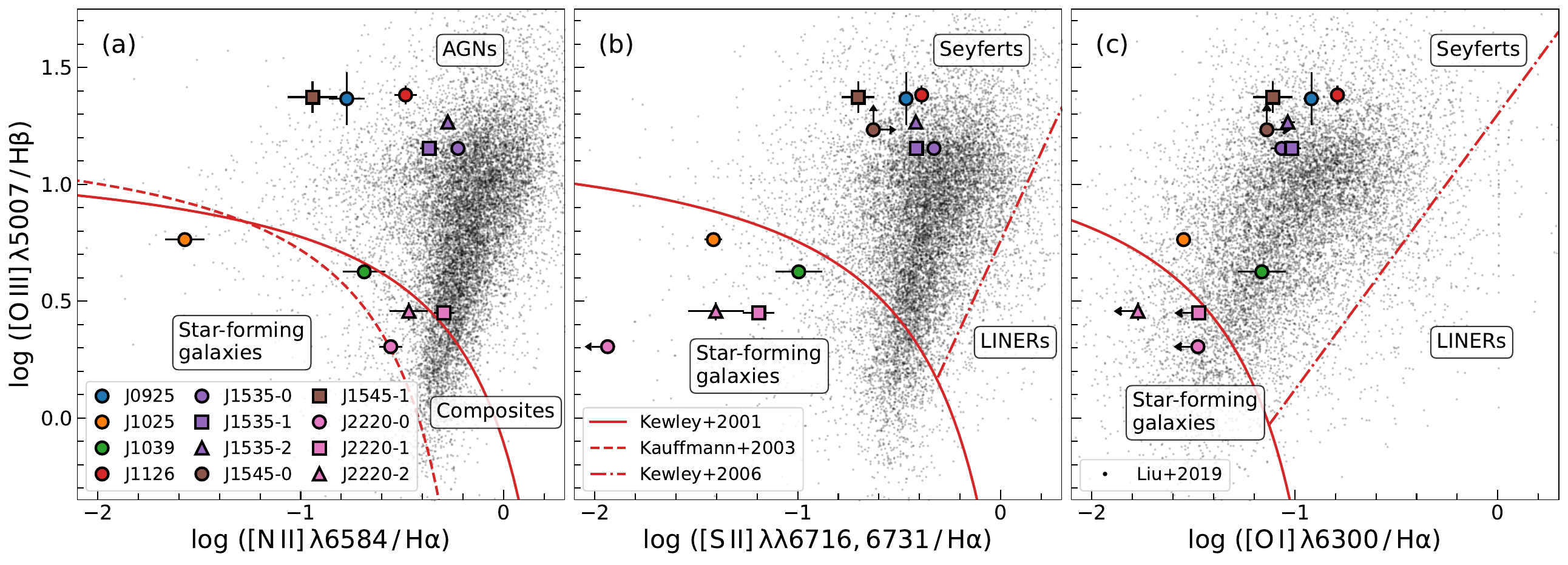}
\caption{The Balmer-absorption AGNs show diverse distribution on the optical line-intensity diagnostic diagrams of [O\,{\sc iii}]~$\lambda 5007$/H$\beta$ versus (a) [N\,{\sc ii}]~$\lambda 6584$/H$\alpha$, (b) [S\,{\sc ii}]~$\lambda\lambda 6716\, 6731$/H$\alpha$, and (c) [O\,{\sc i}]~$\lambda 6300$/H$\alpha$. The model curves of \citet[][red dashed]{Kauffmann2003} and \citet[][solid]{Kewley2001} separate star-forming galaxies from AGNs, with the latter being a more stringent ``maximum starburst'' boundary, while the distinction between high-ionization Seyferts and low-ionization nuclear emission-line regions (LINERs) is denoted by the dash-dotted line \citep{Kewley2006}.}
\label{fig:bpt}
\end{figure*}

Our targets span a wide range of locations on the optical line-intensity diagnostic diagrams (Figure~\ref{fig:bpt}), indicating a diversity in their narrow-line properties. J1126 and J1535 fall on the typical locus of Seyfert galaxies. This is perhaps not surprising, as they are the most ``normal'' objects in our sample in terms of \rfeii, Eddington ratio, the presence of strong forbidden lines, and their very broad Balmer emission.

In contrast, the remaining five sources tend to show very weak \NII, moderate \SII\ and \OI, and a variety of strengths of \OIII. The most extreme case is J1545-0, for which we can only derive an upper limit on \NII. Because we can only measure upper limits for the narrow Balmer lines in J1545-0, we do not plot it in Figure~\ref{fig:bpt}a, as both \NIIa\ and \ha\ are upper limits. The DESI spectrum of J1545-1 has much higher quality and can better characterize the host galaxy, and the SDSS and DESI measurements are largely consistent within the uncertainties. These line ratios suggest that the host galaxies of sources have low gas-phase metallicities. In particular, \NII\ is a sensitive metallicity indicator because nitrogen is a secondary element \citep{VilaCostas1993,Henry2000} and its abundance increases more steeply with metallicity than those of oxygen and sulfur \citep{Nomoto2013}. In this regard, J1025 is the most extreme source: all low-ionization lines, including \NII, \SII, and \OI, are weak, which, together with strong \OIII, suggest a highly excited ionized nebula, one with high ionization parameter and typically higher characteristic electron temperatures than in lower-excitation systems \citep{Groves2006}.

\subsection{Absorption-line Measurements}
\label{ssec:abs}

We can fit the \ha, \hb, and \hg\ absorption troughs simultaneously by tying their optical-depth ratios to the theoretical values and enforcing a common covering factor and line width. This consistency indicates that the Balmer absorption arises predominantly from a single cold absorber along the line-of-sight to the continuum and broad-line emission. The best-fit absorption parameters are listed in Table~\ref{tab:absorption}. The blueshifted absorption velocity offsets are modest: most sources show offsets of $\sim 150$--400~\kms, except for J1039 and J1535, which reach $\sim 700$--850~\kms. Such velocities are inconsistent with classical high-velocity outflows in broad absorption-line quasars and instead favor gas that is nearly stationary, or only mildly moving, in the immediate nuclear environment, plausibly on spatial scales comparable to the BLR. By contrast, the redshifted absorption component in the last two epochs of J2220 reaches the largest velocity offsets in our sample, $\sim 1500$~\kms, suggesting a more dynamic origin for this transient absorber.

Table~\ref{tab:absorption} reports only the \ha\ optical depth because the optical depths of \hb\ and \hg\ are fixed by their theoretical ratios relative to \ha. All sources, except for J2220, have $\taua \gg 1$, implying that \ha\ is highly optically thick. For J0925, J1039, and J1126, we obtain only lower limits, $\taua \gtrsim 20$; in these cases, even \hg\ becomes optically thick, and the data provide limited leverage on the exact value of $\taua$. The absorption troughs are not obviously flat-bottomed, which we attribute to the intrinsically narrow line widths ($\lesssim 50$~\kms) being comparable to, or smaller than, the SDSS instrumental resolution ($\sim 70$~\kms). This underscores the importance of convolving the intrinsic model with the instrumental LSF. It also suggests that the absorbing gas has weak internal kinematics.

\begin{figure}
\centering
\includegraphics[width=0.95\linewidth]{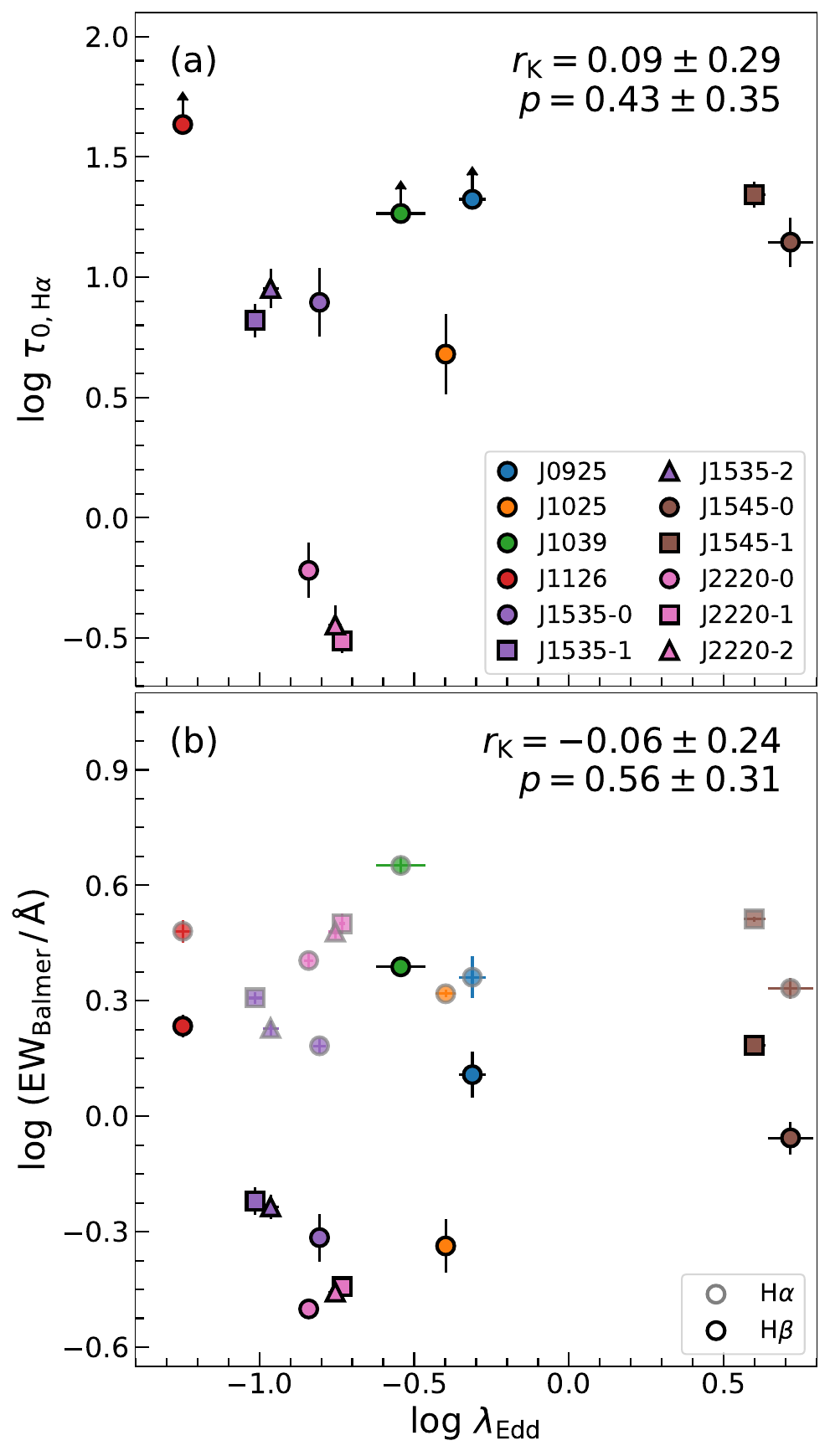}
\caption{The dependence on Eddington ratio of (a) the measured optical depth for \ha\ absorption (\taua) and (b) the EW of \ha\ and \hb\ absorption. The Kendall's rank correlation coefficient ($r_{\rm K}$) and the $p$-value are given in the upper-right corner of each panel. We exclude J2220 in the correlation tests because its absorption-line measurements are likely unreliable. Sources with lower \taua\ show larger EW difference between \ha\ and \hb.}
\label{fig:tau}
\end{figure}

We estimate Balmer-absorption EWs from the posterior samples of the model fits, reporting the median values and uncertainties. As shown in Figure~\ref{fig:tau}, neither \taua\ nor the absorption EWs exhibit a clear trend with Eddington ratio. The \ha\ absorption-line EW is approximately constant with Eddington ratio (Figure~\ref{fig:tau}b), whereas \ewb\ shows substantially larger scatter. A natural explanation is saturation: once \ha\ is highly optically thick, \ewa\ becomes primarily controlled by \cf\ and the line width (see Appendix~\ref{apx:abs}, while \hb\ typically has a lower optical depth and therefore spans a wider dynamic range in EW.

\begin{figure*}
\centering
\includegraphics[width=0.95\textwidth]{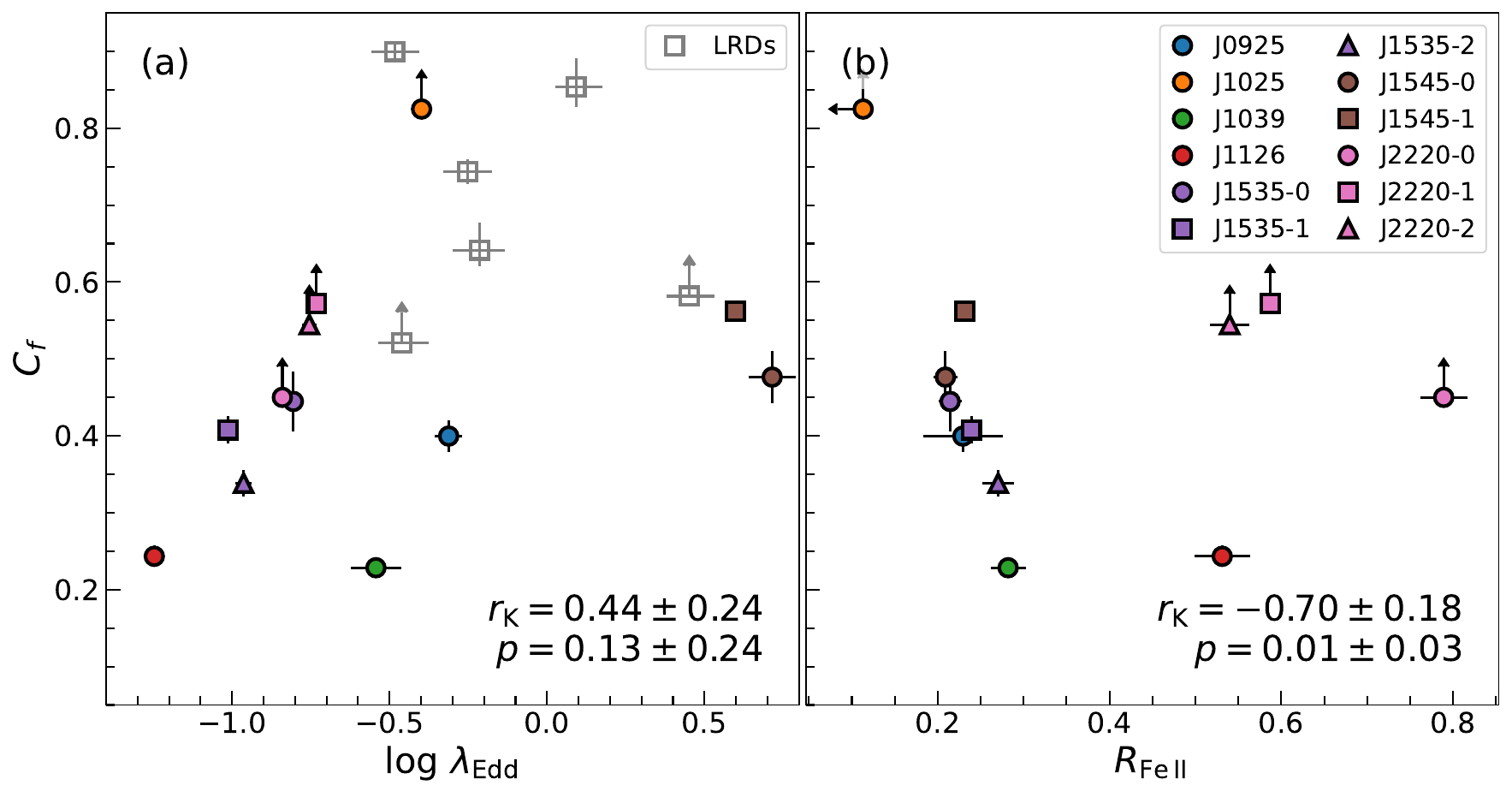}
\caption{The variation of covering factor ($C_f$) with (a) Eddington ratio and (b) \rfeii. The Kendall's rank correlation coefficient ($r_{\rm K}$) and the $p$-value are given in the lower-right corner of each panel. The correlation test in (a) excludes J1025, which shows consistently high covering factor similar to values observed in the LRDs \citep{Chen2026}. We also exclude J2220 in the correlation tests in both (a) and (b) because its absorption-line measurements are likely unreliable.}
\label{fig:cf}
\end{figure*}

By enforcing the Balmer optical-depth ratios, we can constrain the covering factor even in the saturated regime. We find that J1025, the LRD analog, has a high covering factor ($\cf \gtrsim 0.8$), whereas the remaining sources prefer moderate values, $\cf\approx 0.3-0.6$. Excluding J1025, \cf\ shows a tentative positive trend with Eddington ratio (Figure~\ref{fig:cf}a). We also plot the high-$z$ LRD measurements obtained with a similar methodology by \cite{Chen2026}, which fall close to J1025 in Figure~\ref{fig:cf}a. The four LRDs with well-constrained \cf\ have values in the range $\cf\approx 0.6$--$0.9$, while the other two provide only lower limits of $\cf \gtrsim 0.5$. It is worth noting that \citet{DEugenio2025} and \citet{DEugenio2026} reported covering fractions of $\sim 0.55$ and $\gtrsim 0.9$ for the two LRDs Abell2744-QSO1 and 159717, respectively. These values are broadly consistent with our result for J1025 and with the measurements reported by \citet{Chen2026}. We note, however, that in those two studies only the covering fraction, velocity width, and centroid shift were tied between the \ha\ and \hb\ absorption lines, while the central optical depths of the two lines were left as free parameters in the fit. More strikingly, \cf\ correlates with \rfeii. Because \rfeii\ does not correlate with Eddington ratio in the parent sample, the observed \cf--\rfeii\ relation is unlikely to reflect a simple dependence on accretion rate alone. We discuss the implications in Section~\ref{ssec:imp}.

\subsection{Photometric Properties}
\label{ssec:sed}

\begin{figure}
\centering
\includegraphics[width=0.95\linewidth]{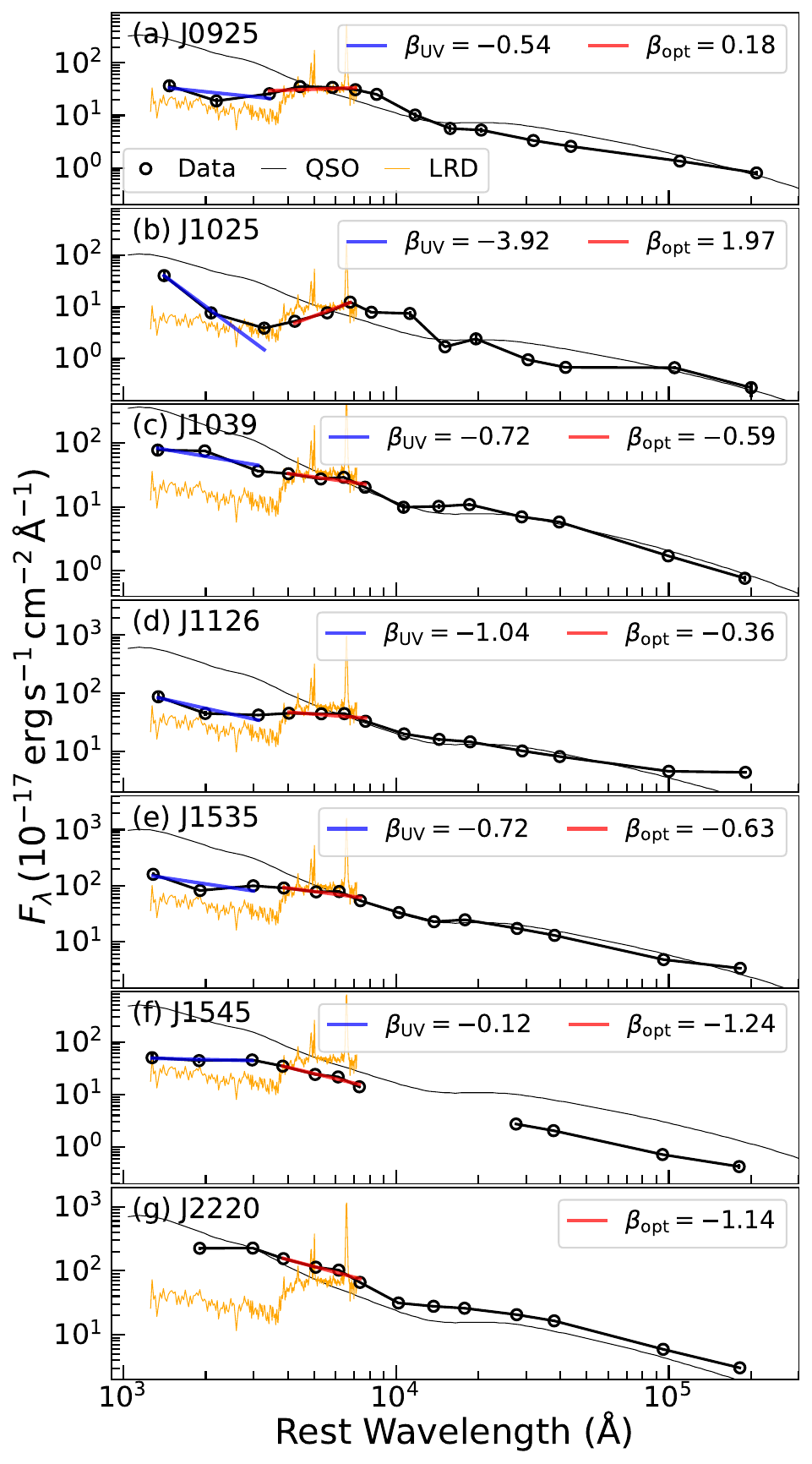}
\caption{Far-UV to mid-infrared SEDs of the Balmer-absorption AGNs. The black open circles show the observed SED. These sources show UV slopes that are redder than that of the average SDSS SED \citep[black thin curves;][]{Richards2006} but comparable to or bluer than that of the typical LRD RUBIES-EGS-49140 \citep[orange;][]{Wang2024}. The fitted power-law slopes of the UV (blue) and optical (red) SEDs are given in each panel.}
\label{fig:sed}
\end{figure}

Broad absorption-line quasars are, on average, redder than regular quasars in the rest-frame UV/optical, commonly interpreted as enhanced dust attenuation associated with their gas-rich outflows \citep{Reichard2003, Richards2003}. It is therefore of interest to compare the broad-band SEDs of Balmer-absorption AGNs to those of typical quasars and to LRDs. For LRDs in particular, a central question is whether Balmer absorption is physically linked to the characteristic ``V-shaped'' UV--optical SED, whose origin remains controversial \citep{Inayoshi2025c}. \cite{Liu2019} compiled UV, optical, near-infrared, and mid-infrared photometric measurements from the GALEX, SDSS, 2MASS, and WISE catalogs.\footnote{J1545 is detected neither in 2MASS nor in the UKIDSS survey. We include the NUV photometry of J2220 from GALEX GR6 \citep{Bianchi2014}; no FUV measurement is reported.} We use these data to examine the broad-band SEDs of our Balmer-absorption AGNs and to compare them with the color properties of LRDs (Figure~\ref{fig:sed}). We measure power-law slopes in the UV and optical bands following the definition adopted in LRD selection studies \citep{Kocevski2025}. Specifically, we fit the 1250--3645~\AA\ range to derive the UV slope ($\beta_\mathrm{UV}$) and the 3300--8000~\AA\ range to derive the optical slope ($\beta_\mathrm{opt}$). The UV wavelength range is chosen to be 100~\AA\ shorter than that in \cite{Kocevski2025} so that the GALEX far-UV band can be included for all sources.

Besides J1025, the low-$z$ LRD analog identified by \cite{Lin2026a}, we find that J0925 also satisfies the LRD color criteria, namely $\beta_\mathrm{opt}>0$ and $\beta_\mathrm{UV}<-0.37$. We note that the UV slope would likely be even steeper if photometry at shorter wavelengths were available. Our spectral decomposition further indicates that the optical continuum of J0925 is dominated by nonstellar emission, although the source is moderately resolved in the SDSS imaging. We therefore cannot exclude a non-negligible host galaxy contribution that could bias $\beta_\mathrm{opt}$ to positive values without a dedicated image decomposition. Overall, the seven objects span a wide range of SED morphologies. Their UV continua are systematically redder than the average SDSS quasar SED \citep{Richards2006} yet bluer than typical LRDs. Beyond the UV, J0925 and J1025 show a tentative flux deficit around $\sim3$~\micron, which may reflect differences in the hot-dust component (e.g., dust temperature distribution and/or covering). In contrast, the remaining sources exhibit mid-infrared SED shapes broadly consistent with those of typical quasars.  Overall, our sample exhibits diverse broad-band properties, and the presence of Balmer absorption does not appear to be uniquely or tightly associated with a V-shaped SED. Determining the intrinsic continuum shapes of Balmer-absorption AGNs will require more detailed host--AGN image decomposition and multi-component SED modeling, which we defer to future work.

\section{Discussion} 
\label{sec:disc}

\subsection{Number Statistics and Selection Effects} 
\label{ssec:stat}

Balmer-absorption AGNs are rare in the overall low-$z$ type~1 AGN population. We identify seven Balmer-absorption AGNs in the $z<0.35$ parent sample of 14{,}584 sources from \cite{Liu2019}. Taken at face value, this corresponds to a miniscule occurrence rate of $\sim 0.05\%$. However, the incidence increases substantially in a specific region of parameter space. Three of our seven sources fall in the low-\rfeii\ and high-$\lambda_\mathrm{Edd}$ locus (Figure~\ref{fig:comp2}c), which is itself uncommon among typical AGNs in the low-$z$ Universe. As discussed in Section~\ref{ssec:agn}, only $\sim 40$ objects in the \cite{Liu2019} sample with robust \feii\ measurement reside in this locus (e.g., $\log\,\lambda_\mathrm{Edd}>-0.45$ and $\rfeii<0.35$). If we restrict the parent population to this subsample, the Balmer-absorption occurrence rate increases to $\sim 10\%$, comparable to the reported occurrence rate of LRDs \citep{Matthee2024,Lin2024,Kocevski2025}. The extreme example is J1025, the low-$z$ LRD analog. The low-\rfeii\ property is also widely observed in JWST AGNs and has been interpreted as a signature of low metallicity \citep{Trefoloni2025}, suggesting that Balmer absorption may be preferentially linked to this broader population.

Even within this locus, the observed incidence can be substantially suppressed by geometric and observational selection effects. For example, J0925 and J1545 have moderate covering factors, $\cf\approx 0.4-0.6$. This seemingly implies only a $\sim 40\%-60$\% probability that the absorber intersects our line-of-sight, even if such gas is generically present. In addition, robust detection typically requires a modest velocity offset ($\gtrsim 100$~\kms) so that the absorption is not strongly blended with narrow Balmer emission; in the \ha\ region, blending with the adjacent \NIIc\ doublet poses further complications. Moreover, in all of the three sources, the inferred absorption is quite narrow ($\sigma \approx 20$--$50~\kms$), so the troughs can be substantially diluted, if not entirely missed, at modest spectral resolution. Collectively, these effects plausibly can reduce an intrinsic incidence of order tens of percent to an observed fraction of $\sim 10\%$, even if absorbing gas is common among AGNs in this region of parameter space.

The recent analysis by \cite{Lin2024} offers a comparatively unbiased measurement of Balmer absorption in high-$z$ AGNs. Using JWST/NIRCam WFSS data from ASPIRE \citep{Wang2023}, they identify 16 broad-line AGNs, three of which ($19\%$) show Balmer absorption troughs. Although their measurement of continuum slope carries substantial uncertainties, two of the three absorbers meet common LRD color criteria ($\beta_\mathrm{UV}<-0.37$ and $\beta_\mathrm{opt}>0$), while the third has a redder UV slope. More generally, $\gtrsim 60\%$ of the AGNs in their sample have $\beta_\mathrm{opt}>0$. While the role of host galaxy contamination remains to be quantified, \cite{Li2026} recently found that the intrinsic SEDs of low-luminosity JWST AGNs are redder than those of typical luminous quasars.

These selection effects are also relevant for LRDs. As shown in Figure~\ref{fig:cf}a, LRDs (including J1025) can exhibit very high covering factors, $\cf \gtrsim 0.6$. If such absorbers are common among LRDs, one may then ask why Balmer absorption is detected in fewer than half of the population. Velocity offsets and intrinsic line widths likely play a central role: absorption close to systemic velocity is easier to conceal underlying narrow emission, and intrinsically narrow troughs are washed out more readily at limited spectral resolution \citep{DEugenio2026}. To illustrate this effect, we degrade our SDSS/DESI spectra ($R \gtrsim 1800$) to the typical resolution of JWST data used to detect Balmer absorption in LRDs ($R\approx 1000$; \citealt{Chen2026}). Under this degradation, the absorption troughs in J0925 and J1025 would be missed entirely. For J1545, the first-epoch spectrum retains only a tentative dip, whereas the second epoch still shows a detectable trough because of its stronger absorber ($\cf\approx  0.6$, $\taua \gtrsim 60$). This experiment suggests that current JWST detections may be biased toward absorbers with large optical depths and/or broader intrinsic profiles, and that higher-resolution JWST spectroscopy could reveal a substantially larger fraction of LRDs with Balmer absorption. Consistent with this picture, Balmer absorption reported in high-$z$ LRDs is often relatively broad ($\gtrsim 50$~\kms), rendering them more easily detectable at moderate resolution \citep{Chen2026}. Interestingly, \citet{Lin2026b} recently reported a sample of 27 low-$z$ LRD analogs selected from DESI spectra and found Balmer absorption in 67\% of them. This high detection fraction is consistent with the above interpretation, suggesting that Balmer absorption may be more common in LRD-like systems than currently inferred from moderate-resolution JWST spectra.

\subsection{Balmer Decrement}
\label{ssec:bd}

The Balmer decrement of narrow recombination lines is commonly used to estimate dust attenuation in the host galaxy interstellar medium, because the intrinsic $\ha/\hb$ ratio is close to the Case~B recombination value and depends only weakly on gas temperature and density \citep[e.g.,][]{Hummer1987,Osterbrock1989}. This approach is widely adopted for H\,{\sc ii} regions and for AGN narrow-line regions when the narrow components can be reliably measured. In contrast, LRDs often show extreme Balmer decrements in their broad components. Early interpretations attributed the large broad-line $\ha/\hb$ to heavy dust attenuation \citep[e.g.,][]{Killi2024}. However, it has long been known that broad-line Balmer decrements in AGNs can deviate significantly from the Case~B prediction on account of intrinsic BLR physics, including high densities, radiative transfer, and collisional effects \citep[e.g.,][]{Dong2008,SchnorrMuller2016,Gaskell2017}. \cite{Yan2026} show that high Balmer decrements can be reproduced by dense, dust-free gas: at $n \gtrsim 10^8~\mathrm{cm^{-3}}$, the emergent Balmer line ratios converge to values closely following the effect of dust reddening. Consistent with this picture, \cite{Chen2026} measure emission and absorption lines for 14 LRDs with NIRSpec Prism and MSA spectra and find that the narrow components show moderate Balmer decrements, while the broad components exhibit remarkably large values. They argue that the dust obscuration is likely moderate, and that the large broad-line decrements more naturally reflect intrinsically dense gas with a low ionization parameter in the central regions.

\begin{figure*}
\centering
\includegraphics[width=0.95\textwidth]{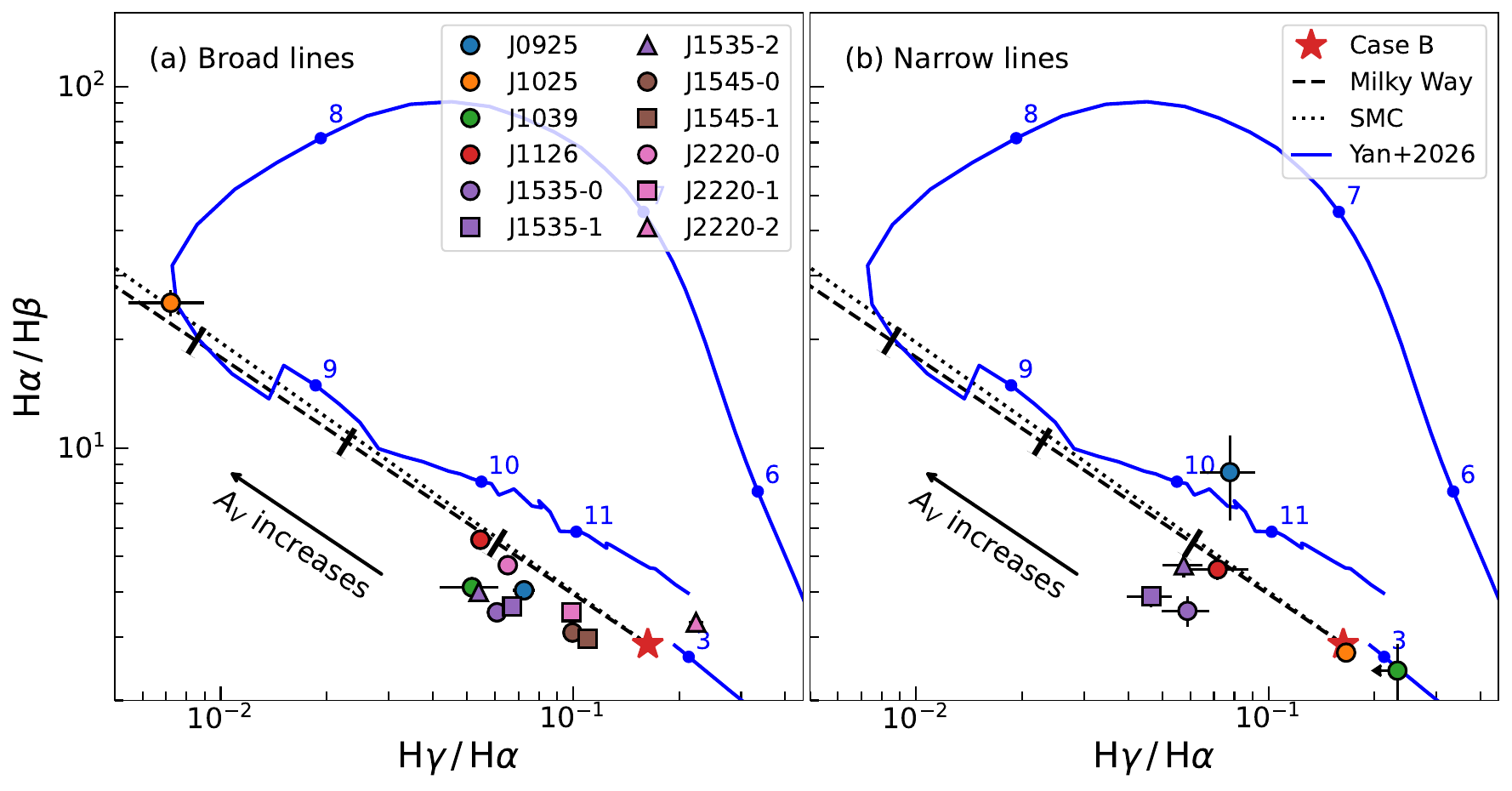}
\caption{The Balmer decrement of the (a) broad and (b) narrow lines. The colored symbols represent the Balmer-absorption AGNs. The red star indicates the line ratios expected for Case~B recombination. The black dashed and dotted lines illustrate the effect of dust attenuation following the Milky Way ($R_V=3.1$; \citealt{Cardelli1989}) and Small Magellanic Cloud \citep{Gordon2024a} attenuation laws, respectively. The black bars perpendicular to the dashed line mark the positions corresponding to $A_V=$2, 4, and 8. The blue solid curve gives the Balmer decrement predicted by \cite{Yan2026} for a typical column density of $N_\mathrm{H}=10^{23}\,\mathrm{cm^{-2}}$, with the blue points mark the volume density from $10^3$ to $10^{11}\,\mathrm{cm^{-3}}$. We exclude J1545 and J2220 in (b) because their narrow Balmer lines are weak, and we fixed their ratio in the fitting.}
\label{fig:bd}
\end{figure*}

Figure~\ref{fig:bd} shows the broad and narrow Balmer decrements of our Balmer-absorption AGNs, using $\ha/\hb$ and $\hg/\ha$. J1025, the low-$z$ LRD analog, exhibits an extreme broad-line decrement, with a very large $\ha/\hb$ and a very small $\hg/\ha$, consistent with the most extreme high-$z$ LRDs. In contrast, its narrow-line decrement is nearly identical to the Case~B expectation, suggesting that the host galaxy attenuation is modest while the broad-line ratios are strongly affected by intrinsic BLR physics. The remaining sources lie much closer to the Case~B values, and their broad and narrow decrements follow a consistent attenuation-like trend, indicating that dust in the host galaxy likely dominates the observed Balmer decrements for both components in these objects. For the multi-epoch sources (J1535, J1545, and J2220), the broad Balmer decrements of J1535 and J1545 remain largely unchanged across epochs, whereas J2220 shows a progressive increase in \hb/\ha\ and \hg/\ha, moving along the trend indicated by the red dashed curve. Such variability has been reported in both typical AGNs and changing-look AGNs \citep[e.g.,][]{Li2022,Son2025}. As discussed in Section~\ref{sssec:v2}, this behavior is unlikely to be caused by dust reddening; instead, it more likely reflects changes in the physical conditions of the BLR \citep{Korista2004,Yan2026}. The narrow Balmer lines of J1535 exhibit consistent decrement ratios across all three epochs, whereas those of J1545 and J2220 are only weakly detected; we therefore fixed their ratios in the fitting and excluded them from Figure~\ref{fig:bd}.

\subsection{Variability}
\label{ssec:var}

\subsubsection{Moderate variability in J1535 and J1545}
\label{sssec:v1}

SDSS observed J1535 three times (June 2001, May 2013, and April 2018), while J1545 has both an SDSS (July 2005) and a DESI (June 2021) spectrum. Modest but noticeable variability is seen in J1535. The source became fainter by $\sim 20\%$ in the second epoch and remained at a similar brightness in the third epoch (Table~\ref{tab:sample}). During the fainter epochs, broad \ha\ and \hb\ became slightly broader (Table~\ref{tab:broadlines}), while the inferred BH mass remained approximately unchanged. Within the measurement uncertainties, the absorption-line properties of J1535 were also consistent across the three epochs.

In contrast, J1545 showed more pronounced changes in the absorption, while the emission-line measurements stayed roughly consistent within the uncertainties. The continuum was slightly fainter in the second epoch (DESI). The broad-line properties did not exhibit obvious variability, but the absorption strengthened. For exmple, the EW of the Balmer lines increased by $\gtrsim 60\%$. The best-fit \cf\ also increased in the second-epoch spectrum. Collectively, these results suggest that both the optical depth and covering factor of the absorber can vary over a timescale of $\sim 16$~yr.

If the absorption variability is driven by changes in the ionization state, the recombination timescale of \HI\ must be shorter than the observed baseline, $\Delta t \approx 16$~yr. This requirement sets a lower limit on the electron density \citep{Barlow1992}, $n_e > (\alpha_\mathrm{rec}\Delta t)^{-1}$. Adopting Case~B recombination at $T=10^4$~K and a recombination coefficient of $\alpha_\mathrm{rec}\approx 2.59\times10^{-13}\, \mathrm{cm}^3\,\mathrm{s}^{-1}$ \citep{Draine2011} implies $n_e \gtrsim 8\times10^3\,\mathrm{cm}^{-3}$. This bound is weak, because Balmer absorption requires much higher densities to maintain a substantial \HI\ population in the $n=2$ level. For example, densities of $n_e \approx 10^6$--$10^8\, \mathrm{cm}^{-3}$ are required if Ly$\alpha$ pumping dominates \citep{Ji2012}, or even higher values are needed for collisional excitation \citep{Hall2002}. At such densities, the recombination timescale of $t_\mathrm{rec} \lesssim 0.1$~yr is far shorter than the multi-year baseline, and the level populations can in principle respond rapidly to changes in the ionizing continuum. It is therefore notable that we observe only modest variability in J1545, and essentially none in J1535, over timescales of a decade. This behavior disfavors a highly transient absorber and instead suggests a dynamically stable configuration whose global properties evolve slowly. In this sense, our findings are qualitatively consistent with the ``enshrouding'' scenarios proposed for LRDs \citep{Inayoshi2025a,Lin2026a,Inayoshi2025c}.

\subsubsection{Strong variability in J2220}
\label{sssec:v2}

Compared with J1535 and J1545, J2220 showed much stronger variability in both the continuum/broad-line emission and the Balmer absorption. Over the $\sim 16$~yr interval between epochs 0 and 1, the blueshifted absorber varied at a level comparable to that seen in J1545, with the most noticeable change being an increase in the absorption-line width from $\sigma \approx 100$ to $\sim 200$~\kms. Unlike J1535 and J1545, however, J2220 also developed a distinct redshifted Balmer absorber that produced clear troughs at least in \ha\ and \hb. More strikingly, the redshifted \ha\ absorption strengthened substantially over the short $\sim 2$ month interval between epochs 1 and 2.

\begin{figure}
\centering
\includegraphics[width=\linewidth]{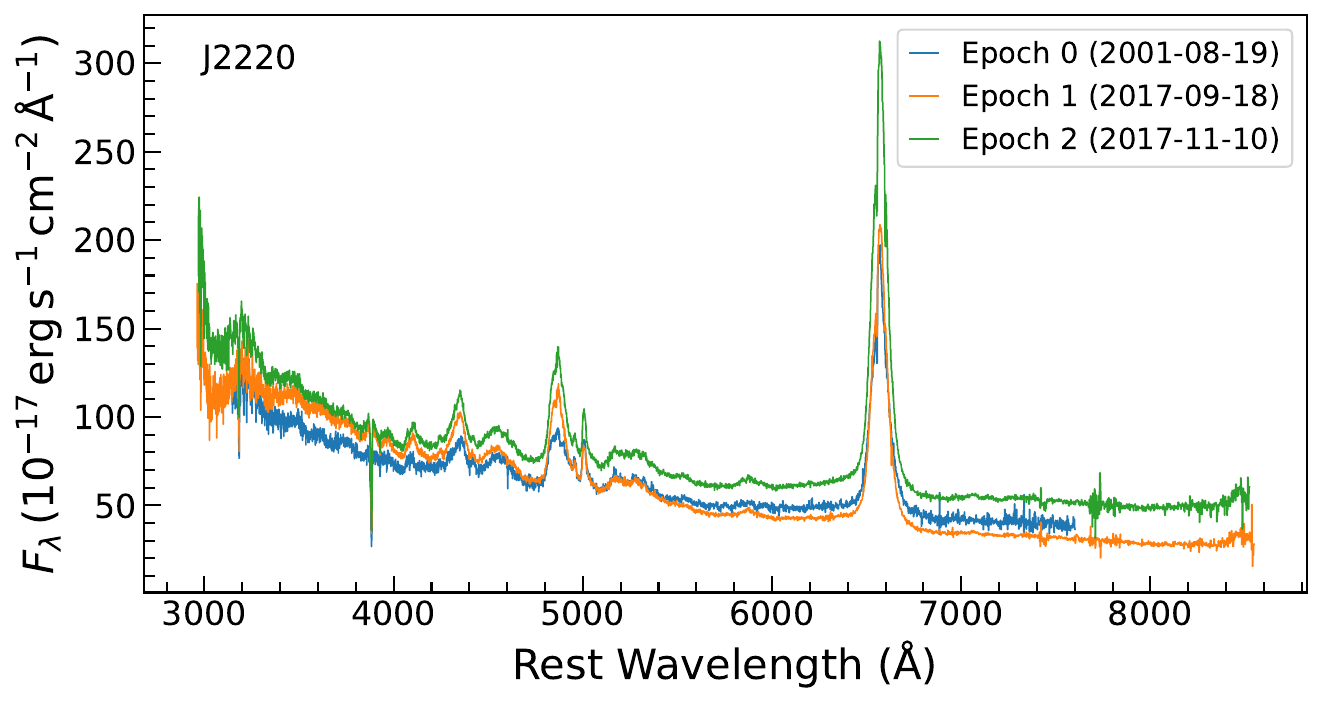}
\caption{Rest-frame optical spectra of J2220 from three SDSS epochs, shown in blue, orange, and green. The spectra illustrate the strong continuum and broad Balmer-line variability discussed in Section~\ref{sssec:v2}.}
\label{fig:spvar}
\end{figure}

\begin{figure}
\centering
\includegraphics[width=\linewidth]{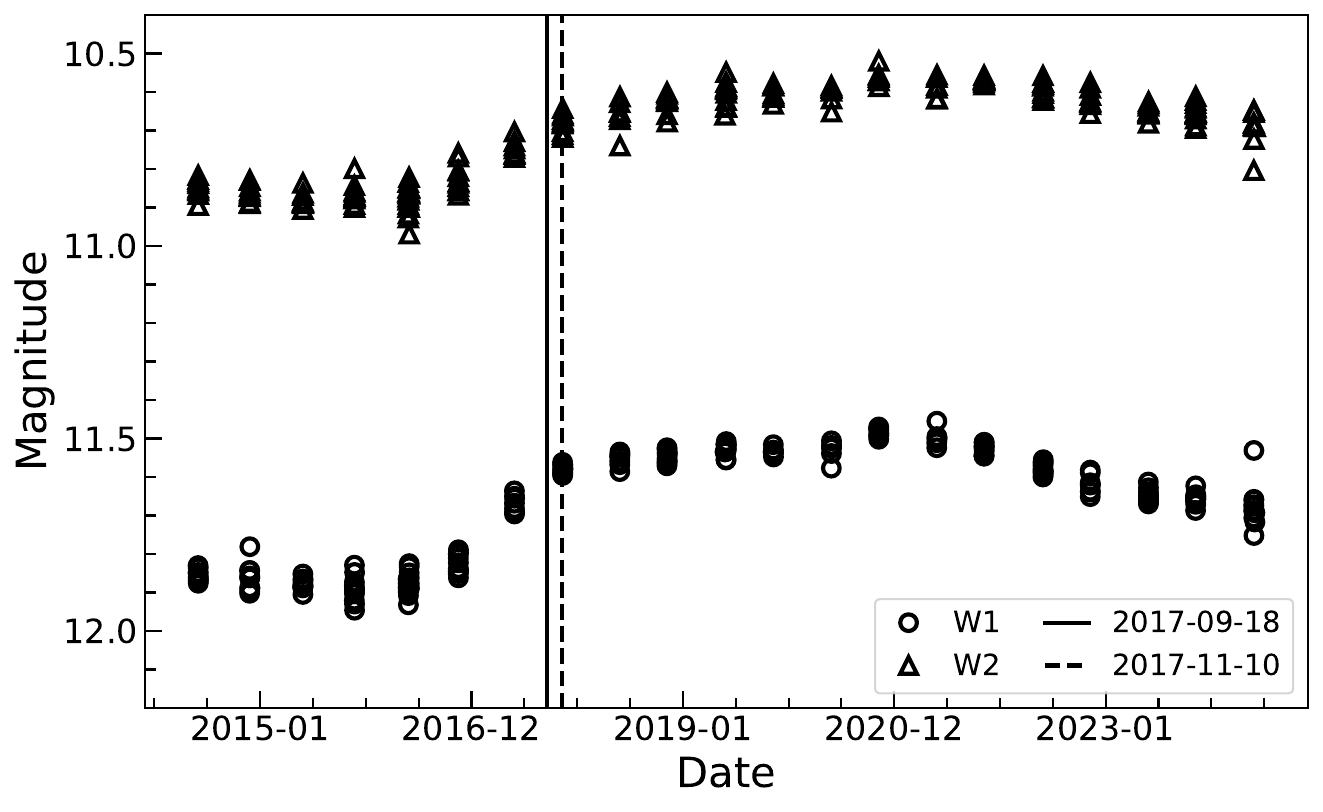}
\caption{WISE mid-infrared light curves of J2220 in the $W1$ (3.4~\micron) and $W2$ (4.6~\micron) bands. The dates of the last two SDSS spectroscopic epochs are marked by the solid and dashed vertical lines, respectively.}
\label{fig:wise}
\end{figure}

Figure~\ref{fig:spvar} shows the rest-frame optical spectra of J2220 in three epochs. The second observation (epoch 1), obtained about 16 years after epoch 0, had a continuum level at 5100~\AA\ comparable to that of epoch 0 but a noticeably bluer power-law slope. Meanwhile, the broad \ha\ flux remained roughly unchanged, whereas the broad \hb\ and \hg\ fluxes were significantly higher in epoch 1. The third observation (epoch 2), taken only about two months after epoch 1, showed a further dramatic brightening: while the UV flux remained similar to epoch 1 up to $\sim 4000$~\AA, the optical-to-near-infrared continuum increased by approximately 70\%, accompanied by clear changes in the broad Balmer-line fluxes. These behaviors indicated rapid spectral variability on both multi-year and month timescales. Figure~\ref{fig:wise} shows the WISE W1 (3.4~\micron) and W2 (4.6~\micron) light curves of J2220 \citep[e.g.,][]{Mainzer2014}. The source exhibited a clear long-term brightening trend, with an increase of $\sim 0.4$ mag in W1, followed by a gradual decline after the peak. The dates of the SDSS observations for epochs 1 and 2 fell near the end of this brightening phase, consistent with the optical spectra having captured the source during a rapidly evolving state.

The rise in the WISE light curve was unlikely to have been caused by a transient dust-clump obscuration event, because the UV continuum in the last two epochs did not show the attenuation expected from increased extinction along the line-of-sight. Instead, the coordinated optical and mid-infrared brightening more likely reflected intrinsic variability of the central engine and its reprocessed dust emission. In this context, the strong variability of the redshifted Balmer absorber may also have been linked to changes in the BH accretion state. Taking the fitted parameters at face value, the inflowing absorber in epoch~1 appeared to be highly optically thick ($\tau_{\mathrm{0,H\alpha}} \gtrsim 9$) but to cover only a small fraction of the emitting region ($\cf \approx 0.05$); by epoch~2, it appeared to evolve into a state with a more moderate optical depth ($\tau_{\mathrm{0,H\alpha}} \approx 1.5$) and a much larger covering factor ($\cf \approx 0.27$). Although these quantitative values should be interpreted with caution, the observed trend qualitatively suggested rapid structural changes in the inflowing absorbing gas.

\subsection{Physical Implications}
\label{ssec:imp}

This work is motivated by the high fraction of LRDs showing Balmer absorption lines. Recent theoretical models propose that the BH accretion flow in LRDs is enshrouded by a dense gas envelope, which can naturally produce a red optical continuum, a prominent Balmer break, and Balmer absorption lines \citep{Inayoshi2025a,Kido2025}. In these scenarios, powerful outflows expected in highly accreting systems may also help to establish and maintain the envelope, and can contribute to the weak variability and X-ray faintness reported for LRDs \citep{Kido2025}. Motivated by these ideas, we seek to test whether Balmer absorption at low redshift is preferentially associated with high accretion rates, and whether its measured properties support an envelope-like geometry.

Our low-$z$ Balmer-absorption AGNs span a range of AGN properties. Four sources (J1039, J1126, J1535, and J2220) resemble more typical broad-line AGNs and have moderate Eddington ratios of $\lambda_\mathrm{Edd} \approx 0.1$. In contrast, three sources (J0925, J1025, and J1545) reside in the low-\rfeii\ and high-$\lambda_\mathrm{Edd}$ locus of the local AGN population (Figure~\ref{fig:comp2}c). Intriguingly, they also possess narrow-line intensity ratios consistent with low-metallicity host galaxies (Figure~\ref{fig:bpt}). This combination is unusual because in the overall parent AGN population \rfeii\ typically increases with Eddington ratio along the eigenvector~1 sequence \citep{Boroson1992,Shen2014}. These three systems are closer in their global properties to the high-$z$ AGNs and LRDs uncovered by JWST \citep{Trefoloni2025}. In particular, J1025 has been identified as a local analog of LRDs \citep{Lin2026a}.

By fitting the \ha, \hb, and \hg\ absorption simultaneously, we find that the Balmer troughs are well reproduced by a single partially covering absorber in which the line-center optical depths are tied to the atomic-physics ratios. This supports a common line-of-sight structure responsible for all Balmer absorption features. Across the sample, excluding J2220, for which the fits are less robust, we measure $\taua \gtrsim 4.5$, and in the most optically thick cases we obtain lower limits of $\taua > 25$. The inferred covering factors span $\cf\approx 0.2-0.6$ for most sources, whereas the LRD analog J1025 requires a substantially higher value of $\cf \gtrsim 0.8$. We find no significant trend between \taua\ and Eddington ratio. In contrast, \cf\ shows a tentative increase with Eddington ratio among the non-LRD sources and, more strikingly, a strong inverse correlation with \rfeii, including J1025 (Figure~\ref{fig:cf}). One plausible interpretation is that the low \rfeii\ in the high-$\lambda_\mathrm{Edd}$ locus reflects low metallicity, instead of weak accretion. In a low-metallicity, high-accretion rate system, sustaining the mass supply may require a larger reservoir of dense neutral gas, which naturally raises the probability of intersecting absorbing material. Meanwhile, lower metallicity reduces the efficiency of line driving, suppressing accretion-disk winds and associated mass loss \citep{Shlosman1985,Proga2000}. If disk winds are less effective at clearing or fragmenting circumnuclear gas, the absorber can retain a high covering factor, qualitatively accounting for the observed inverse correlation between \cf\ and \rfeii.

This interpretation does not imply that iron is absent. In particular, the detection of narrow \feii\ emission and absorption in J1025 \citep{Lin2026a}, J1039, and J1126 demonstrates that iron exists in the gas. Balmer absorption is also \ preferentially found in FeLoBALQs \citep{Leighly2025}. The weak broad \feii\ pseudo-continuum may instead result from a combination of reduced enrichment and different ionization conditions in the BLR. Photoionization calculations show that optical \feii\ emission depends sensitively on metallicity and on the thermodynamic and radiative state of the line-emitting gas \citep[e.g.,][]{Sigut2003,Baldwin2004}. For example, diminished metal-line opacity and self-shielding at low metallicity can shift iron to higher ionization stages and reduce the contribution of broad optical \feii\ emission \citep{Gaskell2022}.

\section{Summary} 
\label{sec:sum}

This work is motivated by the high incidence of Balmer absorption lines reported in JWST observations of LRDs. We search for low-redshift type~1 AGNs exhibiting Balmer absorption and place them in context relative to a homogeneous parent population. Starting from the $z<0.35$ broad-line AGN catalog of \citet{Liu2019}, we visually identify Balmer-absorption candidates whose narrow features are typically unresolved in SDSS-quality spectra. We then perform a joint spectral decomposition of the continuum and emission lines, and model the Balmer absorption with a partially covering absorber model. We consider the instrument broadening effect and fit \ha, \hb, and \hg\ simultaneously with their optical-depth ratios fixed to theoretical values.

\begin{itemize}
\item Balmer-absorption AGNs are exceedingly rare in the overall low-$z$ type~1 population: we identify seven sources among 14{,}584 objects ($\sim 0.05\%$). The incidence increases substantially in the low-\rfeii\ and high-Eddington ratio locus, reaching $\sim 10\%$ when restricting to this rare region of parameter space.

\item Simultaneous fitting of the \ha, \hb, and \hg\ absorption lines shows that the absorption profiles are well described by a single partially covering absorber with tied centroid and width, and with optical depths fixed to the theoretical ratios. This supports a common line-of-sight origin for the Balmer absorption features.

\item All but one source require optically thick \ha\ absorption ($\tau_{\mathrm{H}\alpha}\gtrsim 4.5$), with lower limits $\tau_{\mathrm{H}\alpha}\gtrsim 25$ in the most saturated cases. The inferred covering factors are typically moderate ($C_f\approx 0.2-0.6$), while the local LRD analog J1025 shows a high covering factor ($C_f \gtrsim 0.8$), consistent with high-$z$ LRDs studied in the companion study by \cite{Chen2026}.

\item The absorption lines are intrinsically narrow ($\sigma\approx 20-200$~\kms) and often unresolved in SDSS-quality spectra, requiring convolution with the instrumental line-spread function. Detectability also depends on velocity offset and on blending with narrow Balmer emission and the \NIIc\ doublet, leading to strong selection effects at moderate resolution. In line with this, degrading our SDSS/DESI spectra to JWST prism-like resolution would cause some absorbers to be missed, implying that current JWST detections may be biased toward broader and/or more optically thick systems.

\item Three sources (J0925, J1025, and J1545) occupy an unusual regime of high Eddington ratio but low \rfeii, and their location on the optical line-intensity diagnostic diagrams indicate low-metallicity environments. These systems resemble high-$z$ AGNs and LRDs more closely than typical low-$z$ broad-line AGNs. We find no significant correlation between $\tau_{\mathrm{H}\alpha}$ and Eddington ratio, but $C_f$ increases with Eddington ratio for the non-LRD sources and shows a strong inverse correlation with \rfeii. We suggest that low metallicity suppresses disk-wind clearing, while high accretion requires a large nuclear gas reservoir; together these effects increase the line-of-sight covering factor and hence the likelihood of detecting Balmer absorption in AGNs of low metallicity and high Eddington ratio.

\item For the three objects with available multi-epoch spectroscopy, J1535 shows only modest continuum variability and no significant absorption variability over three SDSS epochs, while J1545 shows strengthened absorption between the 2005 SDSS and 2021 DESI spectra despite only minor changes in the broad emission lines. J2220, in contrast, exhibits much stronger variability: the continuum and broad Balmer lines vary substantially on both $\sim 16$ yr and $\sim 2$ month timescales, and a rapidly varying redshifted Balmer absorber emerges in the 2017 spectra. The limited decade-scale variability in J1535 and J1545, despite short recombination times at the densities required for Balmer absorption, suggests that their absorbers reside in dynamically stable configurations whose global properties evolve slowly, whereas J2220 likely traces a more rapidly evolving accretion/absorption state.
\end{itemize}

\begin{acknowledgments}
JS thanks Zhicheng He for helpful discussions. This work is supported by the China Manned Space Program (CMS-CSST-2025-A09), the National Science Foundation of China (12233001, 12573014), the National Key R\&D Program of China (2022YFF0503401), and ``The Fundamental Research Funds for the Central Universities, Peking University'' (7100604896). This research uses services or data provided by the SPectra Analysis and Retrievable Catalog Lab (SPARCL), which is part of the Community Science and Data Center (CSDC) program at NSF National Optical-Infrared Astronomy Research Laboratory. NOIRLab is operated by the Association of Universities for Research in Astronomy (AURA), Inc. under a cooperative agreement with the National Science Foundation.
\end{acknowledgments}

\facilities{SDSS, DESI, GALEX, WISE, Astro Data Lab}

\software{astropy \citep{Astropy2022}, SPARCL \citep{Juneau2025}, dust\_extinction \citep{Gordon2024b}
          }
\appendix

\section{Spectral Modeling Components}
\label{apx:galspec}

This appendix describes the key spectral modeling components implemented in the \galspec\ package for fitting astronomical spectra, particularly those of AGNs.

\subsection{Multi-Gaussian Line Models}
\label{apx:mgs}

The \texttt{Line\_MultiGauss} and \texttt{Line\_MultiGauss\_doublet} classes model emission lines using multiple Gaussian components, consisting of a core component and optional wind components. The single-line model is defined as

\begin{equation}
F(\lambda) = F_{\rm core}(\lambda) + \sum_{i=0}^{N-1} F_{{\rm wind},i}(\lambda),
\end{equation}

\noindent
where the core component is

\begin{equation}
F_{\rm core}(\lambda) = A_{c} \exp\left[-\frac{1}{2}\left(\frac{v(\lambda) - v_{c}}{\sigma_{c}}\right)^2\right],
\end{equation}

\noindent
and each wind component is

\begin{equation}
F_{{\rm wind},i}(\lambda) = A_{c} \, a_{{\rm w},i} \exp\left[-\frac{1}{2}\left(\frac{v(\lambda) - v_{c} - \Delta v_{{\rm w},i}}{\sigma_{{\rm w},i}}\right)^2\right].
\end{equation}

\noindent
Here, $v(\lambda) = c(\lambda - \lambda_0)/\lambda_0$ is the velocity offset from the line center $\lambda_0$, $A_{c}$ is the core amplitude, $v_{c}$ is the core velocity shift, $\sigma_{c}$ is the core velocity dispersion, and $a_{{\rm w},i}$, $\Delta v_{{\rm w},i}$, and $\sigma_{{\rm w},i}$ are the relative amplitude, velocity offset, and velocity dispersion of the $i$th wind component, respectively. The number of wind components $N$ is user-specified, with $N=0$ corresponding to a single-Gaussian profile.

For doublets (e.g., [O~III] $\lambda\lambda$4959, 5007 or [S~II] $\lambda\lambda$6716, 6731), the \texttt{Line\_MultiGauss\_doublet} class extends the same formalism by summing two kinematically tied line components:

\begin{equation}
F(\lambda) = F_0(\lambda) + F_1(\lambda),
\end{equation}

\noindent
where each line follows the same multi-Gaussian structure,

\begin{equation}
F_j(\lambda) = F_{{\rm core},j}(\lambda) + \sum_{i=0}^{N-1} F_{{\rm wind},j,i}(\lambda),
\end{equation}

\noindent
with

\begin{equation}
F_{{\rm core},j}(\lambda) = A_{{c},j} \exp\left[-\frac{1}{2}\left(\frac{v_j(\lambda) - v_{c}}{\sigma_{c}}\right)^2\right],
\end{equation}

\begin{equation}
F_{{\rm wind},j,i}(\lambda) = A_{{c},j} \, a_{{\rm w},i} \exp\left[-\frac{1}{2}\left(\frac{v_j(\lambda) - v_{c} - \Delta v_{{\rm w},i}}{\sigma_{{\rm w},i}}\right)^2\right].
\end{equation}

\noindent
The velocity offset from each line center is given by $v_j(\lambda) = c(\lambda - \lambda_j)/\lambda_j$ for $j=0,1$, and $A_{{c},0}$ and $A_{{c},1}$ are the core amplitudes of the two lines. All velocity shifts ($v_c$, $\Delta v_{{\rm w},i}$) and velocity dispersions ($\sigma_c$, $\sigma_{{\rm w},i}$) are shared between the two lines, ensuring kinematic coupling while allowing independent amplitudes.

\subsection{Gauss--Hermite models}
\label{apx:gh}

The \texttt{Line\_GaussHermite} and \texttt{Line\_GaussHermite\_doublet} classes model emission lines with a fourth-order Gauss--Hermite expansion, allowing controlled deviations from a pure Gaussian profile.

For a single line, the \texttt{Line\_GaussHermite} model is

\begin{equation}
F(\lambda) = G(\lambda)\left[1 + h_3 H_3\big(w(\lambda)\big) + h_4 H_4\big(w(\lambda)\big)\right],
\end{equation}

with

\begin{equation}
\begin{aligned}
G(\lambda) &= A\exp\left[-\frac{1}{2}w(\lambda)^2\right], \\
w(\lambda) &= \frac{v(\lambda)-v_0}{\sigma}, \\
v(\lambda) &= c\frac{\lambda-\lambda_0}{\lambda_0}.
\end{aligned}
\end{equation}

and Hermite basis functions

\begin{equation}
H_3(w)=\frac{2w^3-3w}{\sqrt{3}},
\qquad
H_4(w)=\frac{4w^4-12w^2+3}{\sqrt{24}}.
\end{equation}

Here, $A$ is the line amplitude, $v_0$ is the velocity shift, and $\sigma$ is the velocity dispersion. The coefficients $h_3$ and $h_4$ control asymmetry (skewness-like distortion) and peak/wing shape (kurtosis-like distortion), respectively. In the implementation, an optional clipping mode sets negative model values to zero.

For doublets, the \texttt{Line\_GaussHermite\_doublet} class extends this formalism by summing two kinematically tied line components:

\begin{equation}
F(\lambda)=F_0(\lambda)+F_1(\lambda),
\end{equation}

where for $j\in\{0,1\}$,

\begin{equation}
\begin{aligned}
F_j(\lambda) &= A_j\exp\left[-\frac{1}{2}w_j(\lambda)^2\right] \\
&\quad \times \left[1+h_3H_3\big(w_j(\lambda)\big)+h_4H_4\big(w_j(\lambda)\big)\right].
\end{aligned}
\end{equation}

\begin{equation}
w_j(\lambda)=\frac{v_j(\lambda)-v_0}{\sigma},
\qquad
v_j(\lambda)=c\frac{\lambda-\lambda_j}{\lambda_j}.
\end{equation}

The two amplitudes $A_0$ and $A_1$ are independent, while $v_0$, $\sigma$, $h_3$, and $h_4$ are shared between the lines, enforcing common kinematics and common non-Gaussian line-shape distortions across the doublet. As with the single-line model, optional clipping can set negative values of the total profile to zero.

\subsection{Line\_Absorption}
\label{apx:abs}

The \texttt{Line\_Absorption} class models an absorption line as a Gaussian optical-depth profile with optional partial covering of the background source. The observed (normalized) flux is given by

\begin{equation}
F_\mathrm{abs}(\lambda) = 1 - C_f + C_f \; e^{-\tau(\lambda)},
\end{equation}

\noindent
where the optical depth profile is a Gaussian in velocity,

\begin{equation}
\tau(\lambda) = \tau_0 \; \exp\left[-\frac{1}{2}\left(\frac{v(\lambda) - v_0}{\sigma}\right)^2\right],
\end{equation}

\noindent
with $v(\lambda) = c(\lambda - \lambda_0)/\lambda_0$ the velocity offset from the line center $\lambda_0$. The model is parameterized by four quantities: $\tau_0$ (dimensionless), the line-center optical depth, which sets the intrinsic absorption strength; $v_0$ (km~s$^{-1}$), the velocity shift of the line center; $\sigma$ (km~s$^{-1}$), the velocity dispersion of the absorbing gas, which sets the intrinsic line width; and $C_f$ (ranging from 0 to 1), the covering fraction. Full covering corresponds to $C_f=1$, in which case the transmitted flux is $F(\lambda)=e^{-\tau(\lambda)}$, whereas $C_f<1$ describes partial coverage where a fraction $1-C_f$ of the background emission remains unabsorbed.

This formulation is widely used to model interstellar, circumgalactic, and intrinsic AGN absorption features when the absorber only partially covers the background emission source, or when line saturation makes it necessary to introduce a covering fraction. The equivalent width (EW) of the partially covering absorber model can be written analytically as

\begin{align}
\mathrm{EW} 
& \equiv \int \left( 1 - F_{\mathrm{abs}}(\lambda) \right)\, d\lambda 
= C_f \int \left[ 1 - e^{-\tau(\lambda)} \right] d\lambda \\
& = C_f \frac{\lambda_c}{c}\,\sqrt{2\pi}\,\sigma
\sum_{n=1}^{\infty}
\frac{(-1)^{n+1}\tau_0^n}{n!\sqrt{n}} .
\end{align}

\noindent
In the optically thin limit (\(\tau_0 \ll 1\)), the EW reduces to
$\mathrm{EW} \approx C_f \frac{\lambda_c}{c}\,\sqrt{2\pi}\,\sigma\,\tau_0$.
In the saturated limit (\(\tau_0 \gg 1\)), it approaches
$\mathrm{EW} \approx 2\sqrt{2}\,C_f\,\frac{\lambda_c}{c}\,\sigma\,\sqrt{\ln (\tau_0 / \ln 2)}$ \citep{Draine2011}.

\subsection{IronTemplate}
\label{apx:feii}

The \texttt{IronTemplate} class models the complex \feii\ pseudo-continuum emission from AGN using empirical templates. Two template options are available: the I~Zwicky~1 template from \citet{Boroson1992} or the Mrk~493 template from \citet{Park2022}. The model is

\begin{equation}
F(\lambda) = A \, T_{\rm Fe\,II}\left[\lambda/(1+z)\right] \otimes G(\sigma),
\end{equation}

\noindent
where $T_{\rm Fe\,II}$ is the normalized iron template, $A$ the amplitude, $z$ the redshift, and $G(\sigma)$ a Gaussian convolution kernel that accounts for the velocity dispersion $\sigma$ of the AGN BLR. The convolution accounts for the difference between the target dispersion and the intrinsic dispersion of the template ($\sigma_{\rm intr} \approx 340-380$~km~s$^{-1}$),

\begin{equation}
\sigma_{\rm conv} = \sqrt{\sigma^2 - \sigma_{\rm intr}^2}.
\end{equation}

\noindent
The template is interpolated to the observed wavelength grid after redshifting and convolution. A lower bound of $\sigma \geq 380$~km~s$^{-1}$ is enforced to prevent unphysical deconvolution.

\subsection{WindowedPowerLaw1D}

The \texttt{WindowedPowerLaw1D} class implements a power-law continuum model that is zero outside a specified wavelength range $[\lambda_{\rm min}, \lambda_{\rm max}]$:

\begin{equation}
F(\lambda) = \begin{cases}
A \left(\frac{\lambda}{\lambda_0}\right)^{-\alpha} & \text{if } \lambda_{\rm min} \leq \lambda \leq \lambda_{\rm max} \\
0 & \text{otherwise},
\end{cases}
\end{equation}

\noindent
where $A$ is the amplitude, $\lambda_0$ is the reference wavelength (typically fixed), and $\alpha$ is the power-law index. This model is useful for fitting localized continuum regions without affecting other spectral windows.

\subsection{BlackBody}
\label{apx:bb}

The \texttt{BlackBody} class models thermal emission using the Planck function

\begin{equation}
F(\lambda) = \begin{cases}
S \, B_\lambda(T) / \max[B_\lambda(T)] & \text{if } \lambda_{\rm min} \leq \lambda \leq \lambda_{\rm max} \\
0 & \text{otherwise},
\end{cases}
\end{equation}

\noindent
with the Planck function

\begin{equation}
B_\lambda(T) = \frac{2hc^2}{\lambda^5} \frac{1}{\exp({hc}/{\lambda k_{\rm B} T}) - 1}.
\end{equation}

\noindent
Here, $T$ is the temperature in Kelvin, $S$ is a scaling factor, $h$ is Planck's constant, $c$ is the speed of light, $k_{\rm B}$ is Boltzmann's constant, and the flux is normalized by its maximum value for numerical convenience. The optional wavelength limits $\lambda_{\rm min}$ and $\lambda_{\rm max}$ allow restricting the blackbody contribution to specific spectral regions, useful for modeling hot dust emission or stellar components in composite spectra.

\subsection{Line-spread Function Convolution}

The \texttt{convolve\_lsf} function accounts for instrumental broadening by convolving spectral models with a Gaussian line-spread function (LSF). This is essential when fitting spectroscopic data where the observed line profiles are broadened by the instrument's finite spectral resolution.

The LSF is characterized by a resolving power $R = \lambda/\Delta\lambda$ at a reference wavelength $\lambda_c$, where $\Delta\lambda$ is the FWHM of the instrumental profile. The Gaussian kernel has a wavelength-space standard deviation

\begin{equation}
\sigma_\lambda = \frac{\lambda_c}{2.3548\,R},
\end{equation}

\noindent
where the factor 2.3548 converts between FWHM and standard deviation.

For any spectral model $F_0(\lambda)$, the convolved model is

\begin{equation}
F(\lambda) = \int_{-\infty}^{\infty} F_0(\lambda') \, G(\lambda - \lambda'; \sigma_\lambda) \, d\lambda',
\end{equation}

\noindent
with the normalized Gaussian kernel

\begin{equation}
G(\lambda; \sigma_\lambda) = \frac{1}{\sqrt{2\pi}\sigma_\lambda} \exp\left(-\frac{\lambda^2}{2\sigma_\lambda^2}\right).
\end{equation}

\noindent
The convolution is implemented using a Gaussian filter on uniformly spaced wavelength grids. The convolved models retain metadata indicating the LSF parameters ($\lambda_c$, $R$, $\sigma_\lambda$), which can be inspected using the \texttt{find\_convolved\_submodels} function. This is particularly useful when working with composite models containing multiple convolved components.


\begin{thebibliography}{}
\expandafter\ifx\csname natexlab\endcsname\relax\def\natexlab#1{#1}\fi
\providecommand{\url}[1]{\href{#1}{#1}}
\providecommand{\dodoi}[1]{doi:~\href{http://doi.org/#1}{\nolinkurl{#1}}}
\providecommand{\doeprint}[1]{\href{http://ascl.net/#1}{\nolinkurl{http://ascl.net/#1}}}
\providecommand{\doarXiv}[1]{\href{https://arxiv.org/abs/#1}{\nolinkurl{https://arxiv.org/abs/#1}}}

\bibitem[{{Ahumada} {et~al.}(2020){Ahumada}, {Allende Prieto}, {Almeida},
  {Anders}, {Anderson}, {Andrews}, {Anguiano}, {Arcodia}, {Armengaud},
  {Aubert}, {Avila}, {Avila-Reese}, {Badenes}, {Balland}, {Barger},
  {Barrera-Ballesteros}, {Basu}, {Bautista}, {Beaton}, {Beers}, {Benavides},
  {Bender}, {Bernardi}, {Bershady}, {Beutler}, {Bidin}, {Bird}, {Bizyaev},
  {Blanc}, {Blanton}, {Boquien}, {Borissova}, {Bovy}, {Brandt}, {Brinkmann},
  {Brownstein}, {Bundy}, {Bureau}, {Burgasser}, {Burtin}, {Cano-D{\'\i}az},
  {Capasso}, {Cappellari}, {Carrera}, {Chabanier}, {Chaplin}, {Chapman},
  {Cherinka}, {Chiappini}, {Doohyun Choi}, {Chojnowski}, {Chung}, {Clerc},
  {Coffey}, {Comerford}, {Comparat}, {da Costa}, {Cousinou}, {Covey}, {Crane},
  {Cunha}, {Ilha}, {Dai}, {Damsted}, {Darling}, {Davidson}, {Davies}, {Dawson},
  {De}, {de la Macorra}, {De Lee}, {Queiroz}, {Deconto Machado}, {de la Torre},
  {Dell'Agli}, {du Mas des Bourboux}, {Diamond-Stanic}, {Dillon}, {Donor},
  {Drory}, {Duckworth}, {Dwelly}, {Ebelke}, {Eftekharzadeh}, {Davis Eigenbrot},
  {Elsworth}, {Eracleous}, {Erfanianfar}, {Escoffier}, {Fan}, {Farr},
  {Fern{\'a}ndez-Trincado}, {Feuillet}, {Finoguenov}, {Fofie},
  {Fraser-McKelvie}, {Frinchaboy}, {Fromenteau}, {Fu}, {Galbany}, {Garcia},
  {Garc{\'\i}a-Hern{\'a}ndez}, {Garma Oehmichen}, {Ge}, {Geimba Maia},
  {Geisler}, {Gelfand}, {Goddy}, {Gonzalez-Perez}, {Grabowski}, {Green},
  {Grier}, {Guo}, {Guy}, {Harding}, {Hasselquist}, {Hawken}, {Hayes}, {Hearty},
  {Hekker}, {Hogg}, {Holtzman}, {Horta}, {Hou}, {Hsieh}, {Huber}, {Hunt}, {Ider
  Chitham}, {Imig}, {Jaber}, {Jimenez Angel}, {Johnson}, {Jones},
  {J{\"o}nsson}, {Jullo}, {Kim}, {Kinemuchi}, {Kirkpatrick}, {Kite}, {Klaene},
  {Kneib}, {Kollmeier}, {Kong}, {Kounkel}, {Krishnarao}, {Lacerna}, {Lan},
  {Lane}, {Law}, {Le Goff}, {Leung}, {Lewis}, {Li}, {Lian}, {Lin}, {Long},
  {Longa-Pe{\~n}a}, {Lundgren}, {Lyke}, {Mackereth}, {MacLeod}, {Majewski},
  {Manchado}, {Maraston}, {Martini}, {Masseron}, {Masters}, {Mathur},
  {McDermid}, {Merloni}, {Merrifield}, {M{\'e}sz{\'a}ros}, {Miglio}, {Minniti},
  {Minsley}, {Miyaji}, {Mohammad}, {Mosser}, {Mueller}, {Muna},
  {Mu{\~n}oz-Guti{\'e}rrez}, {Myers}, {Nadathur}, {Nair}, {Nandra}, {Correa do
  Nascimento}, {Nevin}, {Newman}, {Nidever}, {Nitschelm}, {Noterdaeme},
  {O'Connell}, {Olmstead}, {Oravetz}, {Oravetz}, {Osorio}, {Pace}, {Padilla},
  {Palanque-Delabrouille}, \& {Palicio}}]{Ahumada2020}
{Ahumada}, R., {Allende Prieto}, C., {Almeida}, A., {et~al.} 2020, \apjs, 249,
  3

\bibitem[{{Aoki}(2010)}]{Aoki2010}
{Aoki}, K. 2010, \pasj, 62, 1333

\bibitem[{{Aoki} {et~al.}(2006){Aoki}, {Iwata}, {Ohta}, {Ando}, {Akiyama}, \&
  {Tamura}}]{Aoki2006}
{Aoki}, K., {Iwata}, I., {Ohta}, K., {et~al.} 2006, \apj, 651, 84

\bibitem[{{Astropy Collaboration} {et~al.}(2022){Astropy Collaboration},
  {Price-Whelan}, {Lim}, {Earl}, {Starkman}, {Bradley}, {Shupe}, {Patil},
  {Corrales}, {Brasseur}, {N{\"o}the}, {Donath}, {Tollerud}, {Morris},
  {Ginsburg}, {Vaher}, {Weaver}, {Tocknell}, {Jamieson}, {van Kerkwijk},
  {Robitaille}, {Merry}, {Bachetti}, {G{\"u}nther}, {Aldcroft},
  {Alvarado-Montes}, {Archibald}, {B{\'o}di}, {Bapat}, {Barentsen},
  {Baz{\'a}n}, {Biswas}, {Boquien}, {Burke}, {Cara}, {Cara}, {Conroy},
  {Conseil}, {Craig}, {Cross}, {Cruz}, {D'Eugenio}, {Dencheva}, {Devillepoix},
  {Dietrich}, {Eigenbrot}, {Erben}, {Ferreira}, {Foreman-Mackey}, {Fox},
  {Freij}, {Garg}, {Geda}, {Glattly}, {Gondhalekar}, {Gordon}, {Grant},
  {Greenfield}, {Groener}, {Guest}, {Gurovich}, {Handberg}, {Hart},
  {Hatfield-Dodds}, {Homeier}, {Hosseinzadeh}, {Jenness}, {Jones}, {Joseph},
  {Kalmbach}, {Karamehmetoglu}, {Ka{\l}uszy{\'n}ski}, {Kelley}, {Kern},
  {Kerzendorf}, {Koch}, {Kulumani}, {Lee}, {Ly}, {Ma}, {MacBride}, {Maljaars},
  {Muna}, {Murphy}, {Norman}, {O'Steen}, {Oman}, {Pacifici}, {Pascual},
  {Pascual-Granado}, {Patil}, {Perren}, {Pickering}, {Rastogi}, {Roulston},
  {Ryan}, {Rykoff}, {Sabater}, {Sakurikar}, {Salgado}, {Sanghi}, {Saunders},
  {Savchenko}, {Schwardt}, {Seifert-Eckert}, {Shih}, {Jain}, {Shukla}, {Sick},
  {Simpson}, {Singanamalla}, {Singer}, {Singhal}, {Sinha}, {Sip{\H{o}}cz},
  {Spitler}, {Stansby}, {Streicher}, {{\v{S}}umak}, {Swinbank}, {Taranu},
  {Tewary}, {Tremblay}, {de Val-Borro}, {Van Kooten}, {Vasovi{\'c}}, {Verma},
  {de Miranda Cardoso}, {Williams}, {Wilson}, {Winkel}, {Wood-Vasey}, {Xue},
  {Yoachim}, {Zhang}, {Zonca}, \& {Astropy Project Contributors}}]{Astropy2022}
{Astropy Collaboration}, {Price-Whelan}, A.~M., {Lim}, P.~L., {et~al.} 2022,
  \apj, 935, 167

\bibitem[{{Baldwin} {et~al.}(2004){Baldwin}, {Ferland}, {Korista}, {Hamann}, \&
  {LaCluyz{\'e}}}]{Baldwin2004}
{Baldwin}, J.~A., {Ferland}, G.~J., {Korista}, K.~T., {Hamann}, F., \&
  {LaCluyz{\'e}}, A. 2004, \apj, 615, 610

\bibitem[{{Baldwin} {et~al.}(1981){Baldwin}, {Phillips}, \&
  {Terlevich}}]{Baldwin1981}
{Baldwin}, J.~A., {Phillips}, M.~M., \& {Terlevich}, R. 1981, \pasp, 93, 5

\bibitem[{{Barlow} {et~al.}(1992){Barlow}, {Junkkarinen}, {Burbidge},
  {Weymann}, {Morris}, \& {Korista}}]{Barlow1992}
{Barlow}, T.~A., {Junkkarinen}, V.~T., {Burbidge}, E.~M., {et~al.} 1992, \apj,
  397, 81

\bibitem[{{Bianchi} {et~al.}(2014){Bianchi}, {Conti}, \& {Shiao}}]{Bianchi2014}
{Bianchi}, L., {Conti}, A., \& {Shiao}, B. 2014, Advances in Space Research,
  53, 900

\bibitem[{{Boroson} \& {Green}(1992)}]{Boroson1992}
{Boroson}, T.~A., \& {Green}, R.~F. 1992, \apjs, 80, 109

\bibitem[{{Brazzini} {et~al.}(2025){Brazzini}, {D'Eugenio}, {Maiolino},
  {Juod{\v{z}}balis}, {Ji}, {Scholtz}, \& {Chang}}]{Brazzini2025}
{Brazzini}, M., {D'Eugenio}, F., {Maiolino}, R., {et~al.} 2025, \mnras, 544,
  L167

\bibitem[{{Burke} {et~al.}(2021){Burke}, {Liu}, {Chen}, {Shen}, \&
  {Guo}}]{Burke2021}
{Burke}, C.~J., {Liu}, X., {Chen}, Y.-C., {Shen}, Y., \& {Guo}, H. 2021,
  \mnras, 504, 543

\bibitem[{{Cardelli} {et~al.}(1989){Cardelli}, {Clayton}, \&
  {Mathis}}]{Cardelli1989}
{Cardelli}, J.~A., {Clayton}, G.~C., \& {Mathis}, J.~S. 1989, \apj, 345, 245

\bibitem[{Chen {et~al.}(2026)Chen, Ho, Zhang, Shangguan, Korista, \&
  Li}]{Chen2026}
Chen, C.-H., Ho, L.~C., Zhang, Z., {et~al.} 2026

\bibitem[{{DESI Collaboration} {et~al.}(2025){DESI Collaboration},
  {Abdul-Karim}, {Adame}, {Aguado}, {Aguilar}, {Ahlen}, {Alam}, {Aldering},
  {Alexander}, {Alfarsy}, {Allen}, {Allende Prieto}, {Alves}, {Anand},
  {Andrade}, {Armengaud}, {Avila}, {Aviles}, {Awan}, {Bailey}, {Baleato
  Lizancos}, {Ballester}, {Bault}, {Bautista}, {BenZvi}, {Beraldo e Silva},
  {Bermejo-Climent}, {Beutler}, {Bianchi}, {Blake}, {Blum}, {Bolton}, {Bonici},
  {Brieden}, {Brodzeller}, {Brooks}, {Buckley-Geer}, {Burtin}, {Canning},
  {Carnero Rosell}, {Carr}, {Carrilho}, {Casas}, {Castander}, {Cereskaite},
  {Cervantes-Cota}, {Chaussidon}, {Chaves-Montero}, {Chen}, {Chen},
  {Claybaugh}, {Cole}, {Cooper}, {Cousinou}, {Cuceu}, {Davis}, {Dawson}, {de
  Belsunce}, {de la Cruz}, {de la Macorra}, {de Mattia}, {Deiosso}, {Della
  Costa}, {Demina}, {Demirbozan}, {DeRose}, {Dey}, {Dey}, {Ding}, {Ding},
  {Doel}, {Douglass}, {Dowicz}, {Ebina}, {Edelstein}, {Eisenstein}, {Elbers},
  {Emas}, {Escoffier}, {Fagrelius}, {Fan}, {Fanning}, {Fawcett},
  {Fern\textbackslash'andez-Garc\textbackslash'ia}, {Ferraro}, {Findlay},
  {Font-Ribera}, {Forero-Romero}, {Forero-S\textbackslash'anchez}, {Frenk},
  {G\textbackslash''ansicke}, {Galbany}, {Garc\textbackslash'ia-Bellido},
  {Garcia-Quintero}, {Garrison},
  {Gazta\textbackslash\raisebox{-0.5ex}\textasciitildenaga},
  {Gil-Mar\textbackslash'in}, {Gnedin}, {Gontcho}, {Gonzalez-Morales},
  {Gonzalez-Perez}, {Gordon}, {Graur}, {Green}, {Gruen}, {Gsponer},
  {Guandalin}, {Gutierrez}, {Guy}, {Hahn}, {Han}, {Han}, {He},
  {Herrera-Alcantar}, {Honscheid}, {Hou}, {Howlett}, {Huterer},
  {Ir\textbackslashv\{s\}i\textbackslashv\{c\}}, {Ishak}, {Jacques}, {Jimenez},
  {Jing}, {Joachimi}, {Joudaki}, {Joyce}, {Jullo}, {Juneau},
  {Kara\textbackslashc\{c\}ayl\{\textbackslashi\}}, {Karim}, {Kehoe}, {Kent},
  {Khederlarian}, {Kirkby}, {Kisner}, {Kitaura}, {Kizhuprakkat}, {Kong},
  {Koposov}, {Kremin}, {Krolewski}, {Lahav}, {Lai}, {Lamman}, {Lan},
  {Landriau}, {Lang}, {Lange}, {Lasker}, {Le Goff}, {Le Guillou}, {Leauthaud},
  {Levi}, {Li}, {Li}, {Lodha}, {Lokken}, {Luo}, {Magneville}, {Manera},
  {Manser}, {Margala}, {Martini}, {Maus}, {McCullough}, {McDonald}, {Medina},
  {Medina-Varela}, {Meisner}, {Mena-Fern\textbackslash'andez}, {Menegas},
  {Mezcua}, {Miquel}, {Montero-Camacho}, {Moon}, {Moustakas},
  {Mu\textbackslash\raisebox{-0.5ex}\textasciitildenoz-Guti\textbackslash'errez},
  {Mu\textbackslash\raisebox{-0.5ex}\textasciitildenoz-Santos}, {Myers},
  {Myles}, {Nadathur}, {Najita}, {Napolitano}, {Newman}, {Nikakhtar},
  {Nikutta}, {Niz}, {Noriega}, {Padmanabhan}, {Paillas},
  {Palanque-Delabrouille}, {Palmese}, {Pan}, {Pan}, {Parkinson}, {Peacock},
  {Percival}, {P\textbackslash'erez-Fern\textbackslash'andez},
  {P\textbackslash'erez-R\textbackslash`afols}, \& {Peterson}}]{DESI2025}
{DESI Collaboration}, {Abdul-Karim}, M., {Adame}, A.~G., {et~al.} 2025, arXiv
  e-prints, arXiv:2503.14745

\bibitem[{{D'Eugenio} {et~al.}(2025){D'Eugenio}, {Maiolino}, {Perna}, {Uebler},
  {Ji}, {McClymont}, {Koudmani}, {Sijacki}, {Juod{\v{z}}balis}, {Scholtz},
  {Bennett}, {Bunker}, {Carniani}, {Charlot}, {Cresci}, {Curtis-Lake}, {Dalla
  Bont{\`a}}, {Jones}, {Lyu}, {Marconi}, {Mazzolari}, {Nelson}, {Parlanti},
  {Robertson}, {Schneider}, {Simmonds}, {Tacchella}, {Venturi}, {Willott},
  {Witstok}, \& {Witten}}]{DEugenio2025}
{D'Eugenio}, F., {Maiolino}, R., {Perna}, M., {et~al.} 2025, arXiv e-prints,
  arXiv:2503.11752

\bibitem[{{D'Eugenio} {et~al.}(2026){D'Eugenio}, {Juod{\v{z}}balis}, {Ji},
  {Scholtz}, {Maiolino}, {Carniani}, {Perna}, {Mazzolari}, {{\"U}bler},
  {Arribas}, {Bhatawdekar}, {Bunker}, {Cresci}, {Curtis-Lake}, {Hainline},
  {Inayoshi}, {Isobe}, {Ji}, {Johnson}, {Jones}, {Looser}, {Nelson},
  {Parlanti}, {Pusk{\'a}s}, {Rinaldi}, {Robertson}, {Rodr{\'\i}guez Del Pino},
  {Shivaei}, {Sun}, {Tacchella}, {Venturi}, {Volonteri}, {Williams}, {Willmer},
  {Willott}, \& {Witstok}}]{DEugenio2026}
{D'Eugenio}, F., {Juod{\v{z}}balis}, I., {Ji}, X., {et~al.} 2026, \mnras, 545,
  staf2117

\bibitem[{{Dong} {et~al.}(2008){Dong}, {Wang}, {Wang}, {Yuan}, {Zhou}, {Dai},
  \& {Zhang}}]{Dong2008}
{Dong}, X., {Wang}, T., {Wang}, J., {et~al.} 2008, \mnras, 383, 581

\bibitem[{{Draine}(2011)}]{Draine2011}
{Draine}, B.~T. 2011, {Physics of the Interstellar and Intergalactic Medium}

\bibitem[{{Filippenko}(1985)}]{Filippenko1985}
{Filippenko}, A.~V. 1985, \apj, 289, 475

\bibitem[{{Filippenko} \& {Halpern}(1984)}]{Filippenko1984}
{Filippenko}, A.~V., \& {Halpern}, J.~P. 1984, \apj, 285, 458

\bibitem[{{Foreman-Mackey} {et~al.}(2013){Foreman-Mackey}, {Hogg}, {Lang}, \&
  {Goodman}}]{Foreman-Mackey2013}
{Foreman-Mackey}, D., {Hogg}, D.~W., {Lang}, D., \& {Goodman}, J. 2013, \pasp,
  125, 306

\bibitem[{{Gaskell}(2017)}]{Gaskell2017}
{Gaskell}, C.~M. 2017, \mnras, 467, 226

\bibitem[{{Gaskell} \& {Ferland}(1984)}]{Gaskell1984}
{Gaskell}, C.~M., \& {Ferland}, G.~J. 1984, \pasp, 96, 393

\bibitem[{{Gaskell} {et~al.}(2022){Gaskell}, {Thakur}, {Tian}, \&
  {Saravanan}}]{Gaskell2022}
{Gaskell}, M., {Thakur}, N., {Tian}, B., \& {Saravanan}, A. 2022, Astronomische
  Nachrichten, 343, e210112

\bibitem[{{Gordon}(2024)}]{Gordon2024b}
{Gordon}, K. 2024, The Journal of Open Source Software, 9, 7023

\bibitem[{{Gordon} {et~al.}(2024){Gordon}, {Fitzpatrick}, {Massa}, {Bohlin},
  {Chastenet}, {Murray}, {Clayton}, {Lennon}, {Misselt}, \&
  {Sandstrom}}]{Gordon2024a}
{Gordon}, K.~D., {Fitzpatrick}, E.~L., {Massa}, D., {et~al.} 2024, \apj, 970,
  51

\bibitem[{{Greene} \& {Ho}(2005)}]{Greene2005}
{Greene}, J.~E., \& {Ho}, L.~C. 2005, \apj, 630, 122

\bibitem[{{Greene} {et~al.}(2024){Greene}, {Labbe}, {Goulding}, {Furtak},
  {Chemerynska}, {Kokorev}, {Dayal}, {Volonteri}, {Williams}, {Wang}, \&
  et~al.}]{Greene2024}
{Greene}, J.~E., {Labbe}, I., {Goulding}, A.~D., {et~al.} 2024, \apj, 964, 39

\bibitem[{{Groves} {et~al.}(2006){Groves}, {Heckman}, \&
  {Kauffmann}}]{Groves2006}
{Groves}, B.~A., {Heckman}, T.~M., \& {Kauffmann}, G. 2006, \mnras, 371, 1559

\bibitem[{{Hall}(2007)}]{Hall2007}
{Hall}, P.~B. 2007, \aj, 133, 1271

\bibitem[{{Hall} {et~al.}(2002){Hall}, {Anderson}, {Strauss}, {York},
  {Richards}, {Fan}, {Knapp}, {Schneider}, {Vanden Berk}, {Geballe}, {Bauer},
  {Becker}, {Davis}, {Rix}, {Nichol}, {Bahcall}, {Brinkmann}, {Brunner},
  {Connolly}, {Csabai}, {Doi}, {Fukugita}, {Gunn}, {Haiman}, {Harvanek},
  {Heckman}, {Hennessy}, {Inada}, {Ivezi{\'c}}, {Johnston}, {Kleinman},
  {Krolik}, {Krzesinski}, {Kunszt}, {Lamb}, {Long}, {Lupton}, {Miknaitis},
  {Munn}, {Narayanan}, {Neilsen}, {Newman}, {Nitta}, {Okamura}, {Pentericci},
  {Pier}, {Schlegel}, {Snedden}, {Szalay}, {Thakar}, {Tsvetanov}, {White}, \&
  {Zheng}}]{Hall2002}
{Hall}, P.~B., {Anderson}, S.~F., {Strauss}, M.~A., {et~al.} 2002, \apjs, 141,
  267

\bibitem[{{Henry} {et~al.}(2000){Henry}, {Edmunds}, \&
  {K{\"o}ppen}}]{Henry2000}
{Henry}, R.~B.~C., {Edmunds}, M.~G., \& {K{\"o}ppen}, J. 2000, \apj, 541, 660

\bibitem[{{Ho} {et~al.}(1996){Ho}, {Filippenko}, \& {Sargent}}]{Ho1996}
{Ho}, L.~C., {Filippenko}, A.~V., \& {Sargent}, W. L.~W. 1996, \apj, 462, 183

\bibitem[{{Ho} {et~al.}(1997){Ho}, {Filippenko}, \& {Sargent}}]{Ho1997}
---. 1997, \apjs, 112, 315

\bibitem[{{Ho} \& {Kim}(2009)}]{Ho2009}
{Ho}, L.~C., \& {Kim}, M. 2009, \apjs, 184, 398

\bibitem[{{Hummer} \& {Storey}(1987)}]{Hummer1987}
{Hummer}, D.~G., \& {Storey}, P.~J. 1987, \mnras, 224, 801

\bibitem[{{Hutchings} {et~al.}(2002){Hutchings}, {Crenshaw}, {Kraemer},
  {Gabel}, {Kaiser}, {Weistrop}, \& {Gull}}]{Hutchings2002}
{Hutchings}, J.~B., {Crenshaw}, D.~M., {Kraemer}, S.~B., {et~al.} 2002, \aj,
  124, 2543

\bibitem[{{Inayoshi} \& {Ho}(2025)}]{Inayoshi2025c}
{Inayoshi}, K., \& {Ho}, L.~C. 2025, arXiv e-prints, arXiv:2512.03130

\bibitem[{{Inayoshi} {et~al.}(2025){Inayoshi}, {Kimura}, \&
  {Noda}}]{Inayoshi2025b}
{Inayoshi}, K., {Kimura}, S.~S., \& {Noda}, H. 2025, \pasj, 77, 811

\bibitem[{{Inayoshi} \& {Maiolino}(2025)}]{Inayoshi2025a}
{Inayoshi}, K., \& {Maiolino}, R. 2025, \apjl, 980, L27

\bibitem[{{Ji} {et~al.}(2012){Ji}, {Wang}, {Zhou}, \& {Wang}}]{Ji2012}
{Ji}, T., {Wang}, T.-G., {Zhou}, H.-Y., \& {Wang}, H.-Y. 2012, Research in
  Astronomy and Astrophysics, 12, 369

\bibitem[{{Ji} {et~al.}(2013){Ji}, {Zhou}, {Wang}, \& {Wang}}]{Ji2013}
{Ji}, T., {Zhou}, H.-y., {Wang}, T.-g., \& {Wang}, H.-y. 2013, \caa, 37, 17

\bibitem[{{Ji} {et~al.}(2025{\natexlab{a}}){Ji}, {D'Eugenio},
  {Juod{\v{z}}balis}, {Walton}, {Fabian}, {Maiolino}, {Almeida}, {Pulido},
  {Belokurov}, {Isobe}, {Jones}, {Maraston}, {Scholtz}, {Simmonds},
  {Tacchella}, {Terlevich}, \& {Terlevich}}]{Ji2025b}
{Ji}, X., {D'Eugenio}, F., {Juod{\v{z}}balis}, I., {et~al.} 2025{\natexlab{a}},
  \mnras

\bibitem[{{Ji} {et~al.}(2025{\natexlab{b}}){Ji}, {Maiolino}, {{\"U}bler},
  {Scholtz}, {D'Eugenio}, {Sun}, {Perna}, {Turner}, {Carniani}, {Arribas}, \&
  et~al.}]{Ji2025a}
{Ji}, X., {Maiolino}, R., {{\"U}bler}, H., {et~al.} 2025{\natexlab{b}}, \mnras,
  544, 3900

\bibitem[{{Juneau} {et~al.}(2025){Juneau}, {Jacques}, {Pothier}, {Bolton},
  {Weaver}, {Pucha}, {McManus}, {Nikutta}, \& {Olsen}}]{Juneau2025}
{Juneau}, S., {Jacques}, A., {Pothier}, S., {et~al.} 2025, in Astronomical
  Society of the Pacific Conference Series, Vol. 541, Astronomical Data
  Analysis Software and Systems XXXIII, ed. A.~{Jacques}, R.~{Seaman},
  N.~{Gandilo}, \& T.~{Linder}, 77

\bibitem[{{Kauffmann} {et~al.}(2003){Kauffmann}, {Heckman}, {Tremonti},
  {Brinchmann}, {Charlot}, {White}, {Ridgway}, {Brinkmann}, {Fukugita}, {Hall},
  \& et~al.}]{Kauffmann2003}
{Kauffmann}, G., {Heckman}, T.~M., {Tremonti}, C., {et~al.} 2003, \mnras, 346,
  1055

\bibitem[{{Kewley} {et~al.}(2001){Kewley}, {Dopita}, {Sutherland}, {Heisler},
  \& {Trevena}}]{Kewley2001}
{Kewley}, L.~J., {Dopita}, M.~A., {Sutherland}, R.~S., {Heisler}, C.~A., \&
  {Trevena}, J. 2001, \apj, 556, 121

\bibitem[{{Kewley} {et~al.}(2006){Kewley}, {Groves}, {Kauffmann}, \&
  {Heckman}}]{Kewley2006}
{Kewley}, L.~J., {Groves}, B., {Kauffmann}, G., \& {Heckman}, T. 2006, \mnras,
  372, 961

\bibitem[{{Kido} {et~al.}(2025){Kido}, {Ioka}, {Hotokezaka}, {Inayoshi}, \&
  {Irwin}}]{Kido2025}
{Kido}, D., {Ioka}, K., {Hotokezaka}, K., {Inayoshi}, K., \& {Irwin}, C.~M.
  2025, \mnras, 544, 3407

\bibitem[{{Killi} {et~al.}(2024){Killi}, {Watson}, {Brammer}, {McPartland},
  {Antwi-Danso}, {Newshore}, {Coe}, {Allen}, {Fynbo}, {Gould}, \&
  et~al.}]{Killi2024}
{Killi}, M., {Watson}, D., {Brammer}, G., {et~al.} 2024, \aap, 691, A52

\bibitem[{{Kocevski} {et~al.}(2023){Kocevski}, {Onoue}, {Inayoshi}, {Trump},
  {Arrabal Haro}, {Grazian}, {Dickinson}, {Finkelstein}, {Kartaltepe},
  {Hirschmann}, \& et~al.}]{Kocevski2023}
{Kocevski}, D.~D., {Onoue}, M., {Inayoshi}, K., {et~al.} 2023, \apjl, 954, L4

\bibitem[{{Kocevski} {et~al.}(2025){Kocevski}, {Finkelstein}, {Barro},
  {Taylor}, {Calabr{\`o}}, {Laloux}, {Buchner}, {Trump}, {Leung}, {Yang}, \&
  et~al.}]{Kocevski2025}
{Kocevski}, D.~D., {Finkelstein}, S.~L., {Barro}, G., {et~al.} 2025, \apj, 986,
  126

\bibitem[{{Kokorev} {et~al.}(2024){Kokorev}, {Caputi}, {Greene}, {Dayal},
  {Trebitsch}, {Cutler}, {Fujimoto}, {Labb{\'e}}, {Miller}, {Iani}, \&
  et~al.}]{Kokorev2024}
{Kokorev}, V., {Caputi}, K.~I., {Greene}, J.~E., {et~al.} 2024, \apj, 968, 38

\bibitem[{{Kong} \& {Ho}(2018)}]{Kong2018}
{Kong}, M., \& {Ho}, L.~C. 2018, \apj, 859, 116

\bibitem[{{Korista} \& {Goad}(2004)}]{Korista2004}
{Korista}, K.~T., \& {Goad}, M.~R. 2004, \apj, 606, 749

\bibitem[{{Kuhn} {et~al.}(2024){Kuhn}, {Shangguan}, {Davies}, {Man}, {Cao},
  {Dexter}, {Eisenhauer}, {F{\"o}rster Schreiber}, {Feuchtgruber}, {Genzel}, \&
  et~al.}]{Kuhn2024}
{Kuhn}, L., {Shangguan}, J., {Davies}, R., {et~al.} 2024, \aap, 684, A52

\bibitem[{{Labb{\'e}} {et~al.}(2023){Labb{\'e}}, {van Dokkum}, {Nelson},
  {Bezanson}, {Suess}, {Leja}, {Brammer}, {Whitaker}, {Mathews}, {Stefanon}, \&
  et~al.}]{Labbe2023}
{Labb{\'e}}, I., {van Dokkum}, P., {Nelson}, E., {et~al.} 2023, \nat, 616, 266

\bibitem[{{Leighly} {et~al.}(2025){Leighly}, {Gallagher}, {Choi}, {Terndrup},
  {Voelker}, {Richards}, \& {Morabito}}]{Leighly2025}
{Leighly}, K.~M., {Gallagher}, S.~C., {Choi}, H., {et~al.} 2025, \apj, 993, 129

\bibitem[{{Li} {et~al.}(2026){Li}, {Ho}, {Wang}, \& et~al.}]{Li2026}
{Li}, G., {Ho}, L., {Wang}, R., \& et~al. 2026

\bibitem[{{Li} {et~al.}(2025){Li}, {Ho}, \& {Chen}}]{Li2025}
{Li}, R., {Ho}, L.~C., \& {Chen}, C.-H. 2025, arXiv e-prints, arXiv:2505.12867

\bibitem[{{Li} {et~al.}(2022){Li}, {Ho}, {Ricci}, {Trakhtenbrot}, {Arcavi},
  {Kara}, \& {Hiramatsu}}]{Li2022}
{Li}, R., {Ho}, L.~C., {Ricci}, C., {et~al.} 2022, \apj, 933, 70

\bibitem[{{Lin} {et~al.}(2024){Lin}, {Wang}, {Fan}, {Cai}, {Champagne}, {Sun},
  {Volonteri}, {Yang}, {Hennawi}, {Ba{\~n}ados}, {Barth}, {Eilers}, {Farina},
  {Liu}, {Jin}, {Jun}, {Lupi}, {Kakiichi}, {Mazzucchelli}, {Onoue}, {Pan},
  {Pizzati}, {Rojas-Ruiz}, {Schindler}, {Trakhtenbrot}, {Shen}, {Trebitsch},
  {Zhuang}, {Endsley}, {Meyer}, {Li}, {Li}, {Pudoka}, {Tee}, {Wu}, \&
  {Zhang}}]{Lin2024}
{Lin}, X., {Wang}, F., {Fan}, X., {et~al.} 2024, \apj, 974, 147

\bibitem[{{Lin} {et~al.}(2026){Lin}, {Fan}, {Cai}, {Bian}, {Liu}, {Sun}, {Ma},
  {Greene}, {Strauss}, {Green}, {Lyu}, {Champagne}, {Goulding}, {Inayoshi},
  {Jin}, {Leung}, {Li}, {Liu}, {Liu}, {Mao}, {Pudoka}, {Tee}, {Wang}, {Wang},
  {Wu}, {Yang}, {Zhang}, \& {Zhu}}]{Lin2026a}
{Lin}, X., {Fan}, X., {Cai}, Z., {et~al.} 2026, \apj, 997, 364

\bibitem[{Lin {et~al.}(2026)Lin, Fan, Cai, Liu, Sun, Bian, Li, Mao, Greene,
  Liu, Li, Liu, Ma, Sun, \& Zhang}]{Lin2026b}
Lin, X., Fan, X., Cai, Z., {et~al.} 2026, arXiv e-prints, arXiv:2605.21574

\bibitem[{{Liu} {et~al.}(2025){Liu}, {Jiang}, {Quataert}, {Greene}, \&
  {Ma}}]{Liu2025}
{Liu}, H., {Jiang}, Y.-F., {Quataert}, E., {Greene}, J.~E., \& {Ma}, Y. 2025,
  \apj, 994, 113

\bibitem[{{Liu} {et~al.}(2019){Liu}, {Liu}, {Dong}, {Zhou}, {Wang}, {Lu}, \&
  {Yuan}}]{Liu2019}
{Liu}, H.-Y., {Liu}, W.-J., {Dong}, X.-B., {et~al.} 2019, \apjs, 243, 21

\bibitem[{{Liu} {et~al.}(2015){Liu}, {Zhou}, {Ji}, {Yuan}, {Wang}, {Jian},
  {Shi}, {Zhang}, {Jiang}, {Shu}, \& et~al.}]{Liu2015}
{Liu}, W.-J., {Zhou}, H., {Ji}, T., {et~al.} 2015, \apjs, 217, 11

\bibitem[{{Lu} {et~al.}(2019){Lu}, {Zhao}, {Bai}, \& {Fan}}]{Lu2019}
{Lu}, K.-X., {Zhao}, Y., {Bai}, J.-M., \& {Fan}, X.-L. 2019, \mnras, 483, 1722

\bibitem[{{Mainzer} {et~al.}(2014){Mainzer}, {Bauer}, {Cutri}, {Grav},
  {Masiero}, {Beck}, {Clarkson}, {Conrow}, {Dailey}, {Eisenhardt}, {Fabinsky},
  {Fajardo-Acosta}, {Fowler}, {Gelino}, {Grillmair}, {Heinrichsen}, {Kendall},
  {Kirkpatrick}, {Liu}, {Masci}, {McCallon}, {Nugent}, {Papin}, {Rice},
  {Royer}, {Ryan}, {Sevilla}, {Sonnett}, {Stevenson}, {Thompson}, {Wheelock},
  {Wiemer}, {Wittman}, {Wright}, \& {Yan}}]{Mainzer2014}
{Mainzer}, A., {Bauer}, J., {Cutri}, R.~M., {et~al.} 2014, \apj, 792, 30

\bibitem[{{Martin} {et~al.}(2005){Martin}, {Fanson}, {Schiminovich},
  {Morrissey}, {Friedman}, {Barlow}, {Conrow}, {Grange}, {Jelinsky},
  {Milliard}, {Siegmund}, {Bianchi}, {Byun}, {Donas}, {Forster}, {Heckman},
  {Lee}, {Madore}, {Malina}, {Neff}, {Rich}, {Small}, {Surber}, {Szalay},
  {Welsh}, \& {Wyder}}]{Martin2005}
{Martin}, D.~C., {Fanson}, J., {Schiminovich}, D., {et~al.} 2005, \apjl, 619,
  L1

\bibitem[{{Matthee} {et~al.}(2024){Matthee}, {Naidu}, {Brammer}, {Chisholm},
  {Eilers}, {Goulding}, {Greene}, {Kashino}, {Labbe}, {Lilly}, \&
  et~al.}]{Matthee2024}
{Matthee}, J., {Naidu}, R.~P., {Brammer}, G., {et~al.} 2024, \apj, 963, 129

\bibitem[{{Nomoto} {et~al.}(2013){Nomoto}, {Kobayashi}, \&
  {Tominaga}}]{Nomoto2013}
{Nomoto}, K., {Kobayashi}, C., \& {Tominaga}, N. 2013, \araa, 51, 457

\bibitem[{{Ogle} {et~al.}(1999){Ogle}, {Cohen}, {Miller}, {Tran}, {Goodrich},
  \& {Martel}}]{Ogle1999}
{Ogle}, P.~M., {Cohen}, M.~H., {Miller}, J.~S., {et~al.} 1999, \apjs, 125, 1

\bibitem[{{Osterbrock}(1989)}]{Osterbrock1989}
{Osterbrock}, D.~E. 1989, {Astrophysics of gaseous nebulae and active galactic
  nuclei}

\bibitem[{{Park} {et~al.}(2022){Park}, {Barth}, {Ho}, \& {Laor}}]{Park2022}
{Park}, D., {Barth}, A.~J., {Ho}, L.~C., \& {Laor}, A. 2022, \apjs, 258, 38

\bibitem[{{Planck Collaboration} {et~al.}(2016){Planck Collaboration}, {Ade},
  {Aghanim}, {Arnaud}, {Ashdown}, {Aumont}, {Baccigalupi}, {Banday},
  {Barreiro}, {Bartlett}, \& et~al.}]{Planck2016}
{Planck Collaboration}, {Ade}, P.~A.~R., {Aghanim}, N., {et~al.} 2016, \aap,
  594, A13

\bibitem[{{Proga} {et~al.}(2000){Proga}, {Stone}, \& {Kallman}}]{Proga2000}
{Proga}, D., {Stone}, J.~M., \& {Kallman}, T.~R. 2000, \apj, 543, 686

\bibitem[{{Reichard} {et~al.}(2003){Reichard}, {Richards}, {Hall}, {Schneider},
  {Vanden Berk}, {Fan}, {York}, {Knapp}, \& {Brinkmann}}]{Reichard2003}
{Reichard}, T.~A., {Richards}, G.~T., {Hall}, P.~B., {et~al.} 2003, \aj, 126,
  2594

\bibitem[{{Richards} {et~al.}(2003){Richards}, {Hall}, {Vanden Berk},
  {Strauss}, {Schneider}, {Weinstein}, {Reichard}, {York}, {Knapp}, {Fan}, \&
  et~al.}]{Richards2003}
{Richards}, G.~T., {Hall}, P.~B., {Vanden Berk}, D.~E., {et~al.} 2003, \aj,
  126, 1131

\bibitem[{{Richards} {et~al.}(2006){Richards}, {Lacy}, {Storrie-Lombardi},
  {Hall}, {Gallagher}, {Hines}, {Fan}, {Papovich}, {Vanden Berk}, {Trammell},
  {Schneider}, {Vestergaard}, {York}, {Jester}, {Anderson}, {Budav{\'a}ri}, \&
  {Szalay}}]{Richards2006}
{Richards}, G.~T., {Lacy}, M., {Storrie-Lombardi}, L.~J., {et~al.} 2006, \apjs,
  166, 470

\bibitem[{{Rusakov} {et~al.}(2026){Rusakov}, {Watson}, {Nikopoulos}, {Brammer},
  {Gottumukkala}, {Harvey}, {Heintz}, {Damgaard}, {Sim}, {Sneppen}, \&
  et~al.}]{Rusakov2026}
{Rusakov}, V., {Watson}, D., {Nikopoulos}, G.~P., {et~al.} 2026, \nat, 649, 574

\bibitem[{{Santos} {et~al.}(2025){Santos}, {Shimizu}, {Davies}, {Cao},
  {Dexter}, {de Zeeuw}, {Eisenhauer}, {F{\"o}rster Schreiber}, {Feuchtgruber},
  {Genzel}, \& et~al.}]{Santos2025}
{Santos}, D.~J.~D., {Shimizu}, T., {Davies}, R., {et~al.} 2025, \aap, 696, A30

\bibitem[{{Schnorr-M{\"u}ller} {et~al.}(2016){Schnorr-M{\"u}ller}, {Davies},
  {Korista}, {Burtscher}, {Rosario}, {Storchi-Bergmann}, {Contursi}, {Genzel},
  {Graci{\'a}-Carpio}, {Hicks}, \& et~al.}]{SchnorrMuller2016}
{Schnorr-M{\"u}ller}, A., {Davies}, R.~I., {Korista}, K.~T., {et~al.} 2016,
  \mnras, 462, 3570

\bibitem[{{Schulze} {et~al.}(2018){Schulze}, {Misawa}, {Zuo}, \&
  {Wu}}]{Schulze2018}
{Schulze}, A., {Misawa}, T., {Zuo}, W., \& {Wu}, X.-B. 2018, \apj, 853, 167

\bibitem[{{Shen} \& {Ho}(2014)}]{Shen2014}
{Shen}, Y., \& {Ho}, L.~C. 2014, \nat, 513, 210

\bibitem[{{Shlosman} {et~al.}(1985){Shlosman}, {Vitello}, \&
  {Shaviv}}]{Shlosman1985}
{Shlosman}, I., {Vitello}, P.~A., \& {Shaviv}, G. 1985, \apj, 294, 96

\bibitem[{{Sigut} \& {Pradhan}(2003)}]{Sigut2003}
{Sigut}, T.~A.~A., \& {Pradhan}, A.~K. 2003, \apjs, 145, 15

\bibitem[{{Skrutskie} {et~al.}(2006){Skrutskie}, {Cutri}, {Stiening},
  {Weinberg}, {Schneider}, {Carpenter}, {Beichman}, {Capps}, {Chester},
  {Elias}, {Huchra}, {Liebert}, {Lonsdale}, {Monet}, {Price}, {Seitzer},
  {Jarrett}, {Kirkpatrick}, {Gizis}, {Howard}, {Evans}, {Fowler}, {Fullmer},
  {Hurt}, {Light}, {Kopan}, {Marsh}, {McCallon}, {Tam}, {Van Dyk}, \&
  {Wheelock}}]{Skrutskie2006}
{Skrutskie}, M.~F., {Cutri}, R.~M., {Stiening}, R., {et~al.} 2006, \aj, 131,
  1163

\bibitem[{{Son} {et~al.}(2025){Son}, {Kim}, {Ho}, \& {Li}}]{Son2025}
{Son}, S., {Kim}, M., {Ho}, L.~C., \& {Li}, R. 2025, \apj, 995, 37

\bibitem[{{Trefoloni} {et~al.}(2025){Trefoloni}, {Ji}, {Maiolino}, {D'Eugenio},
  {{\"U}bler}, {Scholtz}, {Marconi}, {Marconcini}, \&
  {Mazzolari}}]{Trefoloni2025}
{Trefoloni}, B., {Ji}, X., {Maiolino}, R., {et~al.} 2025, \aap, 700, A203

\bibitem[{{Vila-Costas} \& {Edmunds}(1993)}]{VilaCostas1993}
{Vila-Costas}, M.~B., \& {Edmunds}, M.~G. 1993, \mnras, 265, 199

\bibitem[{{Wang} {et~al.}(2024){Wang}, {Leja}, {de Graaff}, {Brammer},
  {Weibel}, {van Dokkum}, {Baggen}, {Suess}, {Greene}, {Bezanson}, \&
  et~al.}]{Wang2024}
{Wang}, B., {Leja}, J., {de Graaff}, A., {et~al.} 2024, \apjl, 969, L13

\bibitem[{{Wang} {et~al.}(2023){Wang}, {Yang}, {Hennawi}, {Fan}, {Sun},
  {Champagne}, {Costa}, {Habouzit}, {Endsley}, {Li}, {Lin}, {Meyer},
  {Schindler}, {Wu}, {Ba{\~n}ados}, {Barth}, {Bhowmick}, {Bieri}, {Blecha},
  {Bosman}, {Cai}, {Colina}, {Connor}, {Davies}, {Decarli}, {De Rosa}, {Drake},
  {Egami}, {Eilers}, {Evans}, {Farina}, {Haiman}, {Jiang}, {Jin}, {Jun},
  {Kakiichi}, {Khusanova}, {Kulkarni}, {Li}, {Liu}, {Loiacono}, {Lupi},
  {Mazzucchelli}, {Onoue}, {Pudoka}, {Rojas-Ruiz}, {Shen}, {Strauss}, {Tee},
  {Trakhtenbrot}, {Trebitsch}, {Venemans}, {Volonteri}, {Walter}, {Xie}, {Yue},
  {Zhang}, {Zhang}, \& {Zou}}]{Wang2023}
{Wang}, F., {Yang}, J., {Hennawi}, J.~F., {et~al.} 2023, \apjl, 951, L4

\bibitem[{{Wang} \& {Xu}(2015)}]{Wang2015}
{Wang}, J., \& {Xu}, D.~W. 2015, \aap, 573, A15

\bibitem[{{Wang} {et~al.}(2008){Wang}, {Dai}, \& {Zhou}}]{Wang2008}
{Wang}, T., {Dai}, H., \& {Zhou}, H. 2008, \apj, 674, 668

\bibitem[{{Weymann} {et~al.}(1991){Weymann}, {Morris}, {Foltz}, \&
  {Hewett}}]{Weymann1991}
{Weymann}, R.~J., {Morris}, S.~L., {Foltz}, C.~B., \& {Hewett}, P.~C. 1991,
  \apj, 373, 23

\bibitem[{{Wright} {et~al.}(2010){Wright}, {Eisenhardt}, {Mainzer}, {Ressler},
  {Cutri}, {Jarrett}, {Kirkpatrick}, {Padgett}, {McMillan}, {Skrutskie},
  {Stanford}, {Cohen}, {Walker}, {Mather}, {Leisawitz}, {Gautier}, {McLean},
  {Benford}, {Lonsdale}, {Blain}, {Mendez}, {Irace}, {Duval}, {Liu}, {Royer},
  {Heinrichsen}, {Howard}, {Shannon}, {Kendall}, {Walsh}, {Larsen}, {Cardon},
  {Schick}, {Schwalm}, {Abid}, {Fabinsky}, {Naes}, \& {Tsai}}]{Wright2010}
{Wright}, E.~L., {Eisenhardt}, P. R.~M., {Mainzer}, A.~K., {et~al.} 2010, \aj,
  140, 1868

\bibitem[{{Wu} \& {Shen}(2022)}]{Wu2022}
{Wu}, Q., \& {Shen}, Y. 2022, \apjs, 263, 42

\bibitem[{{Yan} {et~al.}(2025){Yan}, {Inayoshi}, {Chen}, \& {Guo}}]{Yan2025}
{Yan}, Z., {Inayoshi}, K., {Chen}, K., \& {Guo}, J. 2025, arXiv e-prints,
  arXiv:2512.11050

\bibitem[{{Zhang} {et~al.}(2018){Zhang}, {Zhou}, {Shi}, {Pan}, {Ji}, \&
  {Jiang}}]{Zhang2018}
{Zhang}, S., {Zhou}, H., {Shi}, X., {et~al.} 2018, \aj, 156, 4

\bibitem[{{Zhang} {et~al.}(2015){Zhang}, {Zhou}, {Shi}, {Shu}, {Liu}, {Ji},
  {Jiang}, {Sun}, {Zhou}, \& {Pan}}]{Zhang2015}
---. 2015, \apj, 815, 113

\bibitem[{{Zhuang} {et~al.}(2025){Zhuang}, {Li}, {Shen}, {Lin}, {Shapley},
  {Wang}, {Wu}, \& {Yang}}]{Zhuang2025}
{Zhuang}, M.-Y., {Li}, J., {Shen}, Y., {et~al.} 2025, arXiv e-prints,
  arXiv:2505.20393

\end{thebibliography}

\end{document}